\documentclass[twocolumn,prb,citeautoscript,superscriptaddress,8pt]{revtex4-2}

\usepackage{graphicx}
\usepackage{gensymb}
\usepackage{color}
\usepackage[dvipsnames]{xcolor}
\usepackage{float}
\usepackage{upgreek}
\usepackage{bm}
\usepackage{amsmath,amssymb}
\usepackage{mathtools}
\newcommand{\ket}[1]{\ensuremath{\left| #1 \right\rangle}}

\usepackage{hyperref}
\hypersetup{urlcolor=blue, colorlinks=true, linkcolor=blue,citecolor=blue, linktocpage=true}

\begin{document}

\title{Detection of paramagnetic spins with an ultrathin van der Waals quantum sensor}

\author{Islay O. Robertson}
\affiliation{School of Science, RMIT University, Melbourne, VIC 3001, Australia}

\author{Sam C. Scholten}
\affiliation{School of Physics, University of Melbourne, VIC 3010, Australia}
\affiliation{Centre for Quantum Computation and Communication Technology, School of Physics, University of Melbourne, VIC 3010, Australia}

\author{Priya Singh}
\affiliation{School of Science, RMIT University, Melbourne, VIC 3001, Australia}

\author{Alexander J. Healey}
\affiliation{School of Physics, University of Melbourne, VIC 3010, Australia}
\affiliation{Centre for Quantum Computation and Communication Technology, School of Physics, University of Melbourne, VIC 3010, Australia}

\author{Fernando Meneses}
\affiliation{School of Physics, University of Melbourne, VIC 3010, Australia}
\affiliation{Centre for Quantum Computation and Communication Technology, School of Physics, University of Melbourne, VIC 3010, Australia}

\author{Philipp Reineck}
\affiliation{School of Science, RMIT University, Melbourne, VIC 3001, Australia}

\author{Hiroshi Abe} % abe.hiroshi2@qst.go.jp
\affiliation{National Institutes for Quantum Science and Technology (QST), 1233 Watanuki, Takasaki, Gunma 370-1292, Japan}

\author{Takeshi Ohshima} % ohshima.takeshi@qst.go.jp>
\affiliation{National Institutes for Quantum Science and Technology (QST), 1233 Watanuki, Takasaki, Gunma 370-1292, Japan}

\author{Mehran~Kianinia} % Mehran Kianinia
% Mehran.Kianinia@uts.edu.au
\affiliation{School of Mathematical and Physical Sciences, University of Technology Sydney, Ultimo, NSW 2007, Australia}
\affiliation{ARC Centre of Excellence for Transformative Meta-Optical Systems, University of Technology Sydney, Ultimo, NSW 2007, Australia}

\author{Igor~Aharonovich}
%\email{igor.aharonovich@uts.edu.au}
\affiliation{School of Mathematical and Physical Sciences, University of Technology Sydney, Ultimo, NSW 2007, Australia}
\affiliation{ARC Centre of Excellence for Transformative Meta-Optical Systems, University of Technology Sydney, Ultimo, NSW 2007, Australia}

\author{Jean-Philippe Tetienne}
\email{jean-philippe.tetienne@rmit.edu.au}
\affiliation{School of Science, RMIT University, Melbourne, VIC 3001, Australia}

\begin{abstract} 

Detecting magnetic noise from small quantities of paramagnetic spins is a powerful capability for chemical, biochemical, and medical analysis. Quantum sensors based on optically addressable spin defects in bulk semiconductors are typically employed for such purposes, but the 3D crystal structure of the sensor inhibits the sensitivity by limiting the proximity of the defects to the target spins. Here we demonstrate the detection of paramagnetic spins using spin defects hosted in hexagonal boron nitride (hBN), a van der Waals material which can be exfoliated into the 2D regime. We first create negatively charged boron vacancy (V$_{\rm B}^-$) defects in a powder of ultrathin hBN nanoflakes ($<10$~atomic monolayers thick on average) and measure the longitudinal spin relaxation time ($T_1$) of this system. We then decorate the dry hBN nanopowder with paramagnetic Gd$^{3+}$ ions and observe a clear $T_1$ quenching, under ambient conditions, consistent with the added magnetic noise. Finally, we demonstrate the possibility of performing spin measurements including $T_1$ relaxometry using solution-suspended hBN nanopowder. 
Our results highlight the potential and versatility of the hBN quantum sensor for a range of sensing applications, and pave the way towards the realisation of a truly 2D, ultrasensitive quantum sensor.

\end{abstract}

\maketitle 

Optically addressable spin defects in solids have been a significant driver in the advancement of practical applications for highly sensitive quantum measurement and detection devices in a variety of fields including geoscience, materials science, and biology~\cite{Degen2017,Schirhagl2014,Rondin2014,Casola2018}. 
One promising capability of solid-state quantum sensors is being able to probe weak, fluctuating magnetic fields under ambient conditions and at sub-micrometer scales, which cannot typically be accessed by existing techniques such as nuclear magnetic resonance. The most prominent example of such a nanoscale sensor is the nitrogen-vacancy (NV) centre in diamond, which has been successfully employed to interrogate magnetic noise from a variety of sources, from magnons in ferromagnetic materials \cite{Du2017,McCullian2020} to electric current fluctuations in conductors \cite{Kolkowitz2015,Ariyaratne2018} to paramagnetic spins in molecular systems \cite{Steinert2013,Tetienne2013,Ermakova2013,Kaufmanna2013DetectionProbe,Sushkov2014}. The detection of paramagnetic spins, i.e. unpaired electrons that produce a fast fluctuating magnetic noise, is especially relevant in the chemical, biological and medical sciences. For example, NV-based noise sensing has been applied to detect trace amounts of ions in aqueous solutions \cite{Ziem2013,Simpson2017}, monitor the generation of free radicals in cells \cite{Sharmin2022,Nie2022}, determine iron load in ferritin proteins \cite{Grant2023}, and has been proposed as the basis for new methods for rapid and sensitive virus detection \cite{Changhao2022}. In the latter proposal and similar biosensing demonstrations \cite{Changhao2022,Rendler2017}, the biochemical signal of interest (e.g.\ the presence of a specific viral particle or a change of pH) is transduced into magnetic noise using a paramagnetic molecule, typically a gadolinium (Gd) complex, which is attached to the surface of the quantum sensor. 

The method of choice in many of these applications is measuring the longitudinal spin relaxation time ($T_1$) of the defect, which is sensitive to magnetic fluctuations at the defect's spin resonance frequency, $\omega_0$~\cite{Steinert2013,Tetienne2013,Ermakova2013,Kaufmanna2013DetectionProbe,Sushkov2014}.
The change in relaxation rate induced by a nearby paramagnetic spin, $\Gamma_1^{\rm ext}=1/T_1^{\rm ext}$, scales with the distance $d$ between the sensor (spin defect) and the target spin as $\Gamma_1^{\rm ext}\propto d^{-6}$~\cite{Steinert2013,Tetienne2013}. This strong distance dependence necessitates the defects be located close to the surface of the host crystal, in order to maximise sensitivity to spins located outside. However, spin defects in bulk 3D crystals face a practical limit, as the surface is typically plagued with dangling bonds and other sub-bandgap electronic states which compromise the charge- and photo-stability of near-surface defects~\cite{Kaviani2014,Stacey2018,Bluvstein2019}. For instance, an average minimum depth of at best or greater than $d\approx 5$\,nm is typically observed for NV ensembles created near an optimised flat diamond surface~\cite{Ziem2019,HealeyPRA2021,Liu2022}. Additionally, these unwanted surface states are often paramagnetic~\cite{Stacey2018,Sangtawesin2019} and cause a background magnetic noise obscuring the external signal of interest. These effects are especially severe in nanodiamonds with their high surface-to-volume ratio~\cite{Tetienne2013}, limiting their practical sensitivity to external signals. Given nanodiamonds are widely used for chemical and bio-sensing applications owing to their convenience and compatibility with in-solution measurements~\cite{Kaufmanna2013DetectionProbe,Sharmin2022,Nie2022,Changhao2022,Rendler2017,Miller2020}, there is a strong motivation for exploring new quantum sensing systems that may afford shorter sensor-target distances and reduced intrinsic magnetic noise while preserving the versatility of nanodiamonds.

\begin{figure*}[tb!]
\centering
\includegraphics[width=0.9\textwidth]{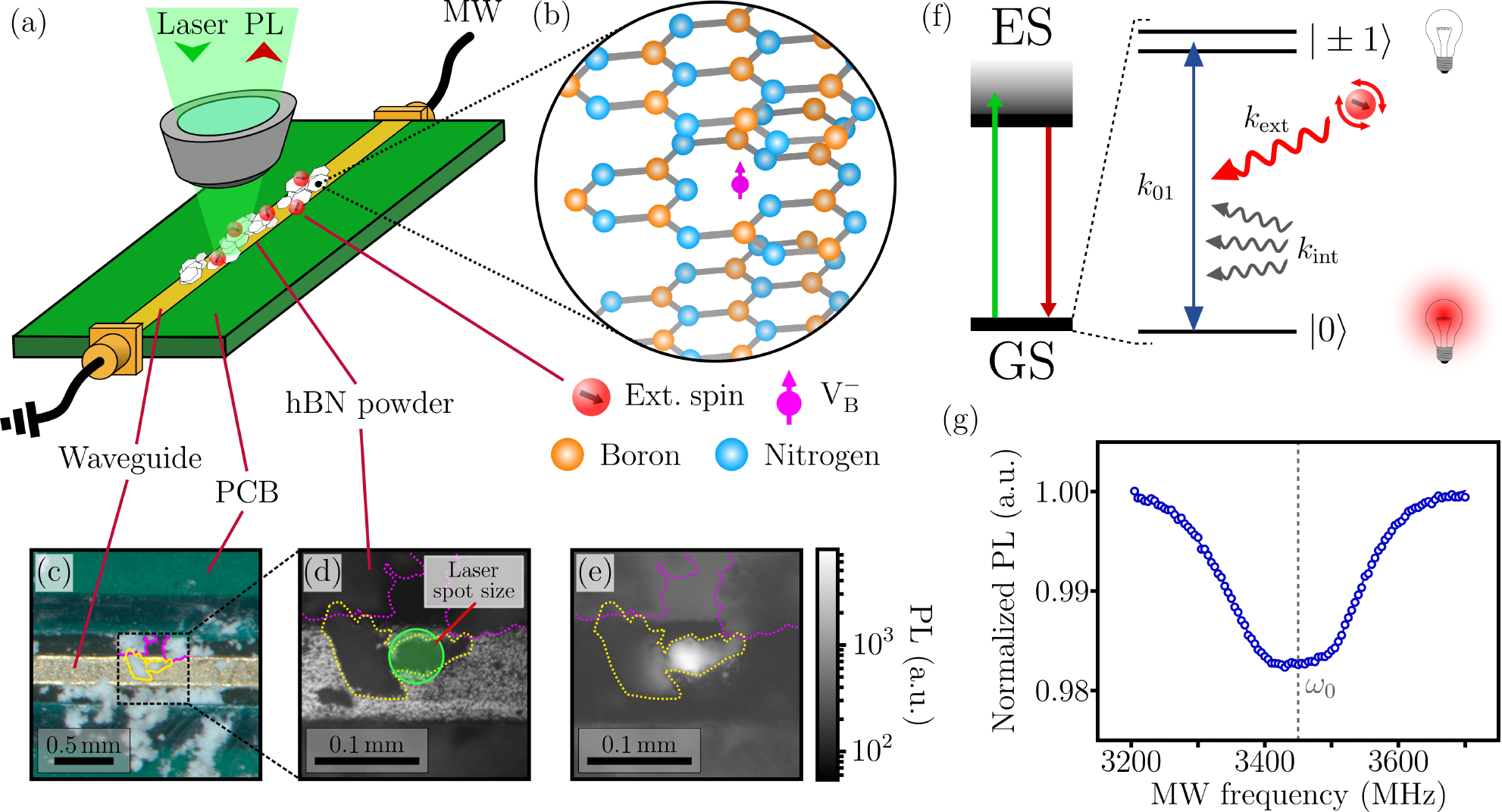}
\caption{{\bf Quantum sensing with hBN nanopowder}. (a) Schematic representation of the experimental setup. Electron-irradiated hBN nanopowder (represented as a collection of white flakes) is deposited onto a printed circuit board (PCB) with a microwave (MW) waveguide. The spin state of the V$_{\rm B}^-$ defects in the powder is optically measured via illumination with a green laser and collection of the photoluminescence (PL), which can be used to sense external paramagnetic spins (represented as red spheres) through their magnetic noise. (b) Crystal structure of the V$_{\rm B}^-$ defect in the hBN lattice. (c) True-colour photograph of hBN powder on the PCB. (d) Reflection micrograph (red channel only, displayed on a grey intensity scale) of a zoomed-in section from (c). Powder on the MW waveguide is outlined in yellow while on the PCB it is highlighted in pink. The approximate position and spot size of the laser used for PL measurements is also shown. (e) Corresponding PL image ($\lambda=750-900$\,nm spectral band) taken under laser illumination ($\lambda=532$\,nm) and displayed on a log scale. (f) Simplified energy level structure of the V$_{\rm B}^-$ defect. Laser illumination populates the excited state (ES) which undergoes spin-dependent PL to return to the ground state (GS). This process selectively populates the GS spin sub-level $\ket{0}$. The population imbalance then decays back to equilibrium with a two-way transition rate $k_{01}=k_{\text{int}}+k_{\text{ext}}$ which has both intrinsic ($k_{\text{int}}$) and external ($k_{\text{ext}}$) contributions. The observed relaxation rate out of $\ket{0}$ is given by $\Gamma_1=3k_{01}$. (g) Typical pulsed-ODMR spectrum measured for the hBN nanopowder in zero magnetic field.}
\label{fig1}
\end{figure*}

Spin defects hosted in a layered van der Waals (vdW) material could provide a solution to both of these problems. Indeed, vdW materials can often be exfoliated into atomically thin flakes while maintaining near-perfect crystallinity and defect-free surfaces. As such, it may be possible to engineer robust spin defects that reside within just a few atomic sites from the surface (i.e. $d\approx1-2$\,nm) and with no surface-induced background magnetic noise, which could open a new frontier in ultrasensitive quantum sensing. Recently, hexagonal boron nitride (hBN) has emerged as a promising material platform to realise such ultrathin quantum sensors~\cite{GottschollNM2020,Gottscholl2021NC,Xingyu2021,Liu2021,HealeyNP2022,Kumar2022,Lyu2022,Yang2022,HuangNC2022}. hBN is an exfoliable, air-stable vdW material and is host to a robust, optically addressable spin defect, the negatively charged boron vacancy (V$_{\rm B}^-$)~\cite{GottschollNM2020,Liu2022review}. The V$_{\rm B}^-$ defect can be introduced in the hBN lattice through a variety of irradiation methods \cite{Kianinia2020,Murzakhanov2021}, and several demonstrations of quantum sensing have subsequently been reported, including the detection and imaging under ambient conditions of static magnetic fields, temperature and strain~\cite{Gottscholl2021NC,HealeyNP2022,Kumar2022,Lyu2022,Yang2022}, and the imaging of magnetic noise from a ferromagnetic material at cryogenic temperatures \cite{HuangNC2022}. 

In this work, we demonstrate the first detection of magnetic noise from paramagnetic spins with an hBN quantum sensor, under ambient conditions. In contrast with previous quantum sensing demonstrations which employed hBN flakes exfoliated from bulk crystals~\cite{Gottscholl2021NC,HealeyNP2022,Kumar2022,Lyu2022,Yang2022}, here we use hBN nanopowder, a convenient and cost-effective alternative which enables sensing in a wider range of environments including in solution. We first characterise the spin properties of the V$_{\rm B}^-$ defects created in hBN nanopowder, and find the $T_1$ time is comparable to that in a bulk hBN crystal. We then dress the nanopowder with Gd$^{3+}$ ions, a common paramagnetic contrast agent used in magnetic resonance imaging, and observe a reliable $T_1$ quenching effect which illustrates the ability to detect external noise sources. Finally, we perform spin measurements of the nanopowder suspended in water and again observe a $T_1$ reduction upon adding Gd$^{3+}$ ions, demonstrating the possibility of in-solution sensing experiments. These results establish hBN nanopowder as a promising platform for magnetic noise sensing applications, and motivate further work to improve the sensitivity of the hBN quantum sensor, including by making hBN flakes approaching the 2D limit. This could make hBN quantum sensors a viable alternative to nanodiamonds for a range of chemical and biosensing applications such as ultrasensitive virus detection and other rapid point-of-care tests.

\section*{Results and Discussion}

Our quantum sensor is based on commercially available hBN nanopowder sourced from Graphene Supermarket. The powder is composed of flake-like particles with a thickness of $6\pm3$\,nm (mean $\pm$ standard deviation) as determined by atomic force microscopy measurements, and a lateral size of order $100$\,nm (see SI, Sec. II). The $6$\,nm mean thickness corresponds to $\approx9$ atomic monolayers. In principle, the size of these nanoflakes can be further reduced through liquid-phase exfoliation~\cite{Zhang2017,Chen2021}, making it an ideal candidate to realise quantum sensors approaching the 2D limit (i.e. made up of a single monolayer of hBN). %, although we did not attempt to reach this limit in the present work. 
The as-received powder was electron irradiated to create a high density of vacancies throughout the entire material, some of which forming the desired V$_{\rm B}^-$ defects~\cite{Murzakhanov2021}. The irradiated powder was used as is, in its dry form, or suspended in water or isopropanol to make powder films by drop-casting or to perform in-solution measurements.   

\subsection*{Spin measurements with hBN nanopowder}

We first characterised the dry hBN nanopowder following electron irradiation, to establish a baseline of its optical and spin properties. To this end, the powder was placed directly onto the surface of a printed circuit board (PCB) with a microwave (MW) waveguide to enable driving of the V$_{\rm B}^-$'s electron spin transitions. A schematic of the experiment is shown in Fig.~\ref{fig1}(a), and the crystal structure of the V$_{\rm B}^-$ defect in Fig.~\ref{fig1}(b). Small clumps of hBN nanopowder are visible in the photograph Fig.~\ref{fig1}(c), which were imaged using a widefield optical microscope [Fig.~\ref{fig1}(d,e)]. The V$_{\rm B}^-$ defects are excited with a green laser ($\lambda=532$\,nm) and the resulting photoluminescence (PL) in the near-infrared ($\lambda=750-900$\,nm) imaged by a camera [Fig.~\ref{fig1}(e)], showing uniform PL across the clump. The ground state of the V$_{\rm B}^-$ defect is a spin triplet ($S=1$) which is polarised into the $\ket{0}$ state upon laser excitation [Fig.~\ref{fig1}(f)]. When left in the dark, this unequal population distribution decays back to thermal equilibrium (an equal mixture of all three spin states $\ket{0,\pm1}$) at a rate $\Gamma_1=1/T_1$ ~\cite{GottschollSA2021}. In general, the total relaxation rate is the sum of intrinsic and extrinsic contributions, $\Gamma_1=\Gamma_1^{\rm int}+\Gamma_1^{\rm ext}$, where $\Gamma_1^{\rm ext}$ may be caused e.g. by paramagnetic spins external to the hBN crystal, as we will demonstrate later. Importantly, the spin state can be read out optically owing to spin-dependent PL~\cite{GottschollNM2020}, which will allow $\Gamma_1$ to be measured. Moreover, by applying a MW field of variable frequency, an optically detected magnetic resonance (ODMR) spectrum can be obtained [Fig.~\ref{fig1}(g)], revealing a broad resonance at $\omega_0\approx 2\pi\times 3.45$~GHz corresponding to the electronic spin transitions $\ket{0}\rightarrow\ket{\pm1}$. The 2\% spin contrast obtained in this pulsed-ODMR spectrum is in line with previous reports on similar powders~\cite{Liu2021}.

\begin{figure}[b!]
\centering
\includegraphics[width=0.9\linewidth]{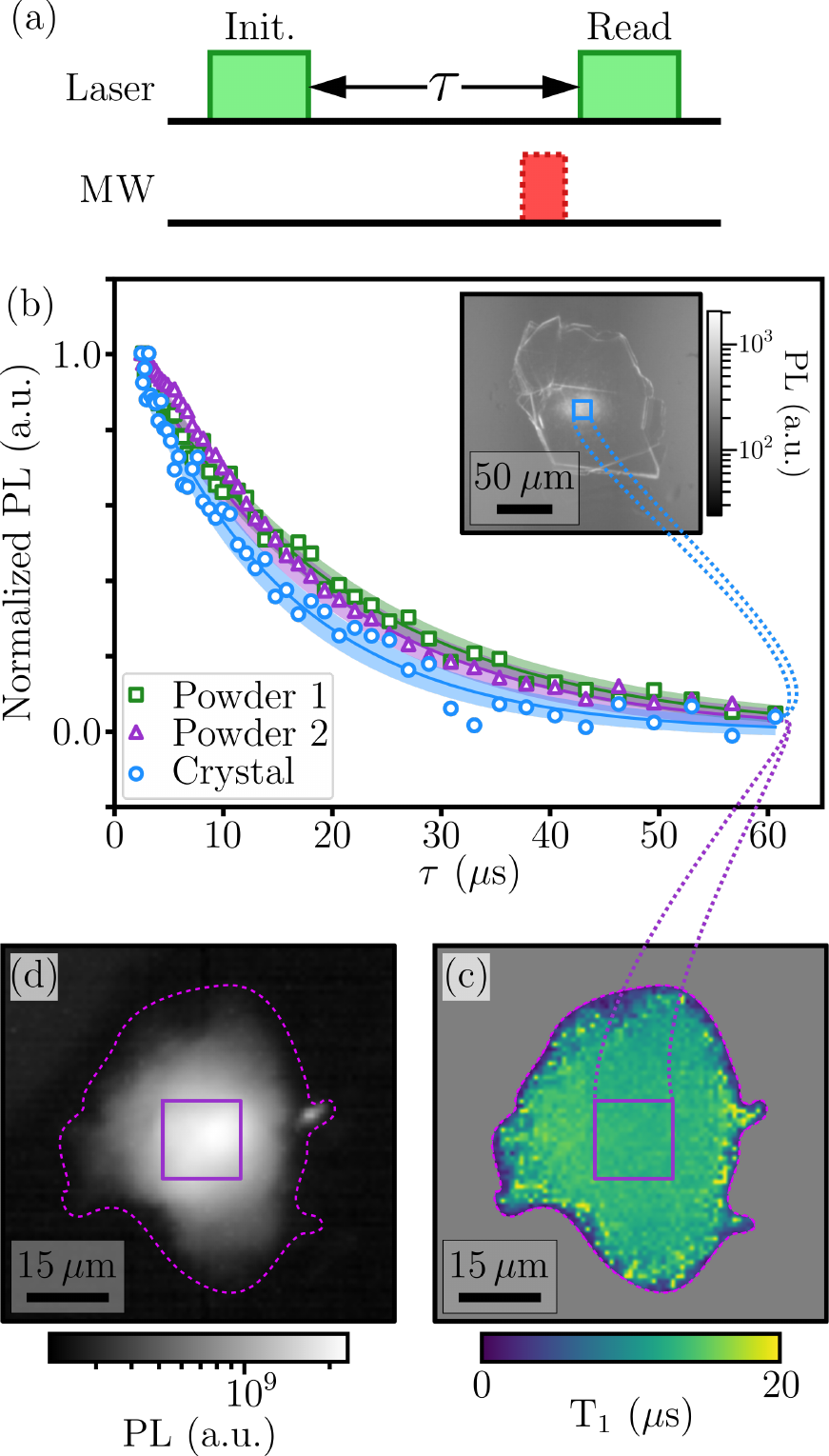}
\caption{{\bf $T_1$ relaxation in hBN nanopowder.} (a) Pulse sequence for $T_1$ measurements. The laser acts both to initialise the spin state and read it out through the PL intensity. For each dark time $\tau$, two PL measurements are performed, with and without the MW pulse, for normalisation purposes. (b)~Spin relaxation curves for the V$_{\rm B}^-$ defects in two nanopowder samples which received different irradiation doses (powder 2 higher than powder 1, see text) and in a bulk crystal flake. Solid lines are single exponential fit; shaded areas indicate one standard error for the fit parameters. Inset: PL image of the bulk crystal flake, displayed on a log scale. (c) Spatial map of $T_1$ for a clump of hBN nanopowder. Area external to the clump is masked as it contains no hBN powder. (d)~The corresponding PL image of the clump plotted with a log scale.}
\label{fig2}
\end{figure} 

\subsection*{Intrinsic $T_1$ time}

\begin{figure*}[tb!]
\centering
\includegraphics[width=1\textwidth]{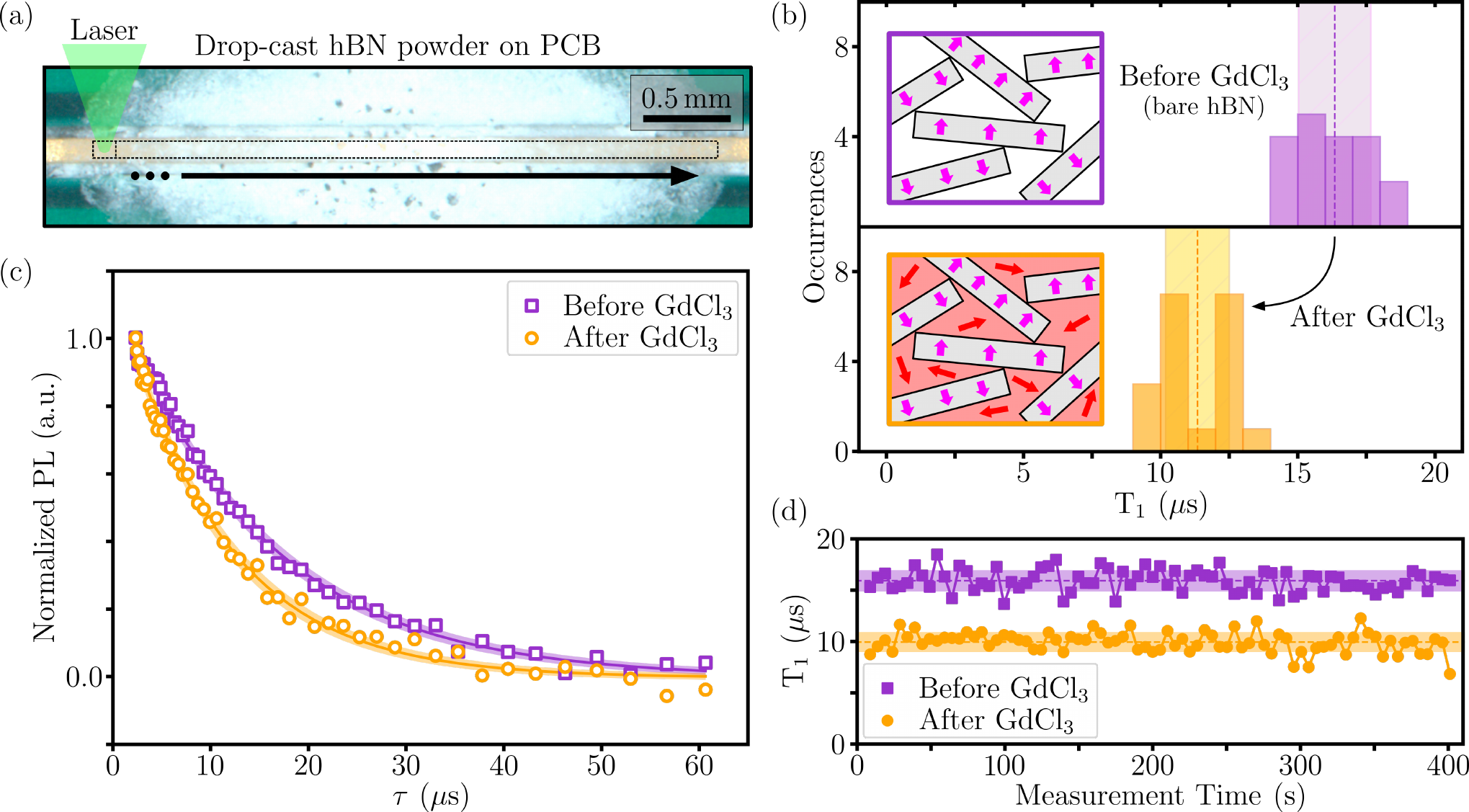}
\caption{{\bf Quenching $T_1$ with external paramagnetic spins}. (a) Photograph of hBN nanopowder film formed by drop-casting. Numerous $T_1$ measurements are taken at random distinct locations along the direction indicated by the arrow. The drop-cast nanopowder is then dosed with a solution of GdCl$_3$, and the $T_1$ measurements repeated. (b) Histograms for 20 $T_1$ measurements taken before (top histogram) and after (bottom) application of the GdCl$_3$ solution ($100$\,mM concentration in water). The vertical dotted lines represent the average recorded $T_1$ and the shaded area is one standard deviation either side of the average. (c) Representative spin relaxation curves selected from each histogram in (b). Solid lines are single exponential fit; shaded areas indicate one standard error for the fit parameters. (d) Time traces of $T_1$ recorded before and after application of the GdCl$_3$ solution. Each data point corresponds to $5$\,s of signal integration.}
\label{fig3}
\end{figure*}

To assess the viability of the hBN nanopowder for $T_1$-based magnetic noise sensing, we first characterised the intrinsic $T_1$ time of the dry powder. The measurement sequence is depicted in Fig.~\ref{fig2}(a) and consists of laser pulses, required to initialise and read out the spin state, separated by a dark time $\tau$ during which a $T_1$ decay takes place. A reference measurement including a resonant MW pulse is performed to remove common-mode variations and extract the spin imbalance, which decays towards zero as $\tau$ is increased, as shown in the example decay curves in Fig.~\ref{fig2}(b). Here, the PL data is integrated from a $15\times15\,\mu$m$^2$ region of powder which is uniformly illuminated by the laser. We found the curves are generally well fit by a mono-exponential function, $e^{-\tau/T_1}$, from which $T_1$ is obtained (see further analysis details in the SI, Sec. IV). 

In Fig.~\ref{fig2}(b), two types of powder are measured and compared to a large ($\sim100\,\mu$m) flake exfoliated from a high-purity bulk hBN crystal (PL image shown as inset of Fig.~\ref{fig2}(b)). Powder 1 received the lowest electron irradiation dose ($2\times10^{18}$\,cm$^{-2}$) and returns $T_1=19.5 \pm 0.8$\,$\mu$s (here the uncertainty corresponds to the standard error from the fit). For powder 2, the irradiation dosage was increased by $2.5\times$ to $5\times10^{18}$~cm$^{-2}$ and $T_1$ was measured as $T_1=17.6 \pm 0.6$\,$\mu$s. For the bulk crystal, which was electron irradiated to the same dose as powder 1, $T_1$ was slightly lower, $T_1=14.0 \pm 0.8$\,$\mu$s. The small differences in $T_1$ may be due to different levels of crystal damage, as was suggested by Guo et al.~\cite{Guo2022GenerationNitride}; here we speculate that the powder may experience less damage than the bulk crystal at the same nominal irradiation dose due to more efficient thermalisation during the irradiation process. Importantly, the $T_1$ values measured for our nanopowders, despite a mean flake thickness of only 6 nm, are comparable to the value reported by Gottschol et al.~\cite{GottschollSA2021} for a neutron-irradiated bulk crystal ($T_1=18\,\mu$s), which is the largest $T_1$ value reported to date for V$_{\rm B}^-$ at room temperature. Thus, nanopowders appear as an ideal platform for $T_1$-based sensing, available in much higher volume and at a lower cost compared to flakes made from bulk crystals.

We also assessed possible $T_1$ spatial variations across a clump of hBN powder. A $T_1$ map is shown in Fig.~\ref{fig2}(c) for a typical clump, with the corresponding PL image shown in Fig.~\ref{fig2}(d). The measured $T_1$ is found to be relatively uniform across the clump, with a standard deviation of about $1$\,$\mu$s comparable with the uncertainty from the fitting of individual pixels. On a larger scale, we observed slightly larger variations up to $\pm10\%$ for the mean $T_1$ of distinct clumps of powder of the same type, and from different spatial locations within a continuous film of powder (as we will see in Fig.~\ref{fig3}). Such variations may be due to local temperature variations (as laser and MW absorption leads to some amount of heating) or reflect measurement and analysis uncertainties, see Sec. IV and V of the SI for further discussions. In the following, we will study $T_1$ changes caused by external magnetic signals, keeping all measurement conditions constant otherwise.

\subsection*{Detection of external paramagnetic spins}

We now test the possibility of detecting external paramagnetic spins with the hBN nanopowder, by adding a solution of gadolinium trichloride (GdCl$_3$). Due to the high spin of the Gd$^{3+}$ ion ($S=\frac{7}{2}$), gadolinium complexes are commonly used as paramagnetic relaxation agents in magnetic resonance imaging. To allow for systematic measurements, we first prepare a uniform film of hBN nanopowder by suspending the powder [powder 2 from Fig.~\ref{fig2}(b)] in isopropanol and drop-casting onto a MW waveguide [Fig.~\ref{fig3}(a)]. A series of $T_1$ measurements are made at multiple random locations along the waveguide to gather a reference distribution of $T_1$ times, giving the top histogram in Fig.~\ref{fig3}(b). The mean value is $T_1=16.3\pm 1.3$\,$\mu$s where here the error bar corresponds to the standard deviation from the histogram. The spread is consistent with the variation of up to $\pm10\%$ observed and discussed previously in relation to Fig.~\ref{fig2}. 

Following these reference measurements, a drop of GdCl$_3$ solution is deposited on the hBN film and allowed to percolate through the powder. Evaporation of the water by heating then leaves the hBN nanoflakes surrounded by a highly concentrated medium of paramagnetic Gd$^{3+}$ ions. Example relaxation curves before and after adding Gd$^{3+}$ are plotted in Fig.~\ref{fig3}(c), showing a clear $T_1$ reduction upon adding Gd$^{3+}$. Repeating the measurements at multiple locations [bottom histogram in Fig.~\ref{fig3}(b)], we find that all $T_1$ values measured are shorter compared to the bare powder, with a mean value of $T_1=11.3\pm 1.2$\,$\mu$s, a 30\% reduction, significantly larger than the $\lesssim10\%$ uncertainty. This $T_1$ reduction corresponds to an additional relaxation rate of $\Gamma_1^{\rm ext}=30\pm10$\,kHz. Furthermore, we performed successive $T_1$ measurements (one every $5$\,s) at a given location and found $T_1$ to be stable over many minutes of monitoring [Fig.~\ref{fig3}(d)], with again a clear offset in the presence of Gd$^{3+}$. Combined, these results indicate a robust $T_1$ quenching effect by the Gd$^{3+}$ spins. Note, we repeated these experiments with higher concentrations of the GdCl$_3$ solution, and by applying multiple doses of the solution, and found a similar level of $T_1$ quenching in all cases, suggesting that the resulting density of Gd$^{3+}$ spins around the hBN flakes has reached a saturation. On the other hand, repeating the process using pure water or a highly diluted GdCl$_3$ solution led to no measurable change in $T_1$.

To verify the observed $T_1$ reduction is consistent with magnetic noise from the Gd$^{3+}$ spins, we assume each hBN nanoflake is surrounded by pure GdCl$_3$ and contain V$_{\rm B}^-$ defects located at a distance $d$ from either surface of the flake of thickness $2d$ (see Sec.~VI of the SI). The Gd$^{3+}$ spins produce a randomly fluctuating magnetic field with a correlation time $\tau_c$, which increases the relaxation rate of the V$_{\rm B}^-$ spins by~\cite{Steinert2013,Tetienne2013}
\begin{equation}
    \Gamma_1^{\rm ext} = 3\gamma_e^2 B_{\perp}^2 \frac{\tau_c}{1 + \omega_0^2 \tau_c^2},
\end{equation}
where $\gamma_e$ is the electron gyromagnetic ratio and $B_{\perp}^2$ is the variance in the transverse magnetic field experienced by the V$_{\rm B}^-$ spins. $\Gamma_1^{\rm ext}$ is maximised when $\tau_c=1/\omega_0\approx~50$\,ps, which is a typical order of magnitude for Gd$^{3+}$ complexes~\cite{Steinert2013}. Assuming this maximum effect, we find the observed $\Gamma_1^{\rm ext}$ is reproduced for a flake thickness of $2d\approx10$\,nm, in good agreement with the $6\pm3$\,nm thickness of our flakes. This supports our claim that the observed $T_1$ quenching is caused by the magnetic noise from the added Gd$^{3+}$ spins.

\subsection*{Spin measurements in solution}

\begin{figure*}[tb!]
\centering
\includegraphics[width=1\textwidth]{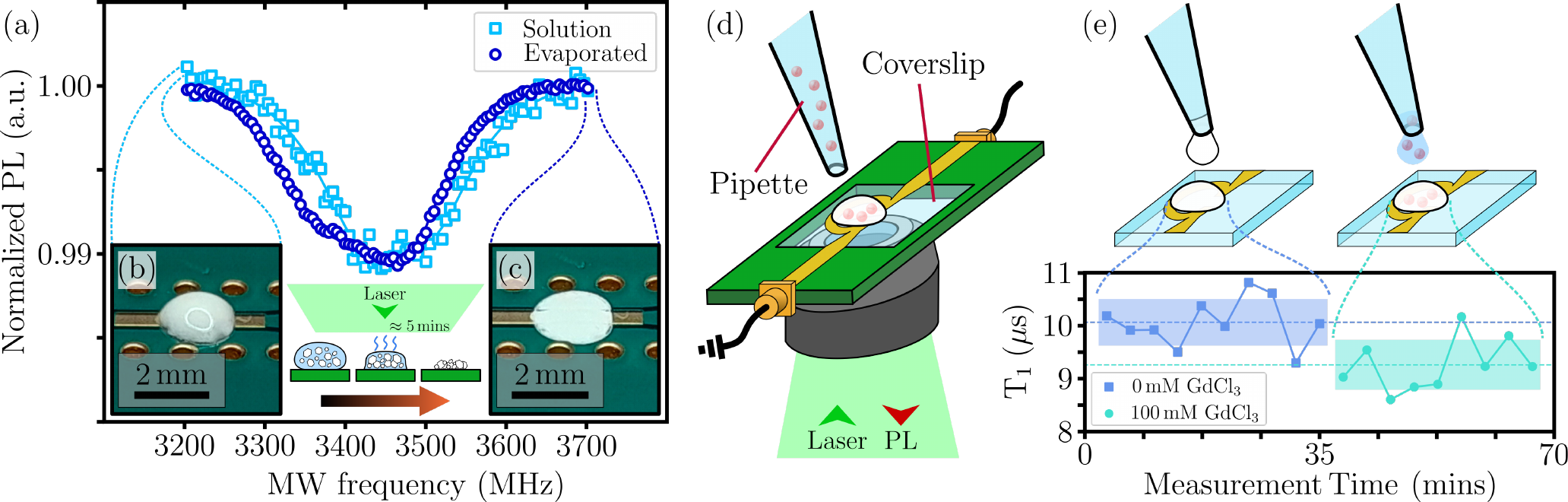}
\caption{{\bf Solution-based spin measurements}. (a) ODMR spectra from hBN nanopowder recorded before and after evaporation of the host water. (b,c) Photographs of the hBN nanopowder on the PCB before (b) and after (c) water evaporation. In (b), the powder forms a colloidal suspension, whereas in (c) the powder forms a dry film as in Fig.~\ref{fig3}(a). (d) Schematic of the experimental setup for in-solution $T_1$ sensing. The inverted geometry facilitates addition of paramagnetic ions and minimises measurement-induced heating allowing extended measurement times. (e) Time trace of $T_1$ measured from the suspension, with $100$\,mM of GdCl$_3$ added after 35 minutes. The horizontal dashed lines represent the average $T_1$ from before and after adding GdCl$_3$, with the shading representing one standard deviation. Note, $T_1$ is shorter even in pure water compared to the dry powder, which may be due to a change in the internal spin or charge environment of the hBN flakes. }
\label{fig4}
\end{figure*}

Many applications of $T_1$-based sensing involve a liquid environment, for instance detecting ions and viruses in aqueous solutions \cite{Ziem2013,Simpson2017,Changhao2022} or free radicals in cells \cite{Sharmin2022,Nie2022}, or to measure the pH of a solution~\cite{Rendler2017}. To test the sensing capabilities of hBN in a wet environment, we performed spin measurements of the nanopowder suspended in water. First, a drop of the hBN solution was placed on a MW waveguide and a pulsed-ODMR spectrum recorded [Fig.~\ref{fig4}(a)]. As shown in the photographs Fig.~\ref{fig4}(b), the solution has a white opaque appearance, such that the collected PL signal must initially come from the top-most part of the colloidal suspension. The water was found to evaporate completely after a  few minutes of continuous measurement, leaving behind a dry hBN powder film [Fig.~\ref{fig4}(c)]. Comparing the ODMR spectra taken before and after evaporation [Fig.~\ref{fig4}(a)], we see that while the contrast remains relatively unchanged ($\approx1\%$), the resonance is shifted towards lower frequencies by $\approx50$~MHz upon evaporation. This decrease in $\omega_0$, which here corresponds to the zero-field splitting parameter of the V$_{\rm B}^-$ spin, implies an increase in local temperature of $\approx70$\,K~\cite{Liu2021,Gottscholl2021NC}. The rapid evaporation followed by a dramatic temperature increase can be explained by the presence of significant heating due to local laser and MW absorption, which is initially dissipated as latent heat during the evaporation phase thus keeping the solution relatively cool, until the water has fully evaporated.

To facilitate in-situ addition of paramagnetic spins to the solution and $T_1$ measurements over extended periods of time, we next moved to a setup based on an inverted microscope configuration [Fig.~\ref{fig4}(d)]. The hBN suspension is deposited on a glass coverslip which comprises an $\Omega$-shaped MW waveguide. Because optical addressing occurs from below, the collected PL signal comes primarily from the hBN flakes that are closest to the glass-solution interface, and so it is possible to address a region located very close to the MW waveguide, thus requiring a significantly reduced MW power. To further reduce the heating induced by the measurements, we used shorter laser pulses and a lower laser peak power. As a result, we were able to perform $T_1$ measurements continuously for up to an hour before complete evaporation of the water, although at the cost of a reduced signal-to-noise ratio -- in future, a cuvette-based setup could be employed to allow optimised, long-term measurements~\cite{Grant2022}. As a test experiment, we monitored the $T_1$ time of the hBN flakes initially suspended in pure water, before adding a solution of $100$\,mM GdCl$_3$ [Fig.~\ref{fig4}(e)]. A small reduction in $T_1$ is observed upon addition of the GdCl$_3$ solution, corresponding to an added relaxation rate of $\Gamma_1^{\rm ext}=10\pm10$\,kHz. While the observed $T_1$ change is close to the measurement uncertainty, these experiments demonstrate the feasibility of in-solution magnetic noise sensing with hBN nanopowder. Note, the induced quenching $\Gamma_1^{\rm ext}$ is smaller than that observed in Fig.~\ref{fig3}, which is expected since here the density of Gd$^{3+}$ ions is reduced due to the presence of solvent molecules. Using higher concentrations of Gd$^{3+}$ ions led to additional background PL which made the $T_1$ measurements unreliable.    

\subsection*{Sensitivity}

The key motivation for developing quantum sensors based on vdW materials is the potential for a drastically reduced minimum distance $d$ between sensor (spin defects) and target (e.g.\ paramagnetic spins), compared to 3D hosts like diamond. In the present work, for our initial demonstration of external paramagnetic spin detection, we used commercially available, off-the-shelf hBN nanopowders which had a flake thickness of $6\pm3$\,nm, meaning the sensing defects are located at most $5$\,nm from the closest surface (for the thickest flakes), which is already an improvement over state-of-the-art experiments based on NV ensembles in diamond~\cite{Ziem2019,HealeyPRA2021,Liu2022}. Furthermore, there is ample room for optimisation of the hBN sensor, in particular liquid-phase exfoliation methods should provide a straightforward way to produce even thinner hBN flakes (a few or even a single monolayer) in large quantities~\cite{Zhang2017,Chen2021}. While there has been no report of V$_{\rm B}^-$ defects in monolayer hBN to date, V$_{\rm B}^-$ defects have been successfully detected in flakes as thin as 4 nm~\cite{Xu2023} while still exhibiting ODMR, implying the defects are within $2$\,nm (i.e. 3 atomic layers) of the closest surface. Compared to an NV centre at $d=5$~nm (typical minimum depth observed~\cite{Ziem2019,HealeyPRA2021,Liu2022}), a V$_{\rm B}^-$ defect at $d=2$~nm would increase the relaxation rate $\Gamma_1^{\rm ext}$ induced by a paramagnetic spin located at the surface by a factor $(5/2)^6=244$, already a dramatic improvement. 

Nevertheless, a potential gain in $\Gamma_1^{\rm ext}$ due to better proximity does not guarantee an improved sensitivity to external spins. In a photon-shot-noise-limited $T_1$-relaxometry experiment, the signal-to-noise ratio scales as~\cite{Steinert2013,Tetienne2013}
\begin{equation}
    {\rm SNR} \propto \frac{\Gamma_1^{\rm ext}}{\sqrt{\Gamma_1^{\rm int}}}{\cal C}\sqrt{I_{\rm PL}},
\end{equation}
where $\Gamma_1^{\rm int}=1/T_1^{\rm int}$ is the intrinsic relaxation rate in the absence of the target spin (assumed to satisfy $\Gamma_1^{\rm int}\gg\Gamma_1^{\rm ext}$), ${\cal C}$ is the relative spin contrast, and $I_{\rm PL}$ is the PL signal from one readout pulse. Currently, the V$_{\rm B}^-$ defect exhibits an inferior contrast and PL output compared to the NV centre, and so will need to be improved through material optimisation or photonics engineering to enhance collection~\cite{Froch2021,Xingyu2021,Xu2023}. On the other hand, the intrinsic relaxation time observed in this work for V$_{\rm B}^-$ in hBN nanopowders ($T_1^{\rm int}\approx15-20\,\mu$s) is similar to that of NVs in nanodiamonds of comparable size~\cite{Tetienne2013}, which do not reach the longer relaxation times exhibited by NVs in bulk diamond but nevertheless have found widespread interest. Considering all these factors together, hBN nanopowders emerge as a viable alternative to nanodiamonds for $T_1$ relaxometry with the potential for an improved sensitivity to external paramagnetic spins and a lower production cost, making it appealing for applications such as high-sensitivity point-of-care diagnostics~\cite{Changhao2022,Miller2020}.

\section*{Conclusions}

In this work, we demonstrated the first detection of magnetic noise from paramagnetic spins under ambient conditions using a van der Waals quantum sensor, namely using optically addressable spin defects (V$_{\rm B}^-$ defects) in hBN nanopowders. We first characterised the intrinsic spin relaxation time ($T_1$) of the V$_{\rm B}^-$ defects in nanopowders, which was found to be comparable to that in bulk hBN crystals. We then observed a reduction in $T_1$ upon the addition of paramagnetic Gd$^{3+}$ ions, in dry conditions, demonstrating the ability to detect external noise sources. Finally, we  performed spin measurements using the hBN nanopowder suspended in water, demonstrating the feasibility of in-solution sensing experiments. The sensitivity of the hBN quantum sensor for the detection of paramagnetic spins was discussed, and with further improvements (in particular, approaching the 2D limit for the host hBN flakes) could be competitive with established quantum sensing platforms such the NV centre in diamond. Our results thus suggest hBN nanopowders could be a viable and potentially more convenient and cost-effective alternative to diamond-based sensors, making it appealing for a range of chemical and biosensing applications such as ultrasensitive virus detection and other rapid point-of-care tests. 

\begin{acknowledgments}
This work was supported by the Australian Research Council (ARC) through grants CE170100012, CE200100010, FT200100073, DE200100279 and FT220100053, and by the Office of Naval Research Global (N62909-22-1-2028). We thank Yongliang Chen for providing AFM data. The work was performed in part at the RMIT Micro Nano Research Facility (MNRF) in the Victorian Node of the Australian National Fabrication Facility (ANFF) and the RMIT Microscopy and Microanalysis Facility (RMMF). I.O.R. and A.J.H. are supported by an Australian Government Research Training Program Scholarship. S.C.S gratefully acknowledges the support of an Ernst and Grace Matthaei scholarship. P.R. acknowledges support through an RMIT University Vice-Chancellor’s Research Fellowship. Part of this study was supported by QST President's Strategic Grant ``QST International Research Initiative''.

\end{acknowledgments}

\bibliography{refs}

\begin{thebibliography}{52}%
\makeatletter
\providecommand \@ifxundefined [1]{%
 \@ifx{#1\undefined}
}%
\providecommand \@ifnum [1]{%
 \ifnum #1\expandafter \@firstoftwo
 \else \expandafter \@secondoftwo
 \fi
}%
\providecommand \@ifx [1]{%
 \ifx #1\expandafter \@firstoftwo
 \else \expandafter \@secondoftwo
 \fi
}%
\providecommand \natexlab [1]{#1}%
\providecommand \enquote  [1]{``#1''}%
\providecommand \bibnamefont  [1]{#1}%
\providecommand \bibfnamefont [1]{#1}%
\providecommand \citenamefont [1]{#1}%
\providecommand \href@noop [0]{\@secondoftwo}%
\providecommand \href [0]{\begingroup \@sanitize@url \@href}%
\providecommand \@href[1]{\@@startlink{#1}\@@href}%
\providecommand \@@href[1]{\endgroup#1\@@endlink}%
\providecommand \@sanitize@url [0]{\catcode `\\12\catcode `\$12\catcode
  `\&12\catcode `\#12\catcode `\^12\catcode `\_12\catcode `\%12\relax}%
\providecommand \@@startlink[1]{}%
\providecommand \@@endlink[0]{}%
\providecommand \url  [0]{\begingroup\@sanitize@url \@url }%
\providecommand \@url [1]{\endgroup\@href {#1}{\urlprefix }}%
\providecommand \urlprefix  [0]{URL }%
\providecommand \Eprint [0]{\href }%
\providecommand \doibase [0]{https://doi.org/}%
\providecommand \selectlanguage [0]{\@gobble}%
\providecommand \bibinfo  [0]{\@secondoftwo}%
\providecommand \bibfield  [0]{\@secondoftwo}%
\providecommand \translation [1]{[#1]}%
\providecommand \BibitemOpen [0]{}%
\providecommand \bibitemStop [0]{}%
\providecommand \bibitemNoStop [0]{.\EOS\space}%
\providecommand \EOS [0]{\spacefactor3000\relax}%
\providecommand \BibitemShut  [1]{\csname bibitem#1\endcsname}%
\let\auto@bib@innerbib\@empty
%</preamble>
\bibitem [{\citenamefont {Degen}\ \emph {et~al.}(2017)\citenamefont {Degen},
  \citenamefont {Reinhard},\ and\ \citenamefont {Cappellaro}}]{Degen2017}%
  \BibitemOpen
  \bibfield  {author} {\bibinfo {author} {\bibfnamefont {C.~L.}\ \bibnamefont
  {Degen}}, \bibinfo {author} {\bibfnamefont {F.}~\bibnamefont {Reinhard}},\
  and\ \bibinfo {author} {\bibfnamefont {P.}~\bibnamefont {Cappellaro}},\
  }\bibfield  {title} {\bibinfo {title} {{Quantum sensing}},\ }\href
  {https://doi.org/10.1103/RevModPhys.89.035002} {\bibfield  {journal}
  {\bibinfo  {journal} {Reviews of Modern Physics}\ }\textbf {\bibinfo {volume}
  {89}},\ \bibinfo {pages} {035002} (\bibinfo {year} {2017})}\BibitemShut
  {NoStop}%
\bibitem [{\citenamefont {Schirhagl}\ \emph {et~al.}(2014)\citenamefont
  {Schirhagl}, \citenamefont {Chang}, \citenamefont {Loretz},\ and\
  \citenamefont {Degen}}]{Schirhagl2014}%
  \BibitemOpen
  \bibfield  {author} {\bibinfo {author} {\bibfnamefont {R.}~\bibnamefont
  {Schirhagl}}, \bibinfo {author} {\bibfnamefont {K.}~\bibnamefont {Chang}},
  \bibinfo {author} {\bibfnamefont {M.}~\bibnamefont {Loretz}},\ and\ \bibinfo
  {author} {\bibfnamefont {C.~L.}\ \bibnamefont {Degen}},\ }\bibfield  {title}
  {\bibinfo {title} {{Nitrogen-Vacancy Centers in Diamond: Nanoscale Sensors
  for Physics and Biology}},\ }\href
  {https://doi.org/10.1146/annurev-physchem-040513-103659} {\bibfield
  {journal} {\bibinfo  {journal} {Annual Review of Physical Chemistry}\
  }\textbf {\bibinfo {volume} {65}},\ \bibinfo {pages} {83} (\bibinfo {year}
  {2014})}\BibitemShut {NoStop}%
\bibitem [{\citenamefont {Rondin}\ \emph {et~al.}(2014)\citenamefont {Rondin},
  \citenamefont {Tetienne}, \citenamefont {Hingant}, \citenamefont {Roch},
  \citenamefont {Maletinsky},\ and\ \citenamefont {Jacques}}]{Rondin2014}%
  \BibitemOpen
  \bibfield  {author} {\bibinfo {author} {\bibfnamefont {L.}~\bibnamefont
  {Rondin}}, \bibinfo {author} {\bibfnamefont {J.-P.}\ \bibnamefont
  {Tetienne}}, \bibinfo {author} {\bibfnamefont {T.}~\bibnamefont {Hingant}},
  \bibinfo {author} {\bibfnamefont {J.-F.}\ \bibnamefont {Roch}}, \bibinfo
  {author} {\bibfnamefont {P.}~\bibnamefont {Maletinsky}},\ and\ \bibinfo
  {author} {\bibfnamefont {V.}~\bibnamefont {Jacques}},\ }\bibfield  {title}
  {\bibinfo {title} {{Magnetometry with nitrogen-vacancy defects in diamond}},\
  }\href {https://doi.org/10.1088/0034-4885/77/5/056503} {\bibfield  {journal}
  {\bibinfo  {journal} {Reports on Progress in Physics}\ }\textbf {\bibinfo
  {volume} {77}},\ \bibinfo {pages} {056503} (\bibinfo {year}
  {2014})}\BibitemShut {NoStop}%
\bibitem [{\citenamefont {Casola}\ \emph {et~al.}(2018)\citenamefont {Casola},
  \citenamefont {van~der Sar},\ and\ \citenamefont {Yacoby}}]{Casola2018}%
  \BibitemOpen
  \bibfield  {author} {\bibinfo {author} {\bibfnamefont {F.}~\bibnamefont
  {Casola}}, \bibinfo {author} {\bibfnamefont {T.}~\bibnamefont {van~der
  Sar}},\ and\ \bibinfo {author} {\bibfnamefont {A.}~\bibnamefont {Yacoby}},\
  }\bibfield  {title} {\bibinfo {title} {{Probing condensed matter physics with
  magnetometry based on nitrogen-vacancy centres in diamond}},\ }\href
  {https://doi.org/10.1038/natrevmats.2017.88} {\bibfield  {journal} {\bibinfo
  {journal} {Nature Reviews Materials}\ }\textbf {\bibinfo {volume} {3}},\
  \bibinfo {pages} {17088} (\bibinfo {year} {2018})}\BibitemShut {NoStop}%
\bibitem [{\citenamefont {Du}\ \emph {et~al.}(2017)\citenamefont {Du},
  \citenamefont {van~der Sar}, \citenamefont {Zhou}, \citenamefont {Upadhyaya},
  \citenamefont {Casola}, \citenamefont {Zhang}, \citenamefont {Onbasli},
  \citenamefont {Ross}, \citenamefont {Walsworth}, \citenamefont
  {Tserkovnyak},\ and\ \citenamefont {Yacoby}}]{Du2017}%
  \BibitemOpen
  \bibfield  {author} {\bibinfo {author} {\bibfnamefont {C.}~\bibnamefont
  {Du}}, \bibinfo {author} {\bibfnamefont {T.}~\bibnamefont {van~der Sar}},
  \bibinfo {author} {\bibfnamefont {T.~X.}\ \bibnamefont {Zhou}}, \bibinfo
  {author} {\bibfnamefont {P.}~\bibnamefont {Upadhyaya}}, \bibinfo {author}
  {\bibfnamefont {F.}~\bibnamefont {Casola}}, \bibinfo {author} {\bibfnamefont
  {H.}~\bibnamefont {Zhang}}, \bibinfo {author} {\bibfnamefont {M.~C.}\
  \bibnamefont {Onbasli}}, \bibinfo {author} {\bibfnamefont {C.~A.}\
  \bibnamefont {Ross}}, \bibinfo {author} {\bibfnamefont {R.~L.}\ \bibnamefont
  {Walsworth}}, \bibinfo {author} {\bibfnamefont {Y.}~\bibnamefont
  {Tserkovnyak}},\ and\ \bibinfo {author} {\bibfnamefont {A.}~\bibnamefont
  {Yacoby}},\ }\bibfield  {title} {\bibinfo {title} {{Control and local
  measurement of the spin chemical potential in a magnetic insulator}},\ }\href
  {https://doi.org/10.1126/science.aak9611} {\bibfield  {journal} {\bibinfo
  {journal} {Science}\ }\textbf {\bibinfo {volume} {357}},\ \bibinfo {pages}
  {195} (\bibinfo {year} {2017})}\BibitemShut {NoStop}%
\bibitem [{\citenamefont {McCullian}\ \emph {et~al.}(2020)\citenamefont
  {McCullian}, \citenamefont {Thabt}, \citenamefont {Gray}, \citenamefont
  {Melendez}, \citenamefont {Wolf}, \citenamefont {Safonov}, \citenamefont
  {Pelekhov}, \citenamefont {Bhallamudi}, \citenamefont {Page},\ and\
  \citenamefont {Hammel}}]{McCullian2020}%
  \BibitemOpen
  \bibfield  {author} {\bibinfo {author} {\bibfnamefont {B.~A.}\ \bibnamefont
  {McCullian}}, \bibinfo {author} {\bibfnamefont {A.~M.}\ \bibnamefont
  {Thabt}}, \bibinfo {author} {\bibfnamefont {B.~A.}\ \bibnamefont {Gray}},
  \bibinfo {author} {\bibfnamefont {A.~L.}\ \bibnamefont {Melendez}}, \bibinfo
  {author} {\bibfnamefont {M.~S.}\ \bibnamefont {Wolf}}, \bibinfo {author}
  {\bibfnamefont {V.~L.}\ \bibnamefont {Safonov}}, \bibinfo {author}
  {\bibfnamefont {D.~V.}\ \bibnamefont {Pelekhov}}, \bibinfo {author}
  {\bibfnamefont {V.~P.}\ \bibnamefont {Bhallamudi}}, \bibinfo {author}
  {\bibfnamefont {M.~R.}\ \bibnamefont {Page}},\ and\ \bibinfo {author}
  {\bibfnamefont {P.~C.}\ \bibnamefont {Hammel}},\ }\bibfield  {title}
  {\bibinfo {title} {Broadband multi-magnon relaxometry using a quantum spin
  sensor for high frequency ferromagnetic dynamics sensing},\ }\href
  {https://doi.org/10.1038/s41467-020-19121-0} {\bibfield  {journal} {\bibinfo
  {journal} {Nat Communications}\ }\textbf {\bibinfo {volume} {11}},\ \bibinfo
  {pages} {5229} (\bibinfo {year} {2020})}\BibitemShut {NoStop}%
\bibitem [{\citenamefont {Kolkowitz}\ \emph {et~al.}(2015)\citenamefont
  {Kolkowitz}, \citenamefont {Safira}, \citenamefont {High}, \citenamefont
  {Devlin}, \citenamefont {Choi}, \citenamefont {Unterreithmeier},
  \citenamefont {Patterson}, \citenamefont {Zibrov}, \citenamefont
  {Manucharyan}, \citenamefont {Park},\ and\ \citenamefont
  {Lukin}}]{Kolkowitz2015}%
  \BibitemOpen
  \bibfield  {author} {\bibinfo {author} {\bibfnamefont {S.}~\bibnamefont
  {Kolkowitz}}, \bibinfo {author} {\bibfnamefont {A.}~\bibnamefont {Safira}},
  \bibinfo {author} {\bibfnamefont {A.~A.}\ \bibnamefont {High}}, \bibinfo
  {author} {\bibfnamefont {R.~C.}\ \bibnamefont {Devlin}}, \bibinfo {author}
  {\bibfnamefont {S.}~\bibnamefont {Choi}}, \bibinfo {author} {\bibfnamefont
  {Q.~P.}\ \bibnamefont {Unterreithmeier}}, \bibinfo {author} {\bibfnamefont
  {D.}~\bibnamefont {Patterson}}, \bibinfo {author} {\bibfnamefont {A.~S.}\
  \bibnamefont {Zibrov}}, \bibinfo {author} {\bibfnamefont {V.~E.}\
  \bibnamefont {Manucharyan}}, \bibinfo {author} {\bibfnamefont
  {H.}~\bibnamefont {Park}},\ and\ \bibinfo {author} {\bibfnamefont {M.~D.}\
  \bibnamefont {Lukin}},\ }\bibfield  {title} {\bibinfo {title} {{Probing
  Johnson noise and ballistic transport in normal metals with a single-spin
  qubit}},\ }\href {https://doi.org/10.1126/science.aaa4298} {\bibfield
  {journal} {\bibinfo  {journal} {Science}\ }\textbf {\bibinfo {volume}
  {347}},\ \bibinfo {pages} {1129} (\bibinfo {year} {2015})}\BibitemShut
  {NoStop}%
\bibitem [{\citenamefont {Ariyaratne}\ \emph {et~al.}(2018)\citenamefont
  {Ariyaratne}, \citenamefont {Bluvstein}, \citenamefont {Myers},\ and\
  \citenamefont {Jayich}}]{Ariyaratne2018}%
  \BibitemOpen
  \bibfield  {author} {\bibinfo {author} {\bibfnamefont {A.}~\bibnamefont
  {Ariyaratne}}, \bibinfo {author} {\bibfnamefont {D.}~\bibnamefont
  {Bluvstein}}, \bibinfo {author} {\bibfnamefont {B.~A.}\ \bibnamefont
  {Myers}},\ and\ \bibinfo {author} {\bibfnamefont {A.~C.~B.}\ \bibnamefont
  {Jayich}},\ }\bibfield  {title} {\bibinfo {title} {Nanoscale electrical
  conductivity imaging using a nitrogen-vacancy center in diamond},\ }\href
  {https://doi.org/10.1038/s41467-018-04798-1} {\bibfield  {journal} {\bibinfo
  {journal} {Nature Communications}\ }\textbf {\bibinfo {volume} {9}},\
  \bibinfo {pages} {2406} (\bibinfo {year} {2018})}\BibitemShut {NoStop}%
\bibitem [{\citenamefont {Steinert}\ \emph {et~al.}(2013)\citenamefont
  {Steinert}, \citenamefont {Ziem}, \citenamefont {Hall}, \citenamefont
  {Zappe}, \citenamefont {Schweikert}, \citenamefont {G{\"{o}}tz},
  \citenamefont {Aird}, \citenamefont {Balasubramanian}, \citenamefont
  {Hollenberg},\ and\ \citenamefont {Wrachtrup}}]{Steinert2013}%
  \BibitemOpen
  \bibfield  {author} {\bibinfo {author} {\bibfnamefont {S.}~\bibnamefont
  {Steinert}}, \bibinfo {author} {\bibfnamefont {F.}~\bibnamefont {Ziem}},
  \bibinfo {author} {\bibfnamefont {L.~T.}\ \bibnamefont {Hall}}, \bibinfo
  {author} {\bibfnamefont {A.}~\bibnamefont {Zappe}}, \bibinfo {author}
  {\bibfnamefont {M.}~\bibnamefont {Schweikert}}, \bibinfo {author}
  {\bibfnamefont {N.}~\bibnamefont {G{\"{o}}tz}}, \bibinfo {author}
  {\bibfnamefont {A.}~\bibnamefont {Aird}}, \bibinfo {author} {\bibfnamefont
  {G.}~\bibnamefont {Balasubramanian}}, \bibinfo {author} {\bibfnamefont
  {L.}~\bibnamefont {Hollenberg}},\ and\ \bibinfo {author} {\bibfnamefont
  {J.}~\bibnamefont {Wrachtrup}},\ }\bibfield  {title} {\bibinfo {title}
  {{Magnetic spin imaging under ambient conditions with sub-cellular
  resolution}},\ }\href {https://doi.org/10.1038/ncomms2588} {\bibfield
  {journal} {\bibinfo  {journal} {Nature Communications}\ }\textbf {\bibinfo
  {volume} {4}},\ \bibinfo {pages} {1607} (\bibinfo {year} {2013})}\BibitemShut
  {NoStop}%
\bibitem [{\citenamefont {Tetienne}\ \emph {et~al.}(2013)\citenamefont
  {Tetienne}, \citenamefont {Hingant}, \citenamefont {Rondin}, \citenamefont
  {Cavaill{\`{e}}s}, \citenamefont {Mayer}, \citenamefont {Dantelle},
  \citenamefont {Gacoin}, \citenamefont {Wrachtrup}, \citenamefont {Roch},\
  and\ \citenamefont {Jacques}}]{Tetienne2013}%
  \BibitemOpen
  \bibfield  {author} {\bibinfo {author} {\bibfnamefont {J.-P.}\ \bibnamefont
  {Tetienne}}, \bibinfo {author} {\bibfnamefont {T.}~\bibnamefont {Hingant}},
  \bibinfo {author} {\bibfnamefont {L.}~\bibnamefont {Rondin}}, \bibinfo
  {author} {\bibfnamefont {A.}~\bibnamefont {Cavaill{\`{e}}s}}, \bibinfo
  {author} {\bibfnamefont {L.}~\bibnamefont {Mayer}}, \bibinfo {author}
  {\bibfnamefont {G.}~\bibnamefont {Dantelle}}, \bibinfo {author}
  {\bibfnamefont {T.}~\bibnamefont {Gacoin}}, \bibinfo {author} {\bibfnamefont
  {J.}~\bibnamefont {Wrachtrup}}, \bibinfo {author} {\bibfnamefont {J.-F.}\
  \bibnamefont {Roch}},\ and\ \bibinfo {author} {\bibfnamefont
  {V.}~\bibnamefont {Jacques}},\ }\bibfield  {title} {\bibinfo {title} {{Spin
  relaxometry of single nitrogen-vacancy defects in diamond nanocrystals for
  magnetic noise sensing}},\ }\href
  {https://doi.org/10.1103/PhysRevB.87.235436} {\bibfield  {journal} {\bibinfo
  {journal} {Physical Review B}\ }\textbf {\bibinfo {volume} {87}},\ \bibinfo
  {pages} {235436} (\bibinfo {year} {2013})}\BibitemShut {NoStop}%
\bibitem [{\citenamefont {Ermakova}\ \emph {et~al.}(2013)\citenamefont
  {Ermakova}, \citenamefont {Pramanik}, \citenamefont {Cai}, \citenamefont
  {Algara-Siller}, \citenamefont {Kaiser}, \citenamefont {Weil}, \citenamefont
  {Tzeng}, \citenamefont {Chang}, \citenamefont {McGuinness}, \citenamefont
  {Plenio}, \citenamefont {Naydenov},\ and\ \citenamefont
  {Jelezko}}]{Ermakova2013}%
  \BibitemOpen
  \bibfield  {author} {\bibinfo {author} {\bibfnamefont {A.}~\bibnamefont
  {Ermakova}}, \bibinfo {author} {\bibfnamefont {G.}~\bibnamefont {Pramanik}},
  \bibinfo {author} {\bibfnamefont {J.~M.}\ \bibnamefont {Cai}}, \bibinfo
  {author} {\bibfnamefont {G.}~\bibnamefont {Algara-Siller}}, \bibinfo {author}
  {\bibfnamefont {U.}~\bibnamefont {Kaiser}}, \bibinfo {author} {\bibfnamefont
  {T.}~\bibnamefont {Weil}}, \bibinfo {author} {\bibfnamefont {Y.~K.}\
  \bibnamefont {Tzeng}}, \bibinfo {author} {\bibfnamefont {H.~C.}\ \bibnamefont
  {Chang}}, \bibinfo {author} {\bibfnamefont {L.~P.}\ \bibnamefont
  {McGuinness}}, \bibinfo {author} {\bibfnamefont {M.~B.}\ \bibnamefont
  {Plenio}}, \bibinfo {author} {\bibfnamefont {B.}~\bibnamefont {Naydenov}},\
  and\ \bibinfo {author} {\bibfnamefont {F.}~\bibnamefont {Jelezko}},\
  }\bibfield  {title} {\bibinfo {title} {{Detection of a few metallo-protein
  molecules using color centers in nanodiamonds}},\ }\href
  {https://doi.org/10.1021/NL4015233/SUPPL{\_}FILE/NL4015233{\_}SI{\_}001.PDF}
  {\bibfield  {journal} {\bibinfo  {journal} {Nano Letters}\ }\textbf {\bibinfo
  {volume} {13}},\ \bibinfo {pages} {3305} (\bibinfo {year}
  {2013})}\BibitemShut {NoStop}%
\bibitem [{\citenamefont {Kaufmanna}\ \emph {et~al.}(2013)\citenamefont
  {Kaufmanna}, \citenamefont {Simpson}, \citenamefont {Hall}, \citenamefont
  {Perunicic}, \citenamefont {Senn}, \citenamefont {Steinert}, \citenamefont
  {McGuinness}, \citenamefont {Johnson}, \citenamefont {Ohshima}, \citenamefont
  {Caruso}, \citenamefont {Wrachtrup}, \citenamefont {Scholten}, \citenamefont
  {Mulvaney},\ and\ \citenamefont {Hollenberg}}]{Kaufmanna2013DetectionProbe}%
  \BibitemOpen
  \bibfield  {author} {\bibinfo {author} {\bibfnamefont {S.}~\bibnamefont
  {Kaufmanna}}, \bibinfo {author} {\bibfnamefont {D.~A.}\ \bibnamefont
  {Simpson}}, \bibinfo {author} {\bibfnamefont {L.~T.}\ \bibnamefont {Hall}},
  \bibinfo {author} {\bibfnamefont {V.}~\bibnamefont {Perunicic}}, \bibinfo
  {author} {\bibfnamefont {P.}~\bibnamefont {Senn}}, \bibinfo {author}
  {\bibfnamefont {S.}~\bibnamefont {Steinert}}, \bibinfo {author}
  {\bibfnamefont {L.~P.}\ \bibnamefont {McGuinness}}, \bibinfo {author}
  {\bibfnamefont {B.~C.}\ \bibnamefont {Johnson}}, \bibinfo {author}
  {\bibfnamefont {T.}~\bibnamefont {Ohshima}}, \bibinfo {author} {\bibfnamefont
  {F.}~\bibnamefont {Caruso}}, \bibinfo {author} {\bibfnamefont
  {J.}~\bibnamefont {Wrachtrup}}, \bibinfo {author} {\bibfnamefont {R.~E.}\
  \bibnamefont {Scholten}}, \bibinfo {author} {\bibfnamefont {P.}~\bibnamefont
  {Mulvaney}},\ and\ \bibinfo {author} {\bibfnamefont {L.}~\bibnamefont
  {Hollenberg}},\ }\bibfield  {title} {\bibinfo {title} {{Detection of atomic
  spin labels in a lipid bilayer using a single-spin nanodiamond probe}},\
  }\href {https://doi.org/10.1073/pnas.1300640110} {\bibfield  {journal}
  {\bibinfo  {journal} {Proceedings of the National Academy of Sciences of the
  United States of America}\ }\textbf {\bibinfo {volume} {110}},\ \bibinfo
  {pages} {10894} (\bibinfo {year} {2013})}\BibitemShut {NoStop}%
\bibitem [{\citenamefont {Sushkov}\ \emph {et~al.}(2014)\citenamefont
  {Sushkov}, \citenamefont {Chisholm}, \citenamefont {Lovchinsky},
  \citenamefont {Kubo}, \citenamefont {Lo}, \citenamefont {Bennett},
  \citenamefont {Hunger}, \citenamefont {Akimov}, \citenamefont {Walsworth},
  \citenamefont {Park},\ and\ \citenamefont {Lukin}}]{Sushkov2014}%
  \BibitemOpen
  \bibfield  {author} {\bibinfo {author} {\bibfnamefont {A.~O.}\ \bibnamefont
  {Sushkov}}, \bibinfo {author} {\bibfnamefont {N.}~\bibnamefont {Chisholm}},
  \bibinfo {author} {\bibfnamefont {I.}~\bibnamefont {Lovchinsky}}, \bibinfo
  {author} {\bibfnamefont {M.}~\bibnamefont {Kubo}}, \bibinfo {author}
  {\bibfnamefont {P.~K.}\ \bibnamefont {Lo}}, \bibinfo {author} {\bibfnamefont
  {S.~D.}\ \bibnamefont {Bennett}}, \bibinfo {author} {\bibfnamefont
  {D.}~\bibnamefont {Hunger}}, \bibinfo {author} {\bibfnamefont
  {A.}~\bibnamefont {Akimov}}, \bibinfo {author} {\bibfnamefont {R.~L.}\
  \bibnamefont {Walsworth}}, \bibinfo {author} {\bibfnamefont {H.}~\bibnamefont
  {Park}},\ and\ \bibinfo {author} {\bibfnamefont {M.~D.}\ \bibnamefont
  {Lukin}},\ }\bibfield  {title} {\bibinfo {title} {All-optical sensing of a
  single-molecule electron spin},\ }\href {https://doi.org/10.1021/nl502988n}
  {\bibfield  {journal} {\bibinfo  {journal} {Nano Letters}\ }\textbf {\bibinfo
  {volume} {14}},\ \bibinfo {pages} {6443} (\bibinfo {year}
  {2014})}\BibitemShut {NoStop}%
\bibitem [{\citenamefont {Ziem}\ \emph {et~al.}(2013)\citenamefont {Ziem},
  \citenamefont {G{\"{o}}tz}, \citenamefont {Zappe}, \citenamefont {Steinert},\
  and\ \citenamefont {Wrachtrup}}]{Ziem2013}%
  \BibitemOpen
  \bibfield  {author} {\bibinfo {author} {\bibfnamefont {F.~C.}\ \bibnamefont
  {Ziem}}, \bibinfo {author} {\bibfnamefont {N.~S.}\ \bibnamefont
  {G{\"{o}}tz}}, \bibinfo {author} {\bibfnamefont {A.}~\bibnamefont {Zappe}},
  \bibinfo {author} {\bibfnamefont {S.}~\bibnamefont {Steinert}},\ and\
  \bibinfo {author} {\bibfnamefont {J.}~\bibnamefont {Wrachtrup}},\ }\bibfield
  {title} {\bibinfo {title} {{Highly Sensitive Detection of Physiological Spins
  in a Microfluidic Device}},\ }\href {https://doi.org/10.1021/nl401522a}
  {\bibfield  {journal} {\bibinfo  {journal} {Nano Letters}\ }\textbf {\bibinfo
  {volume} {13}},\ \bibinfo {pages} {4093} (\bibinfo {year}
  {2013})}\BibitemShut {NoStop}%
\bibitem [{\citenamefont {Simpson}\ \emph {et~al.}(2017)\citenamefont
  {Simpson}, \citenamefont {Ryan}, \citenamefont {Hall}, \citenamefont
  {Panchenko}, \citenamefont {Drew}, \citenamefont {Petrou}, \citenamefont
  {Donnelly}, \citenamefont {Mulvaney},\ and\ \citenamefont
  {Hollenberg}}]{Simpson2017}%
  \BibitemOpen
  \bibfield  {author} {\bibinfo {author} {\bibfnamefont {D.~A.}\ \bibnamefont
  {Simpson}}, \bibinfo {author} {\bibfnamefont {R.~G.}\ \bibnamefont {Ryan}},
  \bibinfo {author} {\bibfnamefont {L.~T.}\ \bibnamefont {Hall}}, \bibinfo
  {author} {\bibfnamefont {E.}~\bibnamefont {Panchenko}}, \bibinfo {author}
  {\bibfnamefont {S.~C.}\ \bibnamefont {Drew}}, \bibinfo {author}
  {\bibfnamefont {S.}~\bibnamefont {Petrou}}, \bibinfo {author} {\bibfnamefont
  {P.~S.}\ \bibnamefont {Donnelly}}, \bibinfo {author} {\bibfnamefont
  {P.}~\bibnamefont {Mulvaney}},\ and\ \bibinfo {author} {\bibfnamefont
  {L.~C.~L.}\ \bibnamefont {Hollenberg}},\ }\bibfield  {title} {\bibinfo
  {title} {{Electron paramagnetic resonance microscopy using spins in diamond
  under ambient conditions}},\ }\href
  {https://doi.org/10.1038/s41467-017-00466-y} {\bibfield  {journal} {\bibinfo
  {journal} {Nature Communications}\ }\textbf {\bibinfo {volume} {8}},\
  \bibinfo {pages} {458} (\bibinfo {year} {2017})}\BibitemShut {NoStop}%
\bibitem [{\citenamefont {Sharmin}\ \emph {et~al.}(2022)\citenamefont
  {Sharmin}, \citenamefont {Nusantara}, \citenamefont {Nie}, \citenamefont
  {Wu}, \citenamefont {Elias~Llumbet}, \citenamefont {Woudstra}, \citenamefont
  {Mzyk},\ and\ \citenamefont {Schirhagl}}]{Sharmin2022}%
  \BibitemOpen
  \bibfield  {author} {\bibinfo {author} {\bibfnamefont {R.}~\bibnamefont
  {Sharmin}}, \bibinfo {author} {\bibfnamefont {A.~C.}\ \bibnamefont
  {Nusantara}}, \bibinfo {author} {\bibfnamefont {L.}~\bibnamefont {Nie}},
  \bibinfo {author} {\bibfnamefont {K.}~\bibnamefont {Wu}}, \bibinfo {author}
  {\bibfnamefont {A.}~\bibnamefont {Elias~Llumbet}}, \bibinfo {author}
  {\bibfnamefont {W.}~\bibnamefont {Woudstra}}, \bibinfo {author}
  {\bibfnamefont {A.}~\bibnamefont {Mzyk}},\ and\ \bibinfo {author}
  {\bibfnamefont {R.}~\bibnamefont {Schirhagl}},\ }\bibfield  {title} {\bibinfo
  {title} {Intracellular {Quantum} {Sensing} of {Free}-{Radical} {Generation}
  {Induced} by {Acetaminophen} ({APAP}) in the {Cytosol}, in {Mitochondria} and
  the {Nucleus} of {Macrophages}},\ }\href
  {https://doi.org/10.1021/acssensors.2c01272} {\bibfield  {journal} {\bibinfo
  {journal} {ACS Sensors}\ }\textbf {\bibinfo {volume} {7}},\ \bibinfo {pages}
  {3326} (\bibinfo {year} {2022})}\BibitemShut {NoStop}%
\bibitem [{\citenamefont {Nie}\ \emph {et~al.}(2022)\citenamefont {Nie},
  \citenamefont {Nusantara}, \citenamefont {Damle}, \citenamefont {Baranov},
  \citenamefont {Chipaux}, \citenamefont {Reyes-San-Martin}, \citenamefont
  {Hamoh}, \citenamefont {Epperla}, \citenamefont {Guricova}, \citenamefont
  {Cigler}, \citenamefont {van~den Bogaart},\ and\ \citenamefont
  {Schirhagl}}]{Nie2022}%
  \BibitemOpen
  \bibfield  {author} {\bibinfo {author} {\bibfnamefont {L.}~\bibnamefont
  {Nie}}, \bibinfo {author} {\bibfnamefont {A.~C.}\ \bibnamefont {Nusantara}},
  \bibinfo {author} {\bibfnamefont {V.~G.}\ \bibnamefont {Damle}}, \bibinfo
  {author} {\bibfnamefont {M.~V.}\ \bibnamefont {Baranov}}, \bibinfo {author}
  {\bibfnamefont {M.}~\bibnamefont {Chipaux}}, \bibinfo {author} {\bibfnamefont
  {C.}~\bibnamefont {Reyes-San-Martin}}, \bibinfo {author} {\bibfnamefont
  {T.}~\bibnamefont {Hamoh}}, \bibinfo {author} {\bibfnamefont {C.~P.}\
  \bibnamefont {Epperla}}, \bibinfo {author} {\bibfnamefont {M.}~\bibnamefont
  {Guricova}}, \bibinfo {author} {\bibfnamefont {P.}~\bibnamefont {Cigler}},
  \bibinfo {author} {\bibfnamefont {G.}~\bibnamefont {van~den Bogaart}},\ and\
  \bibinfo {author} {\bibfnamefont {R.}~\bibnamefont {Schirhagl}},\ }\bibfield
  {title} {\bibinfo {title} {Quantum sensing of free radicals in primary human
  dendritic cells},\ }\href {https://doi.org/10.1021/acs.nanolett.1c03021}
  {\bibfield  {journal} {\bibinfo  {journal} {Nano Letters}\ }\textbf {\bibinfo
  {volume} {22}},\ \bibinfo {pages} {1818} (\bibinfo {year}
  {2022})}\BibitemShut {NoStop}%
\bibitem [{\citenamefont {Grant}\ \emph {et~al.}(2023)\citenamefont {Grant},
  \citenamefont {Hall}, \citenamefont {Hollenberg}, \citenamefont {McColl},\
  and\ \citenamefont {Simpson}}]{Grant2023}%
  \BibitemOpen
  \bibfield  {author} {\bibinfo {author} {\bibfnamefont {E.~S.}\ \bibnamefont
  {Grant}}, \bibinfo {author} {\bibfnamefont {L.~T.}\ \bibnamefont {Hall}},
  \bibinfo {author} {\bibfnamefont {L.~C.~L.}\ \bibnamefont {Hollenberg}},
  \bibinfo {author} {\bibfnamefont {G.}~\bibnamefont {McColl}},\ and\ \bibinfo
  {author} {\bibfnamefont {D.~A.}\ \bibnamefont {Simpson}},\ }\bibfield
  {title} {\bibinfo {title} {Nonmonotonic superparamagnetic behavior of the
  ferritin iron core revealed via quantum spin relaxometry},\ }\href
  {https://doi.org/10.1021/acsnano.2c08698} {\bibfield  {journal} {\bibinfo
  {journal} {ACS Nano}\ }\textbf {\bibinfo {volume} {17}},\ \bibinfo {pages}
  {372} (\bibinfo {year} {2023})}\BibitemShut {NoStop}%
\bibitem [{\citenamefont {Li}\ \emph {et~al.}(2022)\citenamefont {Li},
  \citenamefont {Soleyman}, \citenamefont {Kohandel},\ and\ \citenamefont
  {Cappellaro}}]{Changhao2022}%
  \BibitemOpen
  \bibfield  {author} {\bibinfo {author} {\bibfnamefont {C.}~\bibnamefont
  {Li}}, \bibinfo {author} {\bibfnamefont {R.}~\bibnamefont {Soleyman}},
  \bibinfo {author} {\bibfnamefont {M.}~\bibnamefont {Kohandel}},\ and\
  \bibinfo {author} {\bibfnamefont {P.}~\bibnamefont {Cappellaro}},\ }\bibfield
   {title} {\bibinfo {title} {{SARS-CoV-2 Quantum Sensor Based on
  Nitrogen-Vacancy Centers in Diamond}},\ }\href
  {https://doi.org/10.1021/acs.nanolett.1c02868} {\bibfield  {journal}
  {\bibinfo  {journal} {Nano Letters}\ }\textbf {\bibinfo {volume} {22}},\
  \bibinfo {pages} {43} (\bibinfo {year} {2022})}\BibitemShut {NoStop}%
\bibitem [{\citenamefont {Rendler}\ \emph {et~al.}(2017)\citenamefont
  {Rendler}, \citenamefont {Neburkova}, \citenamefont {Zemek}, \citenamefont
  {Kotek}, \citenamefont {Zappe}, \citenamefont {Chu}, \citenamefont {Cigler},\
  and\ \citenamefont {Wrachtrup}}]{Rendler2017}%
  \BibitemOpen
  \bibfield  {author} {\bibinfo {author} {\bibfnamefont {T.}~\bibnamefont
  {Rendler}}, \bibinfo {author} {\bibfnamefont {J.}~\bibnamefont {Neburkova}},
  \bibinfo {author} {\bibfnamefont {O.}~\bibnamefont {Zemek}}, \bibinfo
  {author} {\bibfnamefont {J.}~\bibnamefont {Kotek}}, \bibinfo {author}
  {\bibfnamefont {A.}~\bibnamefont {Zappe}}, \bibinfo {author} {\bibfnamefont
  {Z.}~\bibnamefont {Chu}}, \bibinfo {author} {\bibfnamefont {P.}~\bibnamefont
  {Cigler}},\ and\ \bibinfo {author} {\bibfnamefont {J.}~\bibnamefont
  {Wrachtrup}},\ }\bibfield  {title} {\bibinfo {title} {Optical imaging of
  localized chemical events using programmable diamond quantum nanosensors},\
  }\href {https://doi.org/10.1038/ncomms14701} {\bibfield  {journal} {\bibinfo
  {journal} {Nature Communications}\ }\textbf {\bibinfo {volume} {8}},\
  \bibinfo {pages} {14701} (\bibinfo {year} {2017})}\BibitemShut {NoStop}%
\bibitem [{\citenamefont {Kaviani}\ \emph {et~al.}(2014)\citenamefont
  {Kaviani}, \citenamefont {Deak}, \citenamefont {Aradi}, \citenamefont
  {Frauenheim}, \citenamefont {Chou},\ and\ \citenamefont
  {Gali}}]{Kaviani2014}%
  \BibitemOpen
  \bibfield  {author} {\bibinfo {author} {\bibfnamefont {M.}~\bibnamefont
  {Kaviani}}, \bibinfo {author} {\bibfnamefont {P.}~\bibnamefont {Deak}},
  \bibinfo {author} {\bibfnamefont {B.}~\bibnamefont {Aradi}}, \bibinfo
  {author} {\bibfnamefont {T.}~\bibnamefont {Frauenheim}}, \bibinfo {author}
  {\bibfnamefont {J.~P.}\ \bibnamefont {Chou}},\ and\ \bibinfo {author}
  {\bibfnamefont {A.}~\bibnamefont {Gali}},\ }\bibfield  {title} {\bibinfo
  {title} {Proper surface termination for luminescent near-surface nv centers
  in diamond},\ }\href {https://doi.org/10.1021/nl501927y} {\bibfield
  {journal} {\bibinfo  {journal} {Nano Letters}\ }\textbf {\bibinfo {volume}
  {14}},\ \bibinfo {pages} {4772} (\bibinfo {year} {2014})}\BibitemShut
  {NoStop}%
\bibitem [{\citenamefont {Stacey}\ \emph {et~al.}(2018)\citenamefont {Stacey},
  \citenamefont {Dontschuk}, \citenamefont {Chou}, \citenamefont {Broadway},
  \citenamefont {Schenk}, \citenamefont {Sear}, \citenamefont {Tetienne},
  \citenamefont {Hoffman}, \citenamefont {Prawer}, \citenamefont {Pakes},
  \citenamefont {Tadich}, \citenamefont {de~Leon}, \citenamefont {Gali},\ and\
  \citenamefont {Hollenberg}}]{Stacey2018}%
  \BibitemOpen
  \bibfield  {author} {\bibinfo {author} {\bibfnamefont {A.}~\bibnamefont
  {Stacey}}, \bibinfo {author} {\bibfnamefont {N.}~\bibnamefont {Dontschuk}},
  \bibinfo {author} {\bibfnamefont {J.}~\bibnamefont {Chou}}, \bibinfo {author}
  {\bibfnamefont {D.~A.}\ \bibnamefont {Broadway}}, \bibinfo {author}
  {\bibfnamefont {A.~K.}\ \bibnamefont {Schenk}}, \bibinfo {author}
  {\bibfnamefont {M.~J.}\ \bibnamefont {Sear}}, \bibinfo {author}
  {\bibfnamefont {J.}~\bibnamefont {Tetienne}}, \bibinfo {author}
  {\bibfnamefont {A.}~\bibnamefont {Hoffman}}, \bibinfo {author} {\bibfnamefont
  {S.}~\bibnamefont {Prawer}}, \bibinfo {author} {\bibfnamefont {C.~I.}\
  \bibnamefont {Pakes}}, \bibinfo {author} {\bibfnamefont {A.}~\bibnamefont
  {Tadich}}, \bibinfo {author} {\bibfnamefont {N.~P.}\ \bibnamefont {de~Leon}},
  \bibinfo {author} {\bibfnamefont {A.}~\bibnamefont {Gali}},\ and\ \bibinfo
  {author} {\bibfnamefont {L.~C.~L.}\ \bibnamefont {Hollenberg}},\ }\bibfield
  {title} {\bibinfo {title} {Evidence for primal sp2 defects at the diamond
  surface: Candidates for electron trapping and noise sources},\ }\href
  {https://doi.org/10.1002/admi.201801449} {\bibfield  {journal} {\bibinfo
  {journal} {Advanced Materials Interfaces}\ }\textbf {\bibinfo {volume} {6}},\
  \bibinfo {pages} {1801449} (\bibinfo {year} {2018})}\BibitemShut {NoStop}%
\bibitem [{\citenamefont {Bluvstein}\ \emph {et~al.}(2019)\citenamefont
  {Bluvstein}, \citenamefont {Zhang},\ and\ \citenamefont
  {Jayich}}]{Bluvstein2019}%
  \BibitemOpen
  \bibfield  {author} {\bibinfo {author} {\bibfnamefont {D.}~\bibnamefont
  {Bluvstein}}, \bibinfo {author} {\bibfnamefont {Z.}~\bibnamefont {Zhang}},\
  and\ \bibinfo {author} {\bibfnamefont {A.~C.~B.}\ \bibnamefont {Jayich}},\
  }\bibfield  {title} {\bibinfo {title} {Identifying and mitigating charge
  instabilities in shallow diamond nitrogen-vacancy centers},\ }\href
  {https://doi.org/10.1103/PhysRevLett.122.076101} {\bibfield  {journal}
  {\bibinfo  {journal} {Phys Rev Lett}\ }\textbf {\bibinfo {volume} {122}},\
  \bibinfo {pages} {076101} (\bibinfo {year} {2019})}\BibitemShut {NoStop}%
\bibitem [{\citenamefont {Ziem}\ \emph {et~al.}(2019)\citenamefont {Ziem},
  \citenamefont {Garsi}, \citenamefont {Fedder},\ and\ \citenamefont
  {Wrachtrup}}]{Ziem2019}%
  \BibitemOpen
  \bibfield  {author} {\bibinfo {author} {\bibfnamefont {F.}~\bibnamefont
  {Ziem}}, \bibinfo {author} {\bibfnamefont {M.}~\bibnamefont {Garsi}},
  \bibinfo {author} {\bibfnamefont {H.}~\bibnamefont {Fedder}},\ and\ \bibinfo
  {author} {\bibfnamefont {J.}~\bibnamefont {Wrachtrup}},\ }\bibfield  {title}
  {\bibinfo {title} {{Quantitative nanoscale MRI with a wide field of view}},\
  }\href {https://doi.org/10.1038/s41598-019-47084-w} {\bibfield  {journal}
  {\bibinfo  {journal} {Scientific Reports}\ }\textbf {\bibinfo {volume} {9}},\
  \bibinfo {pages} {12166} (\bibinfo {year} {2019})}\BibitemShut {NoStop}%
\bibitem [{\citenamefont {Healey}\ \emph {et~al.}(2021)\citenamefont {Healey},
  \citenamefont {Hall}, \citenamefont {White}, \citenamefont {Teraji},
  \citenamefont {Sani}, \citenamefont {Separovic}, \citenamefont {Tetienne},\
  and\ \citenamefont {Hollenberg}}]{HealeyPRA2021}%
  \BibitemOpen
  \bibfield  {author} {\bibinfo {author} {\bibfnamefont {A.~J.}\ \bibnamefont
  {Healey}}, \bibinfo {author} {\bibfnamefont {L.~T.}\ \bibnamefont {Hall}},
  \bibinfo {author} {\bibfnamefont {G.~A.}\ \bibnamefont {White}}, \bibinfo
  {author} {\bibfnamefont {T.}~\bibnamefont {Teraji}}, \bibinfo {author}
  {\bibfnamefont {M.~A.}\ \bibnamefont {Sani}}, \bibinfo {author}
  {\bibfnamefont {F.}~\bibnamefont {Separovic}}, \bibinfo {author}
  {\bibfnamefont {J.~P.}\ \bibnamefont {Tetienne}},\ and\ \bibinfo {author}
  {\bibfnamefont {L.~C.}\ \bibnamefont {Hollenberg}},\ }\bibfield  {title}
  {\bibinfo {title} {{Polarization Transfer to External Nuclear Spins Using
  Ensembles of Nitrogen-Vacancy Centers}},\ }\href
  {https://doi.org/10.1103/PhysRevApplied.15.054052} {\bibfield  {journal}
  {\bibinfo  {journal} {Physical Review Applied}\ }\textbf {\bibinfo {volume}
  {15}},\ \bibinfo {pages} {054052} (\bibinfo {year} {2021})}\BibitemShut
  {NoStop}%
\bibitem [{\citenamefont {Liu}\ \emph {et~al.}(2022{\natexlab{a}})\citenamefont
  {Liu}, \citenamefont {Henning}, \citenamefont {Heindl}, \citenamefont
  {Allert}, \citenamefont {Bartl}, \citenamefont {Sharp}, \citenamefont
  {Rizzato},\ and\ \citenamefont {Bucher}}]{Liu2022}%
  \BibitemOpen
  \bibfield  {author} {\bibinfo {author} {\bibfnamefont {K.~S.}\ \bibnamefont
  {Liu}}, \bibinfo {author} {\bibfnamefont {A.}~\bibnamefont {Henning}},
  \bibinfo {author} {\bibfnamefont {M.~W.}\ \bibnamefont {Heindl}}, \bibinfo
  {author} {\bibfnamefont {R.~D.}\ \bibnamefont {Allert}}, \bibinfo {author}
  {\bibfnamefont {J.~D.}\ \bibnamefont {Bartl}}, \bibinfo {author}
  {\bibfnamefont {I.~D.}\ \bibnamefont {Sharp}}, \bibinfo {author}
  {\bibfnamefont {R.}~\bibnamefont {Rizzato}},\ and\ \bibinfo {author}
  {\bibfnamefont {D.~B.}\ \bibnamefont {Bucher}},\ }\bibfield  {title}
  {\bibinfo {title} {Surface nmr using quantum sensors in diamond},\ }\href
  {https://doi.org/10.1073/pnas.2111607119} {\bibfield  {journal} {\bibinfo
  {journal} {Proc. Natl. Acad. Sci. U.S.A.}\ }\textbf {\bibinfo {volume}
  {119}},\ \bibinfo {pages} {e2111607119} (\bibinfo {year}
  {2022}{\natexlab{a}})}\BibitemShut {NoStop}%
\bibitem [{\citenamefont {Sangtawesin}\ \emph {et~al.}(2019)\citenamefont
  {Sangtawesin}, \citenamefont {Dwyer}, \citenamefont {Srinivasan},
  \citenamefont {Allred}, \citenamefont {Rodgers}, \citenamefont {De~Greve},
  \citenamefont {Stacey}, \citenamefont {Dontschuk}, \citenamefont {O'Donnell},
  \citenamefont {Hu}, \citenamefont {Evans}, \citenamefont {Jaye},
  \citenamefont {Fischer}, \citenamefont {Markham}, \citenamefont {Twitchen},
  \citenamefont {Park}, \citenamefont {Lukin},\ and\ \citenamefont
  {de~Leon}}]{Sangtawesin2019}%
  \BibitemOpen
  \bibfield  {author} {\bibinfo {author} {\bibfnamefont {S.}~\bibnamefont
  {Sangtawesin}}, \bibinfo {author} {\bibfnamefont {B.~L.}\ \bibnamefont
  {Dwyer}}, \bibinfo {author} {\bibfnamefont {S.}~\bibnamefont {Srinivasan}},
  \bibinfo {author} {\bibfnamefont {J.~J.}\ \bibnamefont {Allred}}, \bibinfo
  {author} {\bibfnamefont {L.~V.~H.}\ \bibnamefont {Rodgers}}, \bibinfo
  {author} {\bibfnamefont {K.}~\bibnamefont {De~Greve}}, \bibinfo {author}
  {\bibfnamefont {A.}~\bibnamefont {Stacey}}, \bibinfo {author} {\bibfnamefont
  {N.}~\bibnamefont {Dontschuk}}, \bibinfo {author} {\bibfnamefont {K.~M.}\
  \bibnamefont {O'Donnell}}, \bibinfo {author} {\bibfnamefont {D.}~\bibnamefont
  {Hu}}, \bibinfo {author} {\bibfnamefont {D.~A.}\ \bibnamefont {Evans}},
  \bibinfo {author} {\bibfnamefont {C.}~\bibnamefont {Jaye}}, \bibinfo {author}
  {\bibfnamefont {D.~A.}\ \bibnamefont {Fischer}}, \bibinfo {author}
  {\bibfnamefont {M.~L.}\ \bibnamefont {Markham}}, \bibinfo {author}
  {\bibfnamefont {D.~J.}\ \bibnamefont {Twitchen}}, \bibinfo {author}
  {\bibfnamefont {H.}~\bibnamefont {Park}}, \bibinfo {author} {\bibfnamefont
  {M.~D.}\ \bibnamefont {Lukin}},\ and\ \bibinfo {author} {\bibfnamefont
  {N.~P.}\ \bibnamefont {de~Leon}},\ }\bibfield  {title} {\bibinfo {title}
  {Origins of diamond surface noise probed by correlating single-spin
  measurements with surface spectroscopy},\ }\href
  {https://doi.org/10.1103/PhysRevX.9.031052} {\bibfield  {journal} {\bibinfo
  {journal} {Phys. Rev. X}\ }\textbf {\bibinfo {volume} {9}},\ \bibinfo {pages}
  {031052} (\bibinfo {year} {2019})}\BibitemShut {NoStop}%
\bibitem [{\citenamefont {Miller}\ \emph {et~al.}(2020)\citenamefont {Miller},
  \citenamefont {Bezinge}, \citenamefont {Gliddon}, \citenamefont {Huang},
  \citenamefont {Dold}, \citenamefont {Gray}, \citenamefont {Heaney},
  \citenamefont {Dobson}, \citenamefont {Nastouli}, \citenamefont {Morton},\
  and\ \citenamefont {McKendry}}]{Miller2020}%
  \BibitemOpen
  \bibfield  {author} {\bibinfo {author} {\bibfnamefont {B.~S.}\ \bibnamefont
  {Miller}}, \bibinfo {author} {\bibfnamefont {L.}~\bibnamefont {Bezinge}},
  \bibinfo {author} {\bibfnamefont {H.~D.}\ \bibnamefont {Gliddon}}, \bibinfo
  {author} {\bibfnamefont {D.}~\bibnamefont {Huang}}, \bibinfo {author}
  {\bibfnamefont {G.}~\bibnamefont {Dold}}, \bibinfo {author} {\bibfnamefont
  {E.~R.}\ \bibnamefont {Gray}}, \bibinfo {author} {\bibfnamefont
  {J.}~\bibnamefont {Heaney}}, \bibinfo {author} {\bibfnamefont {P.~J.}\
  \bibnamefont {Dobson}}, \bibinfo {author} {\bibfnamefont {E.}~\bibnamefont
  {Nastouli}}, \bibinfo {author} {\bibfnamefont {J.~J.~L.}\ \bibnamefont
  {Morton}},\ and\ \bibinfo {author} {\bibfnamefont {R.~A.}\ \bibnamefont
  {McKendry}},\ }\bibfield  {title} {\bibinfo {title} {{Spin-enhanced
  nanodiamond biosensing for ultrasensitive diagnostics}},\ }\href
  {https://doi.org/10.1038/s41586-020-2917-1} {\bibfield  {journal} {\bibinfo
  {journal} {Nature}\ }\textbf {\bibinfo {volume} {587}},\ \bibinfo {pages}
  {588} (\bibinfo {year} {2020})}\BibitemShut {NoStop}%
\bibitem [{\citenamefont {Gottscholl}\ \emph {et~al.}(2020)\citenamefont
  {Gottscholl}, \citenamefont {Kianinia}, \citenamefont {Soltamov},
  \citenamefont {Orlinskii}, \citenamefont {Mamin}, \citenamefont {Bradac},
  \citenamefont {Kasper}, \citenamefont {Krambrock}, \citenamefont {Sperlich},
  \citenamefont {Toth}, \citenamefont {Aharonovich},\ and\ \citenamefont
  {Dyakonov}}]{GottschollNM2020}%
  \BibitemOpen
  \bibfield  {author} {\bibinfo {author} {\bibfnamefont {A.}~\bibnamefont
  {Gottscholl}}, \bibinfo {author} {\bibfnamefont {M.}~\bibnamefont
  {Kianinia}}, \bibinfo {author} {\bibfnamefont {V.}~\bibnamefont {Soltamov}},
  \bibinfo {author} {\bibfnamefont {S.}~\bibnamefont {Orlinskii}}, \bibinfo
  {author} {\bibfnamefont {G.}~\bibnamefont {Mamin}}, \bibinfo {author}
  {\bibfnamefont {C.}~\bibnamefont {Bradac}}, \bibinfo {author} {\bibfnamefont
  {C.}~\bibnamefont {Kasper}}, \bibinfo {author} {\bibfnamefont
  {K.}~\bibnamefont {Krambrock}}, \bibinfo {author} {\bibfnamefont
  {A.}~\bibnamefont {Sperlich}}, \bibinfo {author} {\bibfnamefont
  {M.}~\bibnamefont {Toth}}, \bibinfo {author} {\bibfnamefont {I.}~\bibnamefont
  {Aharonovich}},\ and\ \bibinfo {author} {\bibfnamefont {V.}~\bibnamefont
  {Dyakonov}},\ }\bibfield  {title} {\bibinfo {title} {{Initialization and
  read-out of intrinsic spin defects in a van der Waals crystal at room
  temperature}},\ }\href {https://doi.org/10.1038/s41563-020-0619-6} {\bibfield
   {journal} {\bibinfo  {journal} {Nature Materials}\ }\textbf {\bibinfo
  {volume} {19}},\ \bibinfo {pages} {540} (\bibinfo {year} {2020})}\BibitemShut
  {NoStop}%
\bibitem [{\citenamefont {Gottscholl}\ \emph
  {et~al.}(2021{\natexlab{a}})\citenamefont {Gottscholl}, \citenamefont {Diez},
  \citenamefont {Soltamov}, \citenamefont {Kasper}, \citenamefont {Krausse},
  \citenamefont {Sperlich}, \citenamefont {Kianinia}, \citenamefont {Bradac},
  \citenamefont {Aharonovich},\ and\ \citenamefont
  {Dyakonov}}]{Gottscholl2021NC}%
  \BibitemOpen
  \bibfield  {author} {\bibinfo {author} {\bibfnamefont {A.}~\bibnamefont
  {Gottscholl}}, \bibinfo {author} {\bibfnamefont {M.}~\bibnamefont {Diez}},
  \bibinfo {author} {\bibfnamefont {V.}~\bibnamefont {Soltamov}}, \bibinfo
  {author} {\bibfnamefont {C.}~\bibnamefont {Kasper}}, \bibinfo {author}
  {\bibfnamefont {D.}~\bibnamefont {Krausse}}, \bibinfo {author} {\bibfnamefont
  {A.}~\bibnamefont {Sperlich}}, \bibinfo {author} {\bibfnamefont
  {M.}~\bibnamefont {Kianinia}}, \bibinfo {author} {\bibfnamefont
  {C.}~\bibnamefont {Bradac}}, \bibinfo {author} {\bibfnamefont
  {I.}~\bibnamefont {Aharonovich}},\ and\ \bibinfo {author} {\bibfnamefont
  {V.}~\bibnamefont {Dyakonov}},\ }\bibfield  {title} {\bibinfo {title} {Spin
  defects in hbn as promising temperature, pressure and magnetic field quantum
  sensors},\ }\href {https://doi.org/10.1038/s41467-021-24725-1} {\bibfield
  {journal} {\bibinfo  {journal} {Nature Communications}\ }\textbf {\bibinfo
  {volume} {12}},\ \bibinfo {pages} {4480} (\bibinfo {year}
  {2021}{\natexlab{a}})}\BibitemShut {NoStop}%
\bibitem [{\citenamefont {Gao}\ \emph {et~al.}(2021)\citenamefont {Gao},
  \citenamefont {Jiang}, \citenamefont {Llacsahuanga~Allcca}, \citenamefont
  {Shen}, \citenamefont {Sadi}, \citenamefont {Solanki}, \citenamefont {Ju},
  \citenamefont {Xu}, \citenamefont {Upadhyaya}, \citenamefont {Chen},
  \citenamefont {Bhave},\ and\ \citenamefont {Li}}]{Xingyu2021}%
  \BibitemOpen
  \bibfield  {author} {\bibinfo {author} {\bibfnamefont {X.}~\bibnamefont
  {Gao}}, \bibinfo {author} {\bibfnamefont {B.}~\bibnamefont {Jiang}}, \bibinfo
  {author} {\bibfnamefont {A.~E.}\ \bibnamefont {Llacsahuanga~Allcca}},
  \bibinfo {author} {\bibfnamefont {K.}~\bibnamefont {Shen}}, \bibinfo {author}
  {\bibfnamefont {M.~A.}\ \bibnamefont {Sadi}}, \bibinfo {author}
  {\bibfnamefont {A.~B.}\ \bibnamefont {Solanki}}, \bibinfo {author}
  {\bibfnamefont {P.}~\bibnamefont {Ju}}, \bibinfo {author} {\bibfnamefont
  {Z.}~\bibnamefont {Xu}}, \bibinfo {author} {\bibfnamefont {P.}~\bibnamefont
  {Upadhyaya}}, \bibinfo {author} {\bibfnamefont {Y.~P.}\ \bibnamefont {Chen}},
  \bibinfo {author} {\bibfnamefont {S.~A.}\ \bibnamefont {Bhave}},\ and\
  \bibinfo {author} {\bibfnamefont {T.}~\bibnamefont {Li}},\ }\bibfield
  {title} {\bibinfo {title} {High-contrast plasmonic-enhanced shallow spin
  defects in hexagonal boron nitride for quantum sensing},\ }\href
  {https://doi.org/10.1021/acs.nanolett.1c02495} {\bibfield  {journal}
  {\bibinfo  {journal} {Nano Letters}\ }\textbf {\bibinfo {volume} {21}},\
  \bibinfo {pages} {7708} (\bibinfo {year} {2021})}\BibitemShut {NoStop}%
\bibitem [{\citenamefont {Liu}\ \emph {et~al.}(2021)\citenamefont {Liu},
  \citenamefont {Li}, \citenamefont {Yang}, \citenamefont {Yu}, \citenamefont
  {Meng}, \citenamefont {Wang}, \citenamefont {Li}, \citenamefont {Guo},
  \citenamefont {Yan}, \citenamefont {Li}, \citenamefont {Wang}, \citenamefont
  {Xu}, \citenamefont {Wang}, \citenamefont {Tang}, \citenamefont {Li},\ and\
  \citenamefont {Guo}}]{Liu2021}%
  \BibitemOpen
  \bibfield  {author} {\bibinfo {author} {\bibfnamefont {W.}~\bibnamefont
  {Liu}}, \bibinfo {author} {\bibfnamefont {Z.-P.}\ \bibnamefont {Li}},
  \bibinfo {author} {\bibfnamefont {Y.-Z.}\ \bibnamefont {Yang}}, \bibinfo
  {author} {\bibfnamefont {S.}~\bibnamefont {Yu}}, \bibinfo {author}
  {\bibfnamefont {Y.}~\bibnamefont {Meng}}, \bibinfo {author} {\bibfnamefont
  {Z.-A.}\ \bibnamefont {Wang}}, \bibinfo {author} {\bibfnamefont {Z.-C.}\
  \bibnamefont {Li}}, \bibinfo {author} {\bibfnamefont {N.-J.}\ \bibnamefont
  {Guo}}, \bibinfo {author} {\bibfnamefont {F.-F.}\ \bibnamefont {Yan}},
  \bibinfo {author} {\bibfnamefont {Q.}~\bibnamefont {Li}}, \bibinfo {author}
  {\bibfnamefont {J.-F.}\ \bibnamefont {Wang}}, \bibinfo {author}
  {\bibfnamefont {J.-S.}\ \bibnamefont {Xu}}, \bibinfo {author} {\bibfnamefont
  {Y.-T.}\ \bibnamefont {Wang}}, \bibinfo {author} {\bibfnamefont {J.-S.}\
  \bibnamefont {Tang}}, \bibinfo {author} {\bibfnamefont {C.-F.}\ \bibnamefont
  {Li}},\ and\ \bibinfo {author} {\bibfnamefont {G.-C.}\ \bibnamefont {Guo}},\
  }\bibfield  {title} {\bibinfo {title} {Temperature-dependent energy-level
  shifts of spin defects in hexagonal boron nitride},\ }\href
  {https://doi.org/10.1021/acsphotonics.1c00320} {\bibfield  {journal}
  {\bibinfo  {journal} {ACS Photonics}\ }\textbf {\bibinfo {volume} {8}},\
  \bibinfo {pages} {1889} (\bibinfo {year} {2021})}\BibitemShut {NoStop}%
\bibitem [{\citenamefont {Healey}\ \emph {et~al.}(2023)\citenamefont {Healey},
  \citenamefont {Scholten}, \citenamefont {Yang}, \citenamefont {Scott},
  \citenamefont {Abrahams}, \citenamefont {Robertson}, \citenamefont {Hou},
  \citenamefont {Guo}, \citenamefont {Rahman}, \citenamefont {Lu},
  \citenamefont {Kianinia}, \citenamefont {Aharonovich},\ and\ \citenamefont
  {Tetienne}}]{HealeyNP2022}%
  \BibitemOpen
  \bibfield  {author} {\bibinfo {author} {\bibfnamefont {A.~J.}\ \bibnamefont
  {Healey}}, \bibinfo {author} {\bibfnamefont {S.~C.}\ \bibnamefont
  {Scholten}}, \bibinfo {author} {\bibfnamefont {T.}~\bibnamefont {Yang}},
  \bibinfo {author} {\bibfnamefont {J.~A.}\ \bibnamefont {Scott}}, \bibinfo
  {author} {\bibfnamefont {G.~J.}\ \bibnamefont {Abrahams}}, \bibinfo {author}
  {\bibfnamefont {I.~O.}\ \bibnamefont {Robertson}}, \bibinfo {author}
  {\bibfnamefont {X.~F.}\ \bibnamefont {Hou}}, \bibinfo {author} {\bibfnamefont
  {Y.~F.}\ \bibnamefont {Guo}}, \bibinfo {author} {\bibfnamefont
  {S.}~\bibnamefont {Rahman}}, \bibinfo {author} {\bibfnamefont
  {Y.}~\bibnamefont {Lu}}, \bibinfo {author} {\bibfnamefont {M.}~\bibnamefont
  {Kianinia}}, \bibinfo {author} {\bibfnamefont {I.}~\bibnamefont
  {Aharonovich}},\ and\ \bibinfo {author} {\bibfnamefont {J.-P.}\ \bibnamefont
  {Tetienne}},\ }\bibfield  {title} {\bibinfo {title} {{Quantum microscopy with
  van der Waals heterostructures}},\ }\href
  {https://doi.org/10.1038/s41567-022-01815-5} {\bibfield  {journal} {\bibinfo
  {journal} {Nature Physics}\ }\textbf {\bibinfo {volume} {19}},\ \bibinfo
  {pages} {87} (\bibinfo {year} {2023})}\BibitemShut {NoStop}%
\bibitem [{\citenamefont {Kumar}\ \emph {et~al.}(2022)\citenamefont {Kumar},
  \citenamefont {Fabre}, \citenamefont {Durand}, \citenamefont {Clua-Provost},
  \citenamefont {Li}, \citenamefont {Edgar}, \citenamefont {Rougemaille},
  \citenamefont {Coraux}, \citenamefont {Marie}, \citenamefont {Renucci},
  \citenamefont {Robert}, \citenamefont {Robert-Philip}, \citenamefont {Gil},
  \citenamefont {Cassabois}, \citenamefont {Finco},\ and\ \citenamefont
  {Jacques}}]{Kumar2022}%
  \BibitemOpen
  \bibfield  {author} {\bibinfo {author} {\bibfnamefont {P.}~\bibnamefont
  {Kumar}}, \bibinfo {author} {\bibfnamefont {F.}~\bibnamefont {Fabre}},
  \bibinfo {author} {\bibfnamefont {A.}~\bibnamefont {Durand}}, \bibinfo
  {author} {\bibfnamefont {T.}~\bibnamefont {Clua-Provost}}, \bibinfo {author}
  {\bibfnamefont {J.}~\bibnamefont {Li}}, \bibinfo {author} {\bibfnamefont
  {J.}~\bibnamefont {Edgar}}, \bibinfo {author} {\bibfnamefont
  {N.}~\bibnamefont {Rougemaille}}, \bibinfo {author} {\bibfnamefont
  {J.}~\bibnamefont {Coraux}}, \bibinfo {author} {\bibfnamefont
  {X.}~\bibnamefont {Marie}}, \bibinfo {author} {\bibfnamefont
  {P.}~\bibnamefont {Renucci}}, \bibinfo {author} {\bibfnamefont
  {C.}~\bibnamefont {Robert}}, \bibinfo {author} {\bibfnamefont
  {I.}~\bibnamefont {Robert-Philip}}, \bibinfo {author} {\bibfnamefont
  {B.}~\bibnamefont {Gil}}, \bibinfo {author} {\bibfnamefont {G.}~\bibnamefont
  {Cassabois}}, \bibinfo {author} {\bibfnamefont {A.}~\bibnamefont {Finco}},\
  and\ \bibinfo {author} {\bibfnamefont {V.}~\bibnamefont {Jacques}},\
  }\bibfield  {title} {\bibinfo {title} {Magnetic imaging with spin defects in
  hexagonal boron nitride},\ }\href
  {https://doi.org/10.1103/PhysRevApplied.18.L061002} {\bibfield  {journal}
  {\bibinfo  {journal} {Phys. Rev. Appl.}\ }\textbf {\bibinfo {volume} {18}},\
  \bibinfo {pages} {L061002} (\bibinfo {year} {2022})}\BibitemShut {NoStop}%
\bibitem [{\citenamefont {Lyu}\ \emph {et~al.}(2022)\citenamefont {Lyu},
  \citenamefont {Tan}, \citenamefont {Wu}, \citenamefont {Zhang}, \citenamefont
  {Zhang}, \citenamefont {Mu}, \citenamefont {Zúñiga-Pérez}, \citenamefont
  {Cai},\ and\ \citenamefont {Gao}}]{Lyu2022}%
  \BibitemOpen
  \bibfield  {author} {\bibinfo {author} {\bibfnamefont {X.}~\bibnamefont
  {Lyu}}, \bibinfo {author} {\bibfnamefont {Q.}~\bibnamefont {Tan}}, \bibinfo
  {author} {\bibfnamefont {L.}~\bibnamefont {Wu}}, \bibinfo {author}
  {\bibfnamefont {C.}~\bibnamefont {Zhang}}, \bibinfo {author} {\bibfnamefont
  {Z.}~\bibnamefont {Zhang}}, \bibinfo {author} {\bibfnamefont
  {Z.}~\bibnamefont {Mu}}, \bibinfo {author} {\bibfnamefont {J.}~\bibnamefont
  {Zúñiga-Pérez}}, \bibinfo {author} {\bibfnamefont {H.}~\bibnamefont
  {Cai}},\ and\ \bibinfo {author} {\bibfnamefont {W.}~\bibnamefont {Gao}},\
  }\bibfield  {title} {\bibinfo {title} {Strain quantum sensing with spin
  defects in hexagonal boron nitride},\ }\href
  {https://doi.org/10.1021/acs.nanolett.2c01722} {\bibfield  {journal}
  {\bibinfo  {journal} {Nano Letters}\ }\textbf {\bibinfo {volume} {22}},\
  \bibinfo {pages} {6553} (\bibinfo {year} {2022})}\BibitemShut {NoStop}%
\bibitem [{\citenamefont {Yang}\ \emph {et~al.}(2022)\citenamefont {Yang},
  \citenamefont {Mendelson}, \citenamefont {Li}, \citenamefont {Gottscholl},
  \citenamefont {Scott}, \citenamefont {Kianinia}, \citenamefont {Dyakonov},
  \citenamefont {Toth},\ and\ \citenamefont {Aharonovich}}]{Yang2022}%
  \BibitemOpen
  \bibfield  {author} {\bibinfo {author} {\bibfnamefont {T.}~\bibnamefont
  {Yang}}, \bibinfo {author} {\bibfnamefont {N.}~\bibnamefont {Mendelson}},
  \bibinfo {author} {\bibfnamefont {C.}~\bibnamefont {Li}}, \bibinfo {author}
  {\bibfnamefont {A.}~\bibnamefont {Gottscholl}}, \bibinfo {author}
  {\bibfnamefont {J.}~\bibnamefont {Scott}}, \bibinfo {author} {\bibfnamefont
  {M.}~\bibnamefont {Kianinia}}, \bibinfo {author} {\bibfnamefont
  {V.}~\bibnamefont {Dyakonov}}, \bibinfo {author} {\bibfnamefont
  {M.}~\bibnamefont {Toth}},\ and\ \bibinfo {author} {\bibfnamefont
  {I.}~\bibnamefont {Aharonovich}},\ }\bibfield  {title} {\bibinfo {title}
  {Spin defects in hexagonal boron nitride for strain sensing on nanopillar
  arrays},\ }\href {https://doi.org/10.1039/d1nr07919k} {\bibfield  {journal}
  {\bibinfo  {journal} {Nanoscale}\ }\textbf {\bibinfo {volume} {14}},\
  \bibinfo {pages} {5239} (\bibinfo {year} {2022})}\BibitemShut {NoStop}%
\bibitem [{\citenamefont {Huang}\ \emph {et~al.}(2022)\citenamefont {Huang},
  \citenamefont {Zhou}, \citenamefont {Chen}, \citenamefont {Lu}, \citenamefont
  {McLaughlin}, \citenamefont {Li}, \citenamefont {Alghamdi}, \citenamefont
  {Djugba}, \citenamefont {Shi}, \citenamefont {Wang},\ and\ \citenamefont
  {Du}}]{HuangNC2022}%
  \BibitemOpen
  \bibfield  {author} {\bibinfo {author} {\bibfnamefont {M.}~\bibnamefont
  {Huang}}, \bibinfo {author} {\bibfnamefont {J.}~\bibnamefont {Zhou}},
  \bibinfo {author} {\bibfnamefont {D.}~\bibnamefont {Chen}}, \bibinfo {author}
  {\bibfnamefont {H.}~\bibnamefont {Lu}}, \bibinfo {author} {\bibfnamefont
  {N.~J.}\ \bibnamefont {McLaughlin}}, \bibinfo {author} {\bibfnamefont
  {S.}~\bibnamefont {Li}}, \bibinfo {author} {\bibfnamefont {M.}~\bibnamefont
  {Alghamdi}}, \bibinfo {author} {\bibfnamefont {D.}~\bibnamefont {Djugba}},
  \bibinfo {author} {\bibfnamefont {J.}~\bibnamefont {Shi}}, \bibinfo {author}
  {\bibfnamefont {H.}~\bibnamefont {Wang}},\ and\ \bibinfo {author}
  {\bibfnamefont {C.~R.}\ \bibnamefont {Du}},\ }\bibfield  {title} {\bibinfo
  {title} {{Wide field imaging of van der Waals ferromagnet Fe3GeTe2 by spin
  defects in hexagonal boron nitride}},\ }\href
  {https://doi.org/10.1038/s41467-022-33016-2} {\bibfield  {journal} {\bibinfo
  {journal} {Nature Communications}\ }\textbf {\bibinfo {volume} {13}},\
  \bibinfo {pages} {5369} (\bibinfo {year} {2022})}\BibitemShut {NoStop}%
\bibitem [{\citenamefont {Liu}\ \emph {et~al.}(2022{\natexlab{b}})\citenamefont
  {Liu}, \citenamefont {Guo}, \citenamefont {Yu}, \citenamefont {Meng},
  \citenamefont {Li}, \citenamefont {Yang}, \citenamefont {Wang}, \citenamefont
  {Zeng}, \citenamefont {Xie}, \citenamefont {Li}, \citenamefont {Wang},
  \citenamefont {Xu}, \citenamefont {Wang}, \citenamefont {Tang}, \citenamefont
  {Li},\ and\ \citenamefont {Guo}}]{Liu2022review}%
  \BibitemOpen
  \bibfield  {author} {\bibinfo {author} {\bibfnamefont {W.}~\bibnamefont
  {Liu}}, \bibinfo {author} {\bibfnamefont {N.-J.}\ \bibnamefont {Guo}},
  \bibinfo {author} {\bibfnamefont {S.}~\bibnamefont {Yu}}, \bibinfo {author}
  {\bibfnamefont {Y.}~\bibnamefont {Meng}}, \bibinfo {author} {\bibfnamefont
  {Z.-P.}\ \bibnamefont {Li}}, \bibinfo {author} {\bibfnamefont {Y.-Z.}\
  \bibnamefont {Yang}}, \bibinfo {author} {\bibfnamefont {Z.-A.}\ \bibnamefont
  {Wang}}, \bibinfo {author} {\bibfnamefont {X.-D.}\ \bibnamefont {Zeng}},
  \bibinfo {author} {\bibfnamefont {L.-K.}\ \bibnamefont {Xie}}, \bibinfo
  {author} {\bibfnamefont {Q.}~\bibnamefont {Li}}, \bibinfo {author}
  {\bibfnamefont {J.-F.}\ \bibnamefont {Wang}}, \bibinfo {author}
  {\bibfnamefont {J.-S.}\ \bibnamefont {Xu}}, \bibinfo {author} {\bibfnamefont
  {Y.-T.}\ \bibnamefont {Wang}}, \bibinfo {author} {\bibfnamefont {J.-S.}\
  \bibnamefont {Tang}}, \bibinfo {author} {\bibfnamefont {C.-F.}\ \bibnamefont
  {Li}},\ and\ \bibinfo {author} {\bibfnamefont {G.-C.}\ \bibnamefont {Guo}},\
  }\bibfield  {title} {\bibinfo {title} {Spin-active defects in hexagonal boron
  nitride},\ }\href {https://doi.org/10.1088/2633-4356/ac7e9f} {\bibfield
  {journal} {\bibinfo  {journal} {Materials for Quantum Technology}\ }\textbf
  {\bibinfo {volume} {2}},\ \bibinfo {pages} {032002} (\bibinfo {year}
  {2022}{\natexlab{b}})}\BibitemShut {NoStop}%
\bibitem [{\citenamefont {Kianinia}\ \emph {et~al.}(2020)\citenamefont
  {Kianinia}, \citenamefont {White}, \citenamefont {Fröch}, \citenamefont
  {Bradac},\ and\ \citenamefont {Aharonovich}}]{Kianinia2020}%
  \BibitemOpen
  \bibfield  {author} {\bibinfo {author} {\bibfnamefont {M.}~\bibnamefont
  {Kianinia}}, \bibinfo {author} {\bibfnamefont {S.}~\bibnamefont {White}},
  \bibinfo {author} {\bibfnamefont {J.~E.}\ \bibnamefont {Fröch}}, \bibinfo
  {author} {\bibfnamefont {C.}~\bibnamefont {Bradac}},\ and\ \bibinfo {author}
  {\bibfnamefont {I.}~\bibnamefont {Aharonovich}},\ }\bibfield  {title}
  {\bibinfo {title} {Generation of spin defects in hexagonal boron nitride},\
  }\href {https://doi.org/10.1021/acsphotonics.0c00614} {\bibfield  {journal}
  {\bibinfo  {journal} {ACS Photonics}\ }\textbf {\bibinfo {volume} {7}},\
  \bibinfo {pages} {2147} (\bibinfo {year} {2020})}\BibitemShut {NoStop}%
\bibitem [{\citenamefont {Murzakhanov}\ \emph {et~al.}(2021)\citenamefont
  {Murzakhanov}, \citenamefont {Yavkin}, \citenamefont {Mamin}, \citenamefont
  {Orlinskii}, \citenamefont {Mumdzhi}, \citenamefont {Gracheva}, \citenamefont
  {Gabbasov}, \citenamefont {Smirnov}, \citenamefont {Davydov},\ and\
  \citenamefont {Soltamov}}]{Murzakhanov2021}%
  \BibitemOpen
  \bibfield  {author} {\bibinfo {author} {\bibfnamefont {F.~F.}\ \bibnamefont
  {Murzakhanov}}, \bibinfo {author} {\bibfnamefont {B.~V.}\ \bibnamefont
  {Yavkin}}, \bibinfo {author} {\bibfnamefont {G.~V.}\ \bibnamefont {Mamin}},
  \bibinfo {author} {\bibfnamefont {S.~B.}\ \bibnamefont {Orlinskii}}, \bibinfo
  {author} {\bibfnamefont {I.~E.}\ \bibnamefont {Mumdzhi}}, \bibinfo {author}
  {\bibfnamefont {I.~N.}\ \bibnamefont {Gracheva}}, \bibinfo {author}
  {\bibfnamefont {B.~F.}\ \bibnamefont {Gabbasov}}, \bibinfo {author}
  {\bibfnamefont {A.~N.}\ \bibnamefont {Smirnov}}, \bibinfo {author}
  {\bibfnamefont {V.~Y.}\ \bibnamefont {Davydov}},\ and\ \bibinfo {author}
  {\bibfnamefont {V.~A.}\ \bibnamefont {Soltamov}},\ }\bibfield  {title}
  {\bibinfo {title} {Creation of negatively charged boron vacancies in
  hexagonal boron nitride crystal by electron irradiation and mechanism of
  inhomogeneous broadening of boron vacancy-related spin resonance lines},\
  }\href {https://doi.org/10.3390/nano11061373} {\bibfield  {journal} {\bibinfo
   {journal} {Nanomaterials (Basel)}\ }\textbf {\bibinfo {volume} {11}},\
  \bibinfo {pages} {1373} (\bibinfo {year} {2021})}\BibitemShut {NoStop}%
\bibitem [{\citenamefont {Zhang}\ \emph {et~al.}(2017)\citenamefont {Zhang},
  \citenamefont {Feng}, \citenamefont {Wang}, \citenamefont {Yang},\ and\
  \citenamefont {Wang}}]{Zhang2017}%
  \BibitemOpen
  \bibfield  {author} {\bibinfo {author} {\bibfnamefont {K.}~\bibnamefont
  {Zhang}}, \bibinfo {author} {\bibfnamefont {Y.}~\bibnamefont {Feng}},
  \bibinfo {author} {\bibfnamefont {F.}~\bibnamefont {Wang}}, \bibinfo {author}
  {\bibfnamefont {Z.}~\bibnamefont {Yang}},\ and\ \bibinfo {author}
  {\bibfnamefont {J.}~\bibnamefont {Wang}},\ }\bibfield  {title} {\bibinfo
  {title} {Two dimensional hexagonal boron nitride (2d-hbn): synthesis{,}
  properties and applications},\ }\href {https://doi.org/10.1039/C7TC04300G}
  {\bibfield  {journal} {\bibinfo  {journal} {J. Mater. Chem. C}\ }\textbf
  {\bibinfo {volume} {5}},\ \bibinfo {pages} {11992} (\bibinfo {year}
  {2017})}\BibitemShut {NoStop}%
\bibitem [{\citenamefont {Chen}\ \emph {et~al.}(2021)\citenamefont {Chen},
  \citenamefont {Westerhausen}, \citenamefont {Li}, \citenamefont {White},
  \citenamefont {Bradac}, \citenamefont {Bendavid}, \citenamefont {Toth},
  \citenamefont {Aharonovich},\ and\ \citenamefont {Tran}}]{Chen2021}%
  \BibitemOpen
  \bibfield  {author} {\bibinfo {author} {\bibfnamefont {Y.}~\bibnamefont
  {Chen}}, \bibinfo {author} {\bibfnamefont {M.~T.}\ \bibnamefont
  {Westerhausen}}, \bibinfo {author} {\bibfnamefont {C.}~\bibnamefont {Li}},
  \bibinfo {author} {\bibfnamefont {S.}~\bibnamefont {White}}, \bibinfo
  {author} {\bibfnamefont {C.}~\bibnamefont {Bradac}}, \bibinfo {author}
  {\bibfnamefont {A.}~\bibnamefont {Bendavid}}, \bibinfo {author}
  {\bibfnamefont {M.}~\bibnamefont {Toth}}, \bibinfo {author} {\bibfnamefont
  {I.}~\bibnamefont {Aharonovich}},\ and\ \bibinfo {author} {\bibfnamefont
  {T.~T.}\ \bibnamefont {Tran}},\ }\bibfield  {title} {\bibinfo {title}
  {Solvent-exfoliated hexagonal boron nitride nanoflakes for quantum
  emitters},\ }\href {https://doi.org/10.1021/acsanm.1c01974} {\bibfield
  {journal} {\bibinfo  {journal} {ACS Applied Nano Materials}\ }\textbf
  {\bibinfo {volume} {4}},\ \bibinfo {pages} {10449} (\bibinfo {year}
  {2021})}\BibitemShut {NoStop}%
\bibitem [{\citenamefont {Gottscholl}\ \emph
  {et~al.}(2021{\natexlab{b}})\citenamefont {Gottscholl}, \citenamefont {Diez},
  \citenamefont {Soltamov}, \citenamefont {Kasper}, \citenamefont {Sperlich},
  \citenamefont {Kianinia}, \citenamefont {Bradac}, \citenamefont
  {Aharonovich},\ and\ \citenamefont {Dyakonov}}]{GottschollSA2021}%
  \BibitemOpen
  \bibfield  {author} {\bibinfo {author} {\bibfnamefont {A.}~\bibnamefont
  {Gottscholl}}, \bibinfo {author} {\bibfnamefont {M.}~\bibnamefont {Diez}},
  \bibinfo {author} {\bibfnamefont {V.}~\bibnamefont {Soltamov}}, \bibinfo
  {author} {\bibfnamefont {C.}~\bibnamefont {Kasper}}, \bibinfo {author}
  {\bibfnamefont {A.}~\bibnamefont {Sperlich}}, \bibinfo {author}
  {\bibfnamefont {M.}~\bibnamefont {Kianinia}}, \bibinfo {author}
  {\bibfnamefont {C.}~\bibnamefont {Bradac}}, \bibinfo {author} {\bibfnamefont
  {I.}~\bibnamefont {Aharonovich}},\ and\ \bibinfo {author} {\bibfnamefont
  {V.}~\bibnamefont {Dyakonov}},\ }\bibfield  {title} {\bibinfo {title} {{Room
  temperature coherent control of spin defects in hexagonal boron nitride}},\
  }\href {https://doi.org/10.1126/sciadv.abf3630} {\bibfield  {journal}
  {\bibinfo  {journal} {Science Advances}\ }\textbf {\bibinfo {volume} {7}},\
  \bibinfo {pages} {eabf3630} (\bibinfo {year}
  {2021}{\natexlab{b}})}\BibitemShut {NoStop}%
\bibitem [{\citenamefont {Guo}\ \emph {et~al.}(2022)\citenamefont {Guo},
  \citenamefont {Liu}, \citenamefont {Li}, \citenamefont {Yang}, \citenamefont
  {Yu}, \citenamefont {Meng}, \citenamefont {Wang}, \citenamefont {Zeng},
  \citenamefont {Yan}, \citenamefont {Li}, \citenamefont {Wang}, \citenamefont
  {Xu}, \citenamefont {Wang}, \citenamefont {Tang}, \citenamefont {Li},\ and\
  \citenamefont {Guo}}]{Guo2022GenerationNitride}%
  \BibitemOpen
  \bibfield  {author} {\bibinfo {author} {\bibfnamefont {N.-J.}\ \bibnamefont
  {Guo}}, \bibinfo {author} {\bibfnamefont {W.}~\bibnamefont {Liu}}, \bibinfo
  {author} {\bibfnamefont {Z.-P.}\ \bibnamefont {Li}}, \bibinfo {author}
  {\bibfnamefont {Y.-Z.}\ \bibnamefont {Yang}}, \bibinfo {author}
  {\bibfnamefont {S.}~\bibnamefont {Yu}}, \bibinfo {author} {\bibfnamefont
  {Y.}~\bibnamefont {Meng}}, \bibinfo {author} {\bibfnamefont {Z.-A.}\
  \bibnamefont {Wang}}, \bibinfo {author} {\bibfnamefont {X.-D.}\ \bibnamefont
  {Zeng}}, \bibinfo {author} {\bibfnamefont {F.-F.}\ \bibnamefont {Yan}},
  \bibinfo {author} {\bibfnamefont {Q.}~\bibnamefont {Li}}, \bibinfo {author}
  {\bibfnamefont {J.-F.}\ \bibnamefont {Wang}}, \bibinfo {author}
  {\bibfnamefont {J.-S.}\ \bibnamefont {Xu}}, \bibinfo {author} {\bibfnamefont
  {Y.-T.}\ \bibnamefont {Wang}}, \bibinfo {author} {\bibfnamefont {J.-S.}\
  \bibnamefont {Tang}}, \bibinfo {author} {\bibfnamefont {C.-F.}\ \bibnamefont
  {Li}},\ and\ \bibinfo {author} {\bibfnamefont {G.-C.}\ \bibnamefont {Guo}},\
  }\bibfield  {title} {\bibinfo {title} {{Generation of Spin Defects by Ion
  Implantation in Hexagonal Boron Nitride}},\ }\href
  {https://doi.org/10.1021/acsomega.1c04564} {\bibfield  {journal} {\bibinfo
  {journal} {ACS Omega}\ }\textbf {\bibinfo {volume} {7}},\ \bibinfo {pages}
  {1733} (\bibinfo {year} {2022})}\BibitemShut {NoStop}%
\bibitem [{\citenamefont {Grant}\ \emph {et~al.}(2022)\citenamefont {Grant},
  \citenamefont {Olia}, \citenamefont {Walsh}, \citenamefont {Hall},
  \citenamefont {McColl},\ and\ \citenamefont {Simpson}}]{Grant2022}%
  \BibitemOpen
  \bibfield  {author} {\bibinfo {author} {\bibfnamefont {E.~S.}\ \bibnamefont
  {Grant}}, \bibinfo {author} {\bibfnamefont {M.~B.~A.}\ \bibnamefont {Olia}},
  \bibinfo {author} {\bibfnamefont {E.~P.}\ \bibnamefont {Walsh}}, \bibinfo
  {author} {\bibfnamefont {L.~T.}\ \bibnamefont {Hall}}, \bibinfo {author}
  {\bibfnamefont {G.}~\bibnamefont {McColl}},\ and\ \bibinfo {author}
  {\bibfnamefont {D.~A.}\ \bibnamefont {Simpson}},\ }\bibfield  {title}
  {\bibinfo {title} {Method for in-solution, high-throughput t1 relaxometry
  using fluorescent nanodiamonds},\ }\href@noop {} {\bibfield  {journal}
  {\bibinfo  {journal} {Preprint}\ ,\ \bibinfo {pages} {arXiv.2211.14959}}
  (\bibinfo {year} {2022})}\BibitemShut {NoStop}%
\bibitem [{\citenamefont {Xu}\ \emph {et~al.}(2023)\citenamefont {Xu},
  \citenamefont {Solanki}, \citenamefont {Sychev}, \citenamefont {Gao},
  \citenamefont {Peana}, \citenamefont {Baburin}, \citenamefont {Pagadala},
  \citenamefont {Martin}, \citenamefont {Chowdhury}, \citenamefont {Chen},
  \citenamefont {Taniguchi}, \citenamefont {Watanabe}, \citenamefont
  {Rodionov}, \citenamefont {Kildishev}, \citenamefont {Li}, \citenamefont
  {Upadhyaya}, \citenamefont {Boltasseva},\ and\ \citenamefont
  {Shalaev}}]{Xu2023}%
  \BibitemOpen
  \bibfield  {author} {\bibinfo {author} {\bibfnamefont {X.}~\bibnamefont
  {Xu}}, \bibinfo {author} {\bibfnamefont {A.~B.}\ \bibnamefont {Solanki}},
  \bibinfo {author} {\bibfnamefont {D.}~\bibnamefont {Sychev}}, \bibinfo
  {author} {\bibfnamefont {X.}~\bibnamefont {Gao}}, \bibinfo {author}
  {\bibfnamefont {S.}~\bibnamefont {Peana}}, \bibinfo {author} {\bibfnamefont
  {A.~S.}\ \bibnamefont {Baburin}}, \bibinfo {author} {\bibfnamefont
  {K.}~\bibnamefont {Pagadala}}, \bibinfo {author} {\bibfnamefont {Z.~O.}\
  \bibnamefont {Martin}}, \bibinfo {author} {\bibfnamefont {S.~N.}\
  \bibnamefont {Chowdhury}}, \bibinfo {author} {\bibfnamefont {Y.~P.}\
  \bibnamefont {Chen}}, \bibinfo {author} {\bibfnamefont {T.}~\bibnamefont
  {Taniguchi}}, \bibinfo {author} {\bibfnamefont {K.}~\bibnamefont {Watanabe}},
  \bibinfo {author} {\bibfnamefont {I.~A.}\ \bibnamefont {Rodionov}}, \bibinfo
  {author} {\bibfnamefont {A.~V.}\ \bibnamefont {Kildishev}}, \bibinfo {author}
  {\bibfnamefont {T.}~\bibnamefont {Li}}, \bibinfo {author} {\bibfnamefont
  {P.}~\bibnamefont {Upadhyaya}}, \bibinfo {author} {\bibfnamefont
  {A.}~\bibnamefont {Boltasseva}},\ and\ \bibinfo {author} {\bibfnamefont
  {V.~M.}\ \bibnamefont {Shalaev}},\ }\bibfield  {title} {\bibinfo {title}
  {Greatly enhanced emission from spin defects in hexagonal boron nitride
  enabled by a low-loss plasmonic nanocavity},\ }\href
  {https://doi.org/10.1021/acs.nanolett.2c03100} {\bibfield  {journal}
  {\bibinfo  {journal} {Nano Letters}\ }\textbf {\bibinfo {volume} {23}},\
  \bibinfo {pages} {25} (\bibinfo {year} {2023})}\BibitemShut {NoStop}%
\bibitem [{\citenamefont {Froch}\ \emph {et~al.}(2021)\citenamefont {Froch},
  \citenamefont {Spencer}, \citenamefont {Kianinia}, \citenamefont {Totonjian},
  \citenamefont {Nguyen}, \citenamefont {Gottscholl}, \citenamefont {Dyakonov},
  \citenamefont {Toth}, \citenamefont {Kim},\ and\ \citenamefont
  {Aharonovich}}]{Froch2021}%
  \BibitemOpen
  \bibfield  {author} {\bibinfo {author} {\bibfnamefont {J.~E.}\ \bibnamefont
  {Froch}}, \bibinfo {author} {\bibfnamefont {L.~P.}\ \bibnamefont {Spencer}},
  \bibinfo {author} {\bibfnamefont {M.}~\bibnamefont {Kianinia}}, \bibinfo
  {author} {\bibfnamefont {D.~D.}\ \bibnamefont {Totonjian}}, \bibinfo {author}
  {\bibfnamefont {M.}~\bibnamefont {Nguyen}}, \bibinfo {author} {\bibfnamefont
  {A.}~\bibnamefont {Gottscholl}}, \bibinfo {author} {\bibfnamefont
  {V.}~\bibnamefont {Dyakonov}}, \bibinfo {author} {\bibfnamefont
  {M.}~\bibnamefont {Toth}}, \bibinfo {author} {\bibfnamefont {S.}~\bibnamefont
  {Kim}},\ and\ \bibinfo {author} {\bibfnamefont {I.}~\bibnamefont
  {Aharonovich}},\ }\bibfield  {title} {\bibinfo {title} {Coupling spin defects
  in hexagonal boron nitride to monolithic bullseye cavities},\ }\href
  {https://doi.org/10.1021/acs.nanolett.1c01843} {\bibfield  {journal}
  {\bibinfo  {journal} {Nano Letters}\ }\textbf {\bibinfo {volume} {21}},\
  \bibinfo {pages} {6549} (\bibinfo {year} {2021})}\BibitemShut {NoStop}%
\bibitem [{\citenamefont {Myers}\ \emph {et~al.}(2017)\citenamefont {Myers},
  \citenamefont {Ariyaratne},\ and\ \citenamefont {Jayich}}]{Myers2017}%
  \BibitemOpen
  \bibfield  {author} {\bibinfo {author} {\bibfnamefont {B.~A.}\ \bibnamefont
  {Myers}}, \bibinfo {author} {\bibfnamefont {A.}~\bibnamefont {Ariyaratne}},\
  and\ \bibinfo {author} {\bibfnamefont {A.~C.~B.}\ \bibnamefont {Jayich}},\
  }\bibfield  {title} {\bibinfo {title} {Double-quantum spin-relaxation limits
  to coherence of near-surface nitrogen-vacancy centers},\ }\href
  {https://doi.org/10.1103/PhysRevLett.118.197201} {\bibfield  {journal}
  {\bibinfo  {journal} {Phys Rev Lett}\ }\textbf {\bibinfo {volume} {118}},\
  \bibinfo {pages} {197201} (\bibinfo {year} {2017})}\BibitemShut {NoStop}%
\bibitem [{\citenamefont {Gardill}\ \emph {et~al.}(2020)\citenamefont
  {Gardill}, \citenamefont {Cambria},\ and\ \citenamefont
  {Kolkowitz}}]{Gardill2020}%
  \BibitemOpen
  \bibfield  {author} {\bibinfo {author} {\bibfnamefont {A.}~\bibnamefont
  {Gardill}}, \bibinfo {author} {\bibfnamefont {M.}~\bibnamefont {Cambria}},\
  and\ \bibinfo {author} {\bibfnamefont {S.}~\bibnamefont {Kolkowitz}},\
  }\bibfield  {title} {\bibinfo {title} {{Fast Relaxation on Qutrit Transitions
  of Nitrogen-Vacancy Centers in Nanodiamonds}},\ }\href
  {https://doi.org/10.1103/PhysRevApplied.13.034010} {\bibfield  {journal}
  {\bibinfo  {journal} {Physical Review Applied}\ }\textbf {\bibinfo {volume}
  {13}},\ \bibinfo {pages} {034010} (\bibinfo {year} {2020})}\BibitemShut
  {NoStop}%
\bibitem [{\citenamefont {Rosskopf}\ \emph {et~al.}(2014)\citenamefont
  {Rosskopf}, \citenamefont {Dussaux}, \citenamefont {Ohashi}, \citenamefont
  {Loretz}, \citenamefont {Schirhagl}, \citenamefont {Watanabe}, \citenamefont
  {Shikata}, \citenamefont {Itoh},\ and\ \citenamefont {Degen}}]{Rosskopf2014}%
  \BibitemOpen
  \bibfield  {author} {\bibinfo {author} {\bibfnamefont {T.}~\bibnamefont
  {Rosskopf}}, \bibinfo {author} {\bibfnamefont {A.}~\bibnamefont {Dussaux}},
  \bibinfo {author} {\bibfnamefont {K.}~\bibnamefont {Ohashi}}, \bibinfo
  {author} {\bibfnamefont {M.}~\bibnamefont {Loretz}}, \bibinfo {author}
  {\bibfnamefont {R.}~\bibnamefont {Schirhagl}}, \bibinfo {author}
  {\bibfnamefont {H.}~\bibnamefont {Watanabe}}, \bibinfo {author}
  {\bibfnamefont {S.}~\bibnamefont {Shikata}}, \bibinfo {author} {\bibfnamefont
  {K.~M.}\ \bibnamefont {Itoh}},\ and\ \bibinfo {author} {\bibfnamefont
  {C.~L.}\ \bibnamefont {Degen}},\ }\bibfield  {title} {\bibinfo {title}
  {Investigation of surface magnetic noise by shallow spins in diamond},\
  }\href {https://doi.org/10.1103/PhysRevLett.112.147602} {\bibfield  {journal}
  {\bibinfo  {journal} {Phys Rev Lett}\ }\textbf {\bibinfo {volume} {112}},\
  \bibinfo {pages} {147602} (\bibinfo {year} {2014})}\BibitemShut {NoStop}%
\bibitem [{\citenamefont {Barbosa}\ \emph {et~al.}(2023)\citenamefont
  {Barbosa}, \citenamefont {Gutsche},\ and\ \citenamefont
  {Widera}}]{Barbosa2023}%
  \BibitemOpen
  \bibfield  {author} {\bibinfo {author} {\bibfnamefont {I.~C.}\ \bibnamefont
  {Barbosa}}, \bibinfo {author} {\bibfnamefont {J.}~\bibnamefont {Gutsche}},\
  and\ \bibinfo {author} {\bibfnamefont {A.}~\bibnamefont {Widera}},\
  }\bibfield  {title} {\bibinfo {title} {Impact of charge conversion on
  nv-center relaxometry},\ }\href {https://doi.org/10.48550/ARXIV.2301.01063}
  {\bibfield  {journal} {\bibinfo  {journal} {Preprint}\ ,\ \bibinfo {pages}
  {arXiv.2301.01063}} (\bibinfo {year} {2023})}\BibitemShut {NoStop}%
\bibitem [{\citenamefont {Lillie}\ \emph {et~al.}(2020)\citenamefont {Lillie},
  \citenamefont {Broadway}, \citenamefont {Dontschuk}, \citenamefont
  {Scholten}, \citenamefont {Johnson}, \citenamefont {Wolf}, \citenamefont
  {Rachel}, \citenamefont {Hollenberg},\ and\ \citenamefont
  {Tetienne}}]{Lillie2020}%
  \BibitemOpen
  \bibfield  {author} {\bibinfo {author} {\bibfnamefont {S.~E.}\ \bibnamefont
  {Lillie}}, \bibinfo {author} {\bibfnamefont {D.~A.}\ \bibnamefont
  {Broadway}}, \bibinfo {author} {\bibfnamefont {N.}~\bibnamefont {Dontschuk}},
  \bibinfo {author} {\bibfnamefont {S.~C.}\ \bibnamefont {Scholten}}, \bibinfo
  {author} {\bibfnamefont {B.~C.}\ \bibnamefont {Johnson}}, \bibinfo {author}
  {\bibfnamefont {S.}~\bibnamefont {Wolf}}, \bibinfo {author} {\bibfnamefont
  {S.}~\bibnamefont {Rachel}}, \bibinfo {author} {\bibfnamefont {L.~C.~L.}\
  \bibnamefont {Hollenberg}},\ and\ \bibinfo {author} {\bibfnamefont {J.-P.}\
  \bibnamefont {Tetienne}},\ }\bibfield  {title} {\bibinfo {title} {{Laser
  Modulation of Superconductivity in a Cryogenic Wide-field Nitrogen-Vacancy
  Microscope}},\ }\href {https://doi.org/10.1021/acs.nanolett.9b05071}
  {\bibfield  {journal} {\bibinfo  {journal} {Nano Letters}\ }\textbf {\bibinfo
  {volume} {20}},\ \bibinfo {pages} {1855} (\bibinfo {year}
  {2020})}\BibitemShut {NoStop}%
\end{thebibliography}%



\clearpage
\onecolumngrid

\begin{center}

\textbf{\large Supplementary Information for the manuscript ``Detection of paramagnetic spins with an ultrathin van der Waals quantum sensor''}

\end{center}
%%%%%%%%%% Merge with supplemental materials %%%%%%%%%%
%%%%%%%%%% Prefix a "S" to all equations, figures, tables and reset the counter %%%%%%%%%%
\setcounter{equation}{0}
\setcounter{section}{0}
\setcounter{figure}{0}
\setcounter{table}{0}
\setcounter{page}{1}
\makeatletter
\renewcommand{\theequation}{S\arabic{equation}}
\renewcommand{\thefigure}{S\arabic{figure}}

\section{Experimental setup}\label{sec:setup}

The optical and spin measurements reported in the main text were carried out on a custom-built wide-field fluorescence microscope. Optical excitation from a continuous-wave (CW) $\lambda = 532$~nm laser (Laser Quantum Opus 2 W) was gated using an acousto-optic modulator (Gooch \& Housego R35085-5) and focused using a widefield lens ($f=400$~mm) to the back apearture of the objective lens (Nikon S Plan Fluor ELWD 20x, NA = 0.45). The photoluminescence (PL) from the $V_{\rm B}^-$ defects is separated from the excitation light with a dichroic mirror and filtered using a 750~nm longpass filter before being imaged using a tube lens ($f=300$~mm) onto a scientific CMOS camera (Andor Zyla 5.5-W USB3). In all experiments except Fig. 4(d,e) of the main text, the laser spot diameter ($1/e^2$) at the sample was about 50\,$\mu$m and the total CW laser power 500~mW, which gives a maximum intensity of about 0.5\,mW$/\mu$m$^2$ in the centre of the spot. In Fig. 4(d,e), the laser power was reduced to 200 mW.

Microwave (MW) excitation was provided by a signal generator (Windfreak SynthNV PRO) gated using an IQ modulator (Texas Instruments TRF37T05EVM) and amplified (Mini-Circuits HPA-50W-63+). A pulse pattern generator (SpinCore PulseBlasterESR-PRO 500 MHz) was used to gate the excitation laser and MW and to synchronise the image acquisition. The output of the amplifier was connected to the printed circuit board (PCB) which was terminated by a 50\,$\Omega$ termination. In all experiments except Fig. 4(d,e) of the main text, MW driving was achieved via a coplanar waveguide built in the PCB, with the hBN powder deposited directly onto the surface of the waveguide. In Fig. 4(d,e), the MW was delivered by an $\Omega$-shaped resonator fabricated onto a glass coverslip mounted on a PCB, with the hBN suspension deposited on the coverslip. All measurements were performed at room temperature in ambient atmosphere.

\section{Sample preparation}\label{sec:prep}

\begin{figure*}[hb!]
\centering
\includegraphics[width=0.4\textwidth]{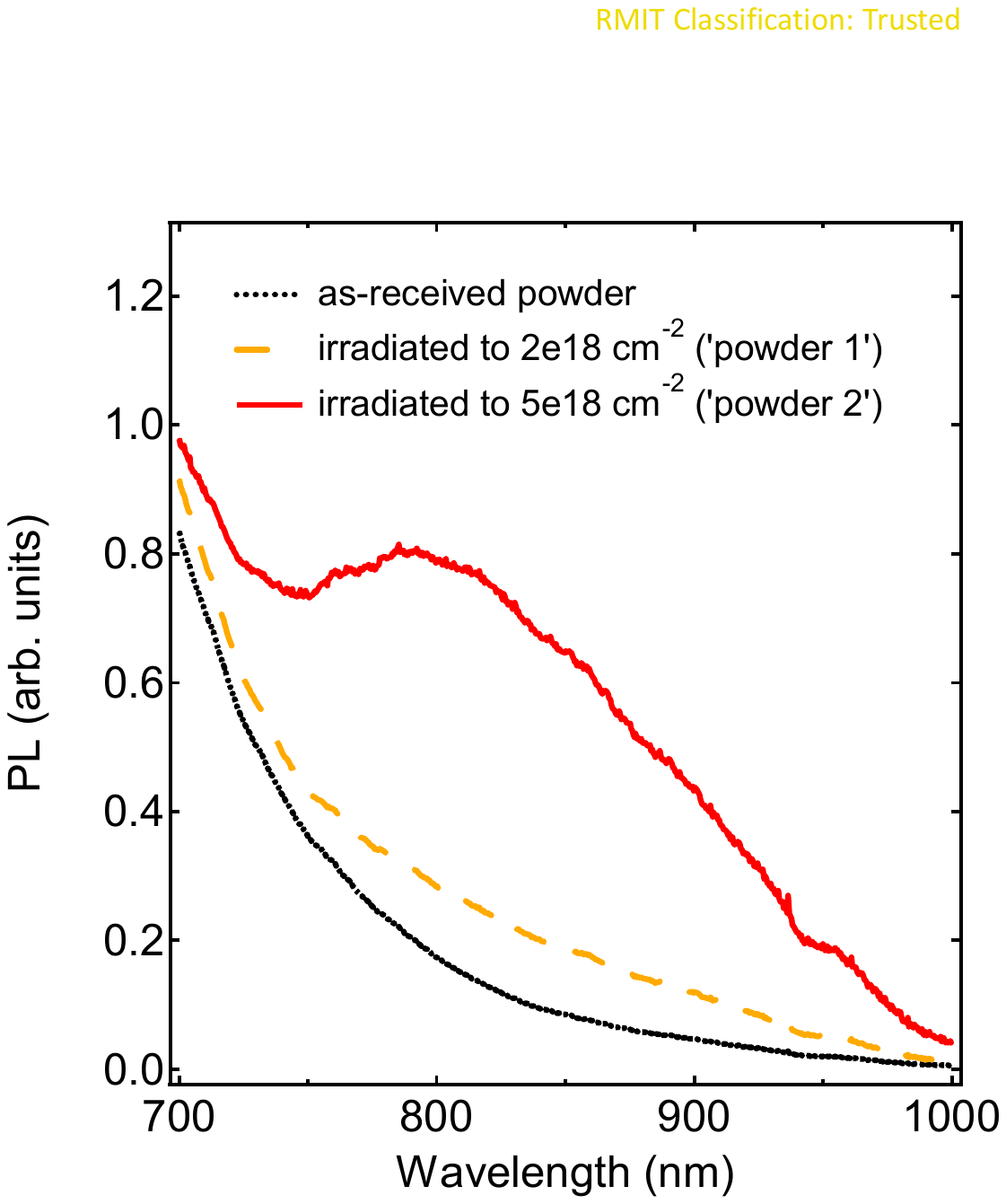}
\caption{PL spectra of the hBN nanopowder following different electron irradiation conditions: no irradiation (as received), $2\times10^{18}$\,cm$^{-2}$ dose (`powder 1'), and $5\times10^{18}$~cm$^{-2}$ (`powder 2'). These spectra were obtained with the same setup as described in Sec.~\ref{sec:setup}, with 500 mW of laser power over a 50\,$\mu$m spot, except that the PL emission was passed through a 550 nm longpass filter to block the laser light and sent to a spectrometer (Ocean Insight Maya2000-Pro).}
\label{Fig_PLspectra}
\end{figure*}

All experiments performed in this work [except the bulk crystal measurement in Fig. 2(b)] used hBN nanopowder sourced from Graphene Supermarket (BN Ultrafine Powder). As the as-received powder contained no measurable concentration of V$_{\rm B}^-$ defects, we subjected the powder to electron irradiation with a beam energy of 2 MeV. The irradiation dose was $2\times10^{18}$\,cm$^{-2}$ (called `powder 1' in the main text) or $5\times10^{18}$~cm$^{-2}$ (`powder 2'). PL spectra of the different powders under $\lambda = 532$~nm excitation [Fig.~\ref{Fig_PLspectra}] show the appearance of the characteristic V$_{\rm B}^-$ emission upon irradiation, i.e. a broad peak centred around 800 nm wavelength~\cite{GottschollNM2020}. 

For the experiments in Fig. 1 and 2 of the main text, the irradiated powder was transferred to the PCB with the tip of a pair of tweezers, forming large clumps (10's of $\mu$m in size).
For the experiments in Fig. 3 of the main text, the irradiated powder was suspended in isopropyl alcohol (IPA) at a concentration of 20 mg/mL and sonicated for 30 min using a horn-sonicator. The sediment from the suspension was drawn using a pipette, then drop cast on the PCB, forming a relatively continuous film. For the experiments in Fig. 4 of the main text, the powder was suspended in water (20 mg/mL) and a drop of the suspension was deposited on the PCB [Fig. 4(a-c)] or on the coverslip [Fig. 4(d,e)].

\begin{figure*}[ht!]
\centering
\includegraphics[width=0.8\textwidth]{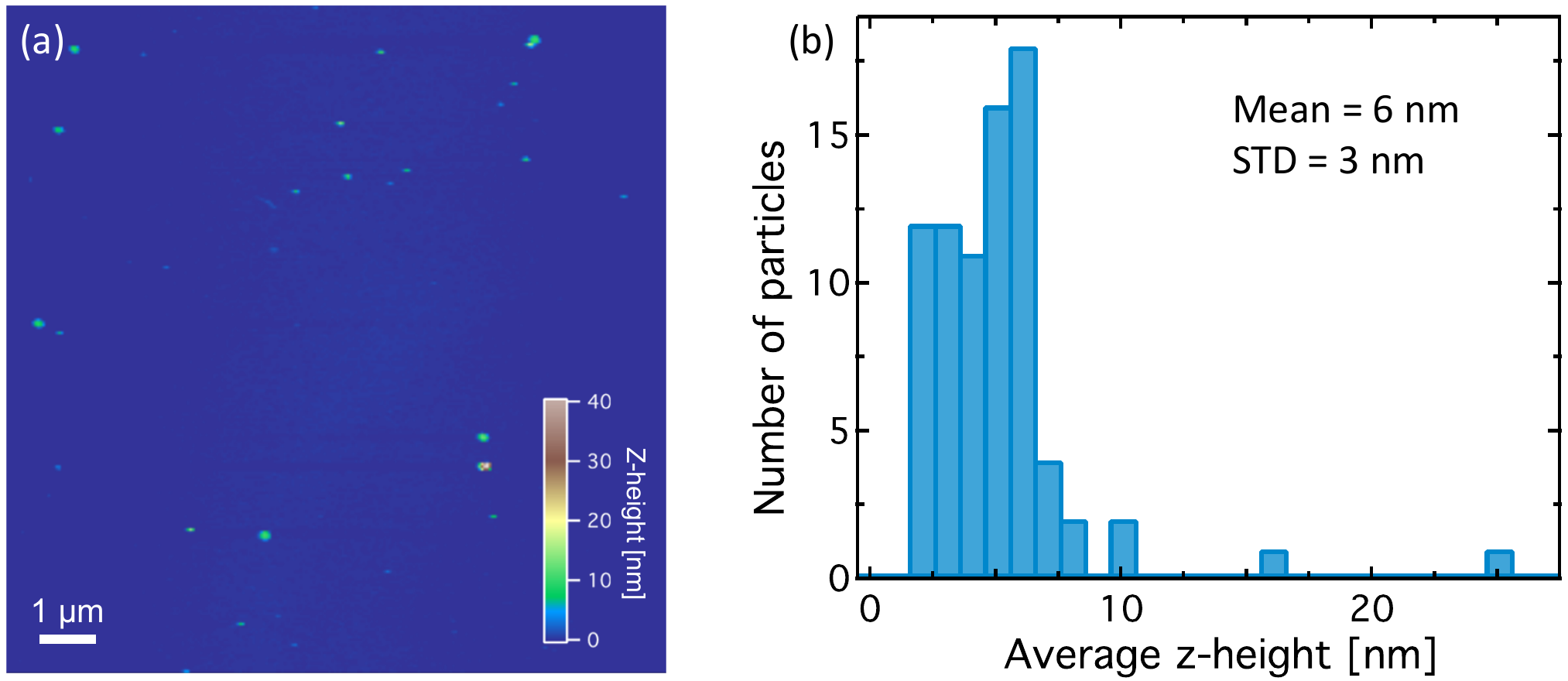}
\caption{(a) Typical AFM image of hBN nanoflakes spin-coated onto a silicon wafer. (b) Histogram of the average hBN particle height.}
\label{Fig_AFM}
\end{figure*}

To characterise the size of the hBN nanoflakes, we performed atomic force microscope (AFM) measurements of powder 2. The particle stock suspension (20 mg/mL in IPA) was diluted to a concentration of 1 mg/mL, sonicated for 10 min in a sonication bath, centrifuged at 1000 rcf for 1 min to remove large aggregates, and the supernatant was spin coated at 2000 rpm onto a clean silicon wafer. AFM images were acquired using an Oxford Instruments Asylum Research MFP-3D Infinity AFM in AC mode using a Budget Sensors Tap300AL-G probe. The images were collected at a scan rate of 1 Hz. A typical AFM image is shown in Fig.~\ref{Fig_AFM}(a), where isolated flakes are visibe. The images were analysed to extract the average thickness of each individual flake, excluding all large particle aggregates, giving the histogram Fig.~\ref{Fig_AFM}(b). The flake thickness ranges mainly between 2 nm and 9 nm, with a mean value of 6 nm and a standard deviation of 3 nm. The lateral size of the flakes is of the order of 100 nm. These values are consistent with previous measurements of powder from the same manufacturer~\cite{Chen2021}. They are also consistent with the specific surface area of $\sim20$\,m$^2/$g measured by the manufacturer. 

For the $T_1$ quenching experiments reported in Fig. 3 and 4 of the main text, a solution of GdCl$_3$ was first prepared by dissolving GdCl$_3$ solid powder (Sigma-Aldrich, 99.99\% purity) in water to a concentration of $100$\,mM. A drop of the solution was then deposited on the hBN film. The PCB was then heated immediately to $80^\circ$C for 5 min to evaporate any remaining water. We note that the density of the hBN nanopowder in dry form is 0.30\,g/cm$^3$ according to the manufacturer, which is 7 times less than the density of bulk hBN (2.1\,g/cm$^3$). This indicates that the dry powder forms a largely porous medium. We conjecture that the GdCl$_3$ solution percolates through the hBN film in such a way that most individual hBN flakes are completely surrounded by GdCl$_3$ upon solvent evaporation, as depicted in the inset of Fig. 3(b). 

Finally, the bulk crystal measurement in Fig. 2(b) was done using a high-purity crystal purchased from HQ Graphene. The crystal was electron irradiated to a dose of $2\times10^{18}$\,cm$^{-2}$ with a beam energy of 2 MeV. Following irradiation, a $\sim1\,\mu$m-thick flake was exfoliated using scotch tape and transferred to a quartz coverslip placed on a PCB with a MW waveguide. To avoid possible edge effects, the $T_1$ time was measured near the centre of the flake where the PL from the V$_{\rm B}^-$ defects was uniform. 

\section{Data acquisition} \label{sec:acquisition}

The spin measurements reported in the main text were obtained using the wide-field setup described in Sec~\ref{sec:setup}. For optically detected magnetic resonance (ODMR) spectra, the pulse sequence was typically: $0.5\,\mu$s laser pulse, $0.5\,\mu$s wait time, $30$\,ns MW pulse (corresponding to a $\pi$ spin flip as estimated from a Rabi measurement). This sequence was repeated thousands of times to fill the exposure time of the camera, generally set to 10\,ms. The MW frequency was swept such that one camera frame is recorded for each MW frequency value, with a reference frame with the MW off recorded every other frame to remove common-mode noise -- forming a normalised ODMR spectrum (PL with MW on divided by PL with MW off) such as that shown in Fig. 1(g). The MW sweep was repeated tens to hundreds of times to improve the signal-to-noise ratio (SNR), corresponding to several minutes of acquisition per spectrum.

$T_1$ measurements were obtained by applying a series of laser pulses separated by a variable dark time $\tau$. For each $\tau$ value, the base pulse sequence (laser pulse, wait time $\tau$, optional MW pulse) was repeated $N = t_{\rm cam} / t_0$ times where $t_{0}$ is the duration of the base sequence with the shortest $\tau$, and $t_{\rm cam} = 10$\,ms is the minimum camera exposure time. As $\tau$ is increased, we keep $N$ fixed and increase the camera exposure time accordingly. For each $\tau$ value, a camera frame is acquired without the MW pulse (`signal'), followed by a camera frame with the MW pulse (`reference'). Example PL traces $S(\tau)$ and $R(\tau)$ thus obtained (without and with the MW pulse, respectively), are shown in Fig.~\ref{Fig_PLnorm}(a). The reference trace serves to remove common-mode variations and normalise the data, as will be discussed in Sec.~\ref{sec:PLnorm}. The entire $\tau$ sweep is repeated over several minutes to improve the SNR in the same way we do for ODMR measurements.

The laser pulse acts as both initialisation and readout of the V$_{\rm B}^-$ spin state. Its duration (between $0.5$\,$\mu$s and $4$\,$\mu$s in our experiments) is chosen long enough to initialise the spin in the $\ket{0}$ state with sufficient fidelity as appropriate for $T_1$ measurements while preserving a good spin contrast, as will be discussed in Sec.~\ref{sec:fidelity}. The MW pulse was around $30$\,ns long typically, corresponding to a $\pi$ spin flip as estimated from a Rabi measurement. The dark time $\tau$ was defined as the time between consecutive laser pulses, inclusive of the MW pulse time when present. The timing of the MW pulse was adjusted to account for the delay incurred by the AOM, such that the MW pulse occurred immediately prior to the laser pulse. For all $T_1$ measurements presented in the paper, 51 $\tau$ points are collected between $2$\,$\mu$s to $60$\,$\mu$s with exponentially weighted spacing, where we excluded the $\tau=0-2\,\mu$s range to allow the system to fully relax to its electronic ground state. 

Except in Fig. 2(c) where the $T_1$ data was analysed pixel by pixel to form a spatial map of $T_1$, all ODMR and $T_1$ measurements reported integrated the data over a $15\times15\,\mu$m$^2$ area where the laser intensity was approximately uniform.

\section{Analysis of $T_1$ measurements}

\subsection{Population dynamics in the dark}
\label{sec:dark}

In order to analyse the $T_1$ measurements, we consider a simple three-level system corresponding to the three spin states $\ket{0,\pm1}$ of the spin-$1$ system [see main text Fig. 1(f)]. In the dark (no laser), $\ket{0}$ and $\ket{\pm1}$ are coupled by a two-way transition rate $k_{01}=k_{\text{int}}+k_{\text{ext}}$ which captures both intrinsic ($k_{\text{int}}$) and external ($k_{\text{ext}}$) contributions~\cite{Tetienne2013,Kolkowitz2015}. For simplicity we assume that there is no coupling between $\ket{+1}$ and $\ket{-1}$, which is valid if the relaxation dynamics is dominated by magnetic effects rather than electric or phonon processes~\cite{Kolkowitz2015,Myers2017,Gardill2020}. In Ref.~\cite{GottschollSA2021}, a strong temperature dependence of $T_1$ was observed for the V$_{\rm B}^-$ defect indicating a thermally activated process, but the exact mechanism remains unclear -- one possibility is that $T_1$ relaxation is primarily due to magnetic noise from the electron spin bath with a temperature-dependent correlation time~\cite{Rosskopf2014}.

To determine the population dynamics in the dark, we solve the matrix rate equation $\partial_t \rho = \Phi \rho$ where $\Phi$ is the transition rate matrix and $\rho$ is the population density matrix with elements $\rho_0(t)$, $\rho_{+1}(t)$, and $\rho_{-1}(t)$, which describe the time-dependent populations of the individual states in the spin-$1$ system. Under the above-mentioned assumption, the transition rate matrix is, 
\begin{align}
    \Phi = \begin{bmatrix}
                -k_{01} & k_{01} & 0\\
                k_{01} & -2k_{01} & k_{01}\\
                0 & k_{01} & -k_{01}
    \end{bmatrix}.
\end{align}
 Solving the matrix rate equation with initial populations $\rho_0(0)$, $\rho_{+1}(0)$, and $\rho_{0}(0)$, gives 
\begin{align}
    \rho_0(t) &= \frac{1}{3} + \bigg[\rho_0(0) - \frac{1}{3} \bigg] e^{-3k_{01}t}\\
    \rho_{\pm1}(t) &= \frac{1}{3} - \frac{1}{2}\bigg[\rho_0(0) - \frac{1}{3} \bigg] e^{-3k_{01}t} \pm \frac{1}{2} \bigg[\rho_{+1}(0) - \rho_{-1}(0) \bigg] e^{-k_{01}t}~.
\end{align}

By convention, $T_1$ is defined as the characteristic decay time out of the $\ket{0}$ state, i.e. the decay in the population of $\rho_0(t)$; thus, $\frac{1}{T_1} =3k_{01}$. 

\subsection{PL normalisation}\label{sec:PLnorm}

\begin{figure*}[t!]
\centering
\includegraphics[width=0.9\textwidth]{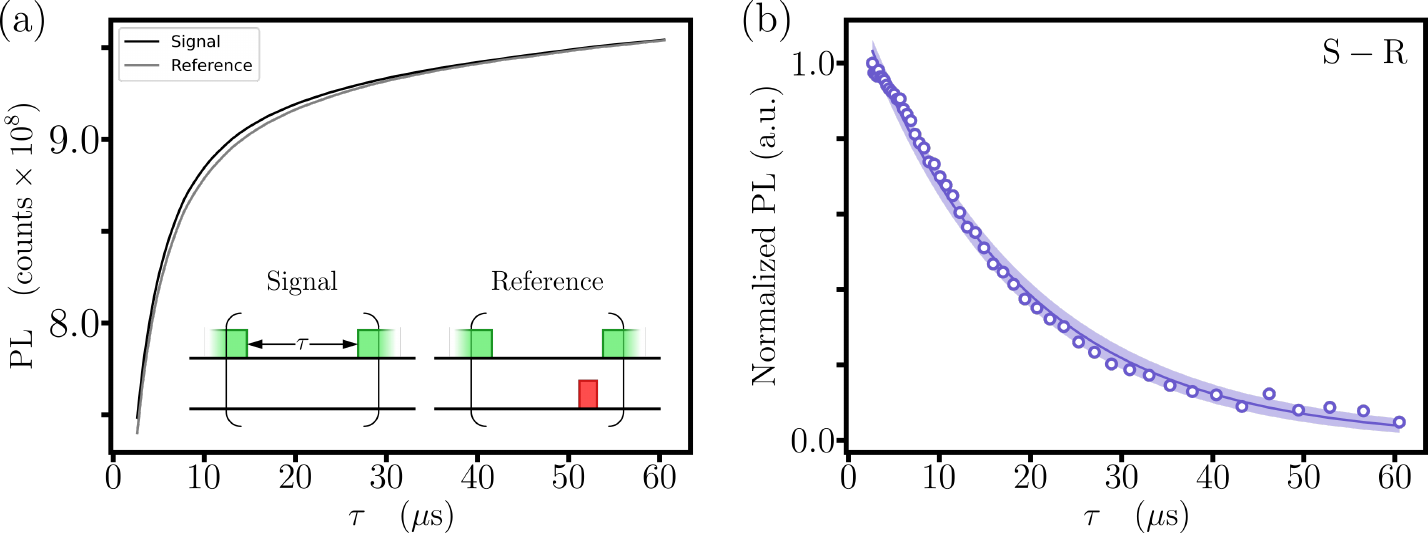}
\caption{(a) Typical raw traces $S(\tau)$ and $R(\tau)$ (called signal and reference, respectively) obtained from a $T_1$ measurement. The inset shows the pulse sequence in each case, where the reference measurement includes a MW $\pi$ pulse prior to the laser readout pulse. (b) Normalised relaxation curve deduced from the raw traces in (a) using Eq.~\ref{eq:N1}. The solid line is a mono-exponential fit used to extract a value for $T_1$.}
\label{Fig_PLnorm}
\end{figure*}

Experimentally, $T_1$ is estimated via the PL emitted by the V$_{\rm B}^-$ defects following a dark time $\tau$, as the PL depends on the spin state owing to an intersystem crossing~\cite{GottschollNM2020}. One can write the PL intensity measured in a $T_1$ experiment as the sum 
\begin{equation}
\label{PL_eqn}
    I(\tau) = I_{\text{B}}(\tau) + A_0(\tau) \rho_0(\tau) + A_1(\tau) \big[\rho_{-1}(\tau) + \rho_{+1}(\tau) \big]
\end{equation}
where $I_{\text{B}}$ is the background (spin-independent) PL, and $A_0$ and $A_1$ are the PL rates associated with the $\ket{0}$ and $\ket{\pm1}$ states, respectively. In Eq.~\ref{PL_eqn}, we included a $\tau$ dependence in all the constants to highlight the importance of the choice of normalisation method~\cite{Bluvstein2019,Barbosa2023}, discussed below.

As described in Sec.~\ref{sec:acquisition}, the experimental data set comprises a `signal' trace $S(\tau)$ and a `reference' trace $R(\tau)$, corresponding to the PL measured without and with a MW $\pi$ pulse following the dark time $\tau$, respectively. To obtain equations for $S(\tau)$ and $R(\tau)$, we substitute in the expressions for the population densities with corresponding initial populations which assume the laser perfectly initialises the $\ket{0}$ state and the MW pulse perfectly swaps the $\ket{0}$ and $\ket{\pm 1}$ populations: 
\begin{align}
    S(\tau) &= I_{\text{B}}(\tau) + \bigg( \frac{A_0(\tau) + 2A_1(\tau)}{3} \bigg) \bigg[ 1 + 2 \bigg( \frac{A_0(\tau) - A_1(\tau)}{A_0(\tau) + 2A_1(\tau)} \bigg) e^{-\frac{\tau}{T_1}} \bigg]\\
    R(\tau) &= I_{\text{B}}(\tau) + \bigg( \frac{A_0(\tau) + 2A_1(\tau)}{3} \bigg) \bigg[ 1 - \bigg( \frac{A_0(\tau) - A_1(\tau)}{A_0(\tau) + 2A_1(\tau)} \bigg) e^{-\frac{\tau}{T_1}} \bigg]
\end{align}
To simplify and generalise we rewrite in terms of the spin-related PL $I_{\text{S}}=[A_0 + 2A_1]/3$ and the contrast $\mathcal{C}_1=-2\mathcal{C}_2=2[A_0 - A_1]/[A_0 + 2A_1]$: 

\begin{align}
    S(\tau) &= I_{\text{B}}(\tau) + I_{\text{S}}(\tau) \bigg[ 1 + \mathcal{C}_1(\tau) e^{-\frac{\tau}{T_1}} \bigg]\\
    R(\tau) &= I_{\text{B}}(\tau) + I_{\text{S}}(\tau) \bigg[ 1 + \mathcal{C}_2(\tau) e^{-\frac{\tau}{T_1}} \bigg]~.
\end{align}

Two relevant cases allow $T_1$ to be extracted exactly~\cite{Bluvstein2019}. First, if the product $I_{\text{S}}\mathcal{C}_1=2[A_0 - A_1]/3$ is constant (i.e. independent of $\tau$), then subtraction of $R(\tau)$ from $S(\tau)$ gives a simple exponential decay
\begin{equation} \label{eq:N1}
    {\cal N}(\tau)\coloneqq S(\tau) - R(\tau) = I_{\text{S}} \mathcal{C} e^{\frac{-\tau}{T_1}}
\end{equation}
where $\mathcal{C} = \mathcal{C}_1 - \mathcal{C}_2$. The $T_1$ time can thus be estimated via an exponential fit to the normalised data ${\cal N}(\tau)$. This method is valid provided the absolute spin contrast $A_0 - A_1$ is independent of $\tau$, which may not be the case in the presence of charge population relaxation during the dark time $\tau$. Such charge dynamics have been reported in experiments with diamond NV centres~\cite{Bluvstein2019,Barbosa2023}, but there has been no report of charge relaxation in the V$_{\rm B}^-$ system. We note that Eq.~\ref{eq:N1} remains valid even if the MW pulse is imperfect (i.e. it causes only a partial swap of populations) and for arbitrary initial populations, as these effects simply change the values of $\mathcal{C}_1$ and $\mathcal{C}_2$.      

Alternatively, if the background PL is negligible ($I_{\text{B}} \ll I_{\text{S}}$), then we can normalise the data as follows:
\begin{equation} \label{eq:N2}
    {\cal N}'(\tau)\coloneqq\frac{S(\tau) - R(\tau)}{S(\tau) + R(\tau)} = \frac{\mathcal{C}}{2} e^{\frac{-\tau}{T_1}}
\end{equation}
where we assumed $\mathcal{C}_{1,2}\ll 1$. In this case, $T_1$ can be estimated via an exponential fit to the normalised data ${\cal N}'(\tau)$ provided the relative spin contrast $\mathcal{C}$ is independent of $\tau$. The latter condition may be expected to hold even in the presence of charge relaxation. 

Outside these two scenarios, it is generally not possible to estimate $T_1$ without knowledge of the functional forms of $I_{\text{B}}(\tau)$, $I_{\text{S}}(\tau)$ and $\mathcal{C}_{1,2}(\tau)$. In our experiments, the measured PL contains a non-negligible background contribution (non V$_{\rm B}^-$ spin related), as evidenced by the PL spectrum Fig.~\ref{Fig_PLspectra} where the non-irradiated powder emits a significant amount of PL overlapping with the V$_{\rm B}^-$ emission. We also observed that this background PL tends to increase following a dark time which is seen as a common-mode increase in the raw traces $S,R(\tau)$ [Fig.~\ref{Fig_PLnorm}(a)]. Furthermore, adding GdCl$_3$ whether dry or in solution adds a small but non-negligible background contribution. Therefore, the second normalisation method ${\cal N}'(\tau)$ cannot be used here. Instead, we used the first method ${\cal N}(\tau)$ as it only assumes that the absolute V$_{\rm B}^-$ spin contribution to the total PL, precisely the product $I_{\text{S}}\mathcal{C}$, is constant independent of $\tau$. 

The normalised data is relatively well fit with a mono-exponential function [Fig.~\ref{Fig_PLnorm}(b)], here giving $T_1 = 18$\,$\mu$s. A stretched exponential fit $\exp[-(\tau/T_1)^n]$, often used in NV $T_1$ sensing experiments~\cite{Steinert2013,Simpson2017}, returns a similar $T_1$ value but does not seem to fit the data better (not shown). We thus choose to fit all data with a mono-exponential as to reduce the number of fitting parameters. We note that a small deviation from the exponential behaviour is sometimes observed at small $\tau$ values, as can be seen in the example Fig.~\ref{Fig_PLnorm}(b). This deviation may be due to a decrease in the sample temperature as $\tau$ is increased (see Sec.~\ref{sec:temp}) or to residual relaxation effects that extend beyond the $\tau=0-2\,\mu$s window. The variability of this deviation across different experiments adds an uncertainty to the measured $T_1$ values, which we estimate (e.g. by changing the spacing between $\tau$ points in the $T_1$ curve or restricting the fit to larger $\tau$ values) to be about $\pm10\%$.        

\subsection{Effect of finite spin initialisation fidelity}
\label{sec:fidelity}

In this section, we develop a simple model to quantify the effect of finite spin initialisation (due to a finite laser pulse duration) on the estimated $T_1$ time. Indeed, in general insufficient repumping from the laser during a $T_1$ measurement leads to an apparent decay time $T_d$ that is shorter than the true spin decay time $T_1$.

\begin{figure*}[t!]
\centering
\includegraphics[width=1\textwidth]{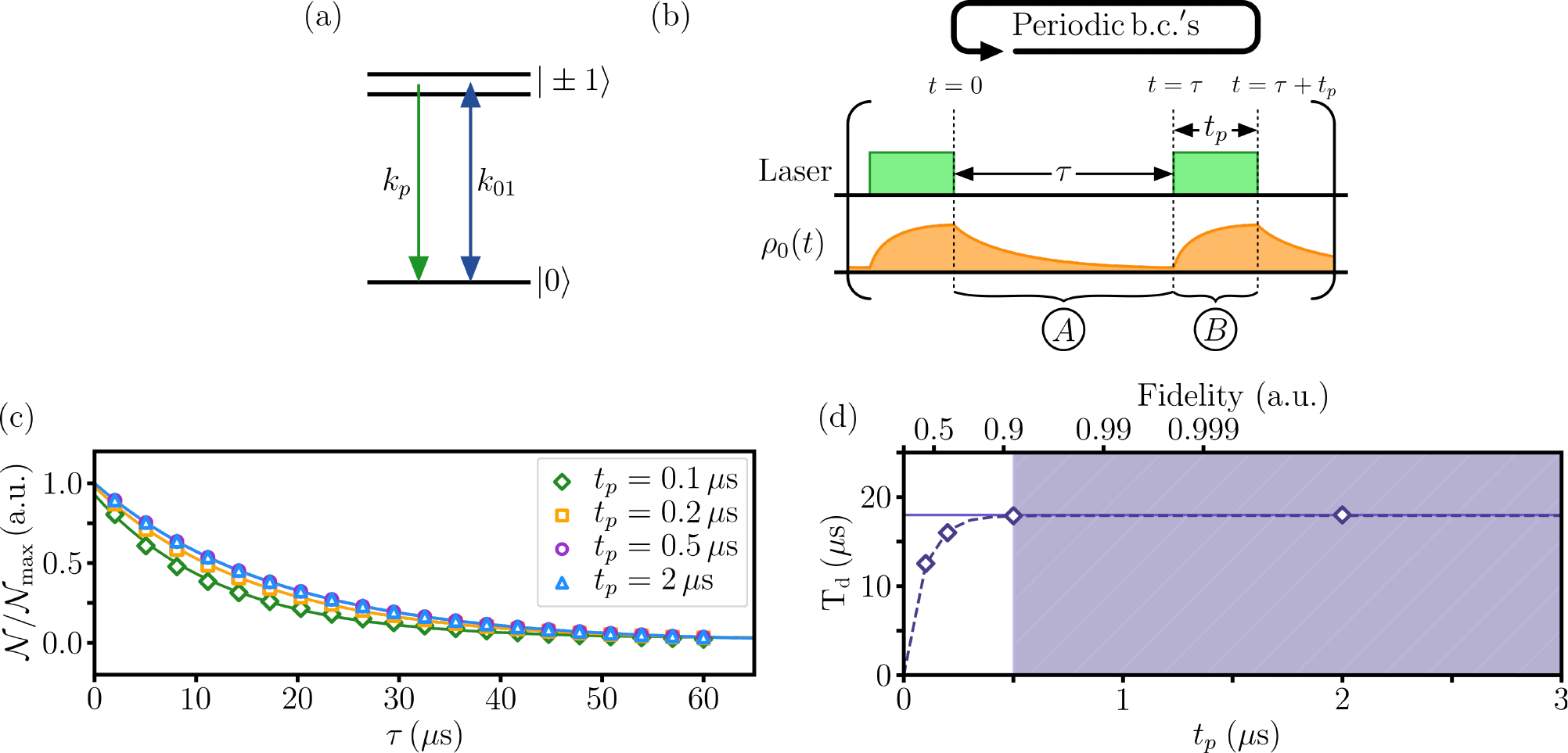}
\caption{(a) Simple two-level model used to describe the effect of finite spin initialisation. The two levels are coupled by the two-way transition rate $k_{01}$ defining the spontaneous spin relaxation rate and thus $T_1$. When the laser is on there is an additional pumping rate $k_p$ which drives the transition $\ket{\pm 1} \rightarrow \ket{0}$. (b) Schematic representation of the timing for the laser pulses and the effect on the population of the $\ket{0}$ state $\rho_0(t)$. The regions labelled $A$ and $B$ are the areas where we solve the dynamical rate equations by employing the indicated periodic boundary conditions. (c) Simulated relaxation curves plotted as $\mathcal{N}/\mathcal{N}_{\text{max}}$ versus $\tau$ for different values of the laser pulse duration $t_p$. (d) Plot of the decay time $T_d$ versus $t_p$ (bottom axis) and calculated initialisation fidelity (top axis) with $T_d$ being extracted from the curves in (c) by fitting with a mono-exponential. The solid line indicates the nominal $T_1$ value input into the model ($T_1 = 18$\,$\mu$s) while the dotted line is a separate exponential fit which acts as a guide to the eye which shows how $T_d$ converges on $T_1$ as $t_p$ is increased. Values for $t_p$ used in experiments were all within the shaded area.}
\label{Fig_LaserInit}
\end{figure*}

We consider a two-level model corresponding to $\ket{0}$ and the doublet $\ket{\pm 1}$ treated as a single level. When the laser is off, these two levels are coupled by the transition rate $k_{01}$ which mediates spontaneous spin relaxation and defines $T_1$ as discussed in Sec.~\ref{sec:dark}, where here the relation between the two is $1/T_1=2k_{01}$. Turning the laser on adds a pumping rate $k_p$ which drives the $\ket{\pm 1} \rightarrow \ket{0}$ transition [Fig.~\ref{Fig_LaserInit}(a)]. This pumping rate scales with the laser intensity used. We apply this model to calculate the population dynamics during our measurement sequence, which alternates a free decay part ($k_p=0$, duration $\tau$) and a pumping part ($k_p\neq 0$, duration $t_p$) where $t_p$ is the laser pulse duration [Fig.~\ref{Fig_LaserInit}(b)]. Note, while the model omits the overall photodynamics of the defect (e.g., the intersystem crossing) as well as possible charge dynamics effects, we expect it to provide a useful guide to choose an appropriate pulse duration $t_p$ allowing a faithful estimation of $T_1$.  

We first treat the two stages of the sequence [$A$ and $B$, see Fig.~\ref{Fig_LaserInit}(b)] separately, defining rate matrices $\Phi_A$ and $\Phi_B$ and solving the corresponding linear matrix rate equations to determine the time-dependent functions for the populations $\rho_0(t)$ and $\rho_{\pm 1}(t)$:
\begin{align}
  \begin{split}
    \Phi_A &= \begin{bmatrix}
-k_{01} & k_{01}\\
k_{01} & -k_{01}
\end{bmatrix}\\
&\implies
    \begin{cases}
      \rho_{0,A}(t) = C_0 + C_1 e^{-2k_{01}t}\\
      \rho_{\pm 1,A}(t) = C_0 - C_1 e^{-2k_{01}t}
    \end{cases}
  \end{split}
  \begin{split}
    \Phi_B &= \begin{bmatrix}
-k_{01} & k_{01} + k_p\\
k_{01} & -(k_{01} + k_p)
\end{bmatrix}\\
&\implies
    \begin{cases}
      \rho_{0,B}(t) = C_2\frac{k_{01} + k_p}{k_{01}} - C_3 e^{-(2k_{01} + k_p)t}\\
      \rho_{\pm 1,B}(t) = C_2 + C_3 e^{-(2k_{01} + k_p)t}
    \end{cases}
  \end{split}
\end{align}

To solve for the coefficients $C_n$, we first use the fact that the system is closed, such that $\rho_{0,A}(t) + \rho_{\pm 1,A}(t) = \rho_{0,B}(t) + \rho_{\pm 1,B}(t) = 1$. Second, since the sequence is repeated $N\sim1000$ times, we can apply periodic boundary conditions $\rho_{0,A}(0) = \rho_{0,B}(\tau + t_p)$, combined with the trivial condition $\rho_{0,A}(\tau) = \rho_{0,B}(\tau)$ [see Fig.~\ref{Fig_LaserInit}(b)].

We obtain:

\begin{align}
    \big[ \rho_{0,A}(t) \big]_{\text{sig.}} &= \frac{1}{2} + \frac{k_p}{2(2k_{01} + k_p)} \bigg(\frac{e^{-(2k_{01} + k_p)t_p} - 1}{e^{-(2k_{01} + k_p)t_p - 2k_{01}\tau} - 1} \bigg) e^{-2k_{01}t}\\
    \big[ \rho_{\pm 1,A}(t) \big]_{\text{sig.}} &= \frac{1}{2} - \frac{k_p}{2(2k_{01} + k_p)} \bigg(\frac{e^{-(2k_{01} + k_p)t_p} - 1}{e^{-(2k_{01} + k_p)t_p - 2k_{01}\tau} - 1} \bigg) e^{-2k_{01}t}\\
    \big[ \rho_{0,B}(t) \big]_{\text{sig.}} &= \frac{k_{01} + k_p}{2k_{01} + k_p} - \frac{k_p}{2(2k_{01} + k_p)} \Bigg( \frac{e^{-2k_{01}\tau} - 1}{e^{-(2k_{01} + k_p)\tau} \big(e^{-(2k_{01} + k_p)t_p - 2k_{01}\tau} - 1 \big) } \Bigg) e^{-(2k_{01} + k_p)t}\\
    \big[ \rho_{\pm 1,B}(t) \big]_{\text{sig.}} &= \frac{k_{01}}{2k_{01} + k_p} + \frac{k_p}{2(2k_{01} + k_p)} \Bigg( \frac{e^{-2k_{01}\tau} - 1}{e^{-(2k_{01} + k_p)\tau} \big(e^{-(2k_{01} + k_p)t_p - 2k_{01}\tau} - 1 \big) } \Bigg) e^{-(2k_{01} + k_p)t}
\end{align}

Note these equations are representative of the `signal' collection where no MW pulse is applied. Similarly, to obtain equations for the `reference' measurement, the MW pulse is assumed to perfectly swap the populations of the $\ket{0}$ and $\ket{\pm1}$ states. For simplicity it is assumed this process is instantaneous and occurs at $t=\tau$. Re-solving for the coefficients with the now new condition $\rho_{0,B}(\tau) = \rho_{\pm 1,A}(\tau)$, new rate equations which describe the reference measurement are obtained, giving:
\begin{align}
    \big[\rho_{0,B}(t) \big]_{\text{ref.}} &= \frac{k_{01} + k_p}{2k_{01} + k_p} - \frac{k_p}{2(2k_{01} + k_p)} \Bigg( \frac{e^{-2k_{01}\tau} + 1}{e^{-(2k_{01} + k_p)\tau} \big(e^{-(2k_{01} + k_p)t_p - 2k_{01}\tau} + 1 \big) } \Bigg) e^{-(2k_{01} + k_p)t}\\
    \big[\rho_{\pm 1,B}(t) \big]_{\text{ref.}} &= \frac{k_{01}}{2k_{01} + k_p} + \frac{k_p}{2(2k_{01} + k_p)} \Bigg( \frac{e^{-2k_{01}\tau} + 1}{e^{-(2k_{01} + k_p)\tau} \big(e^{-(2k_{01} + k_p)t_p - 2k_{01}\tau} + 1 \big) } \Bigg) e^{-(2k_{01} + k_p)t}
\end{align}

Now, we express the PL traces $S(\tau)$ and $R(\tau)$ in accordance with Eq.~\ref{PL_eqn}, but here integrating the populations over the laser pulse duration:
\begin{align}
    S(\tau) &= I_{\text{B}} + \frac{1}{t_p}\int_0^{t_p}dt\left( A_0 \big[\rho_{0,B}(t) \big]_{\text{sig.}} + A_1 \big[\rho_{\pm 1,B}(t) \big]_{\text{sig.}} \right)\\
    R(\tau) &= I_{\text{B}} + \frac{1}{t_p}\int_0^{t_p}dt\left(A_0 \big[\rho_{0,B}(t) \big]_{\text{ref.}} + A_1 \big[\rho_{\pm 1,B}(t) \big]_{\text{ref.}}\right)
\end{align}

Finally, we can express the normalised PL as defined by Eq.~\ref{eq:N1}, i.e. the PL difference ${\cal N}=S-R$. This gives: 
\begin{align}
    {\cal N}(\tau) &= \frac{k_p(A_0 - A_1)}{t_p(2k_{01} + k_p)^2}\frac{\big(1 - e^{-(2k_{01} + k_p)t_p} \big)^2}{1 - e^{-2(2k_{01} + k_p)t_p - 4k_{01}\tau}} e^{-2k_{01}\tau}
\end{align}

To numerically test the influence of the laser pulse duration $t_p$, in Fig.~\ref{Fig_LaserInit}(c) we plot ${\cal N}(\tau)$ curves with different values of $t_p$. The parameters were chosen as $T_1=1/2k_{01}=18\,\mu$s and  $k_p=5.1$\,MHz, the latter corresponding to a typical experimental value, as we will see below. By fitting the curves in Fig.~\ref{Fig_LaserInit}(c) with an exponential decay $\exp(-\tau/T_d)$, we can extract the apparent characteristic decay time $T_d$, which is plotted against $t_p$ in Fig.~\ref{Fig_LaserInit}(d). As $t_p$ is increased the value for $T_d$ increases until it reaches a maximal asymptote equal to the true $T_1$.  Thus, the measured $T_d$ will always be less than $T_1$ unless $k_p t_p$ is sufficiently large. However, for $t_p\geq0.5$\,$\mu$s as used experimentally, the relative error $\varepsilon=\frac{T_1-T_d}{T_1}$ is negligible, $\varepsilon\leq0.01\%$.

\begin{figure*}[b!]
\centering
\includegraphics[width=0.4\textwidth]{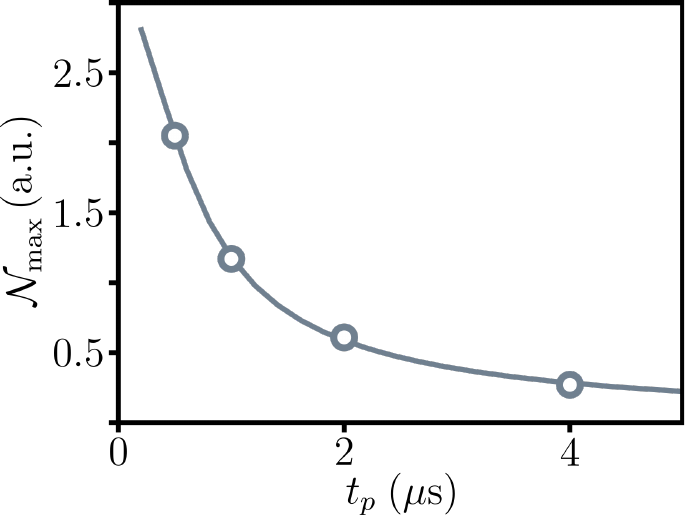}
\caption{Experimental data of the spin contrast measured at different laser pulse durations. The data is fit with Eq.~\ref{eq:contrast} in order to extract a value for the pumping rate $k_p$.}
\label{Fig_Contrast}
\end{figure*}

To estimate the experimental pumping rate $k_p$ at a given laser intensity, we use the fact that the maximum PL difference ${\cal N}_{\rm max}={\cal N}(\tau=0)$ scales with $t_p$ as
\begin{align}
\label{eq:contrast}
    {\cal N}_{\rm max} = \frac{A_0 - A_1}{k_p t_p} \frac{1 - e^{-k_p t_p}}{1 + e^{-k_p t_p}},
\end{align}
where to obtain a workable expression, we assume that $k_p \gg k_{01}$ which amounts to neglecting spin relaxation during laser pumping, which is a valid assumption as will be justified below. Thus, by measuring ${\cal N}_{\rm max}$ as a function of $t_p$ and fitting Eq.~\ref{eq:contrast} to the data, we can estimate $k_p$ for our experimental conditions. An example measurement of ${\cal N}_{\rm max}$ vs $t_p$ taken in the conditions of main text Fig. 2 is shown in Fig.~\ref{Fig_Contrast}. Here ${\cal N}_{\rm max}$ was taken as the PL difference at $\tau=2\,\mu$s rather than at $\tau=0$ to ensure full relaxation to the electronic ground state, but numerically calculating ${\cal N}(\tau=2\,\mu$s) shows the correction is negligible. Fitting Eq.~\ref{eq:contrast} to the data in Fig.~\ref{Fig_Contrast} gives $k_p=5.1\pm0.5$\,MHz, which satisfies $k_p\gg k_{01}$ validating the above assumption. This is the value of $k_p$ that was used in the prediction Fig.~\ref{Fig_LaserInit}(d).
        
We can also define an initialisation fidelity 
\begin{align}
    \mathcal{F} & \coloneqq 2\rho_{0,B}(t=t_p) - 1\\
    &= 1-e^{-k_p t_p}
\end{align}
where $\rho_{0,B}(t=t_p)$ is the population at the end of the laser pulse when starting from an unpolarised state, i.e. $\rho_0(t=0)=1/2$. With $k_p=5.1$\,MHz as determined before, we obtain ${\cal F}=0.93$ for $t_p=0.5\,\mu$s, for instance. This initialisation fidelity directly dictates the relative error $\varepsilon$ made in the estimation of $T_1$. In all the $T_1$ measurements presented in the paper, we used a fidelity of ${\cal F}>0.9$, ensuring that the associated error was $\varepsilon<0.01\%$, well below the statistical noise (i.e. the uncertainty from the fit).

\section{$T_1$ temperature dependence}
\label{sec:temp}

\begin{figure*}[b!]
\centering
\includegraphics[width=0.8\textwidth]{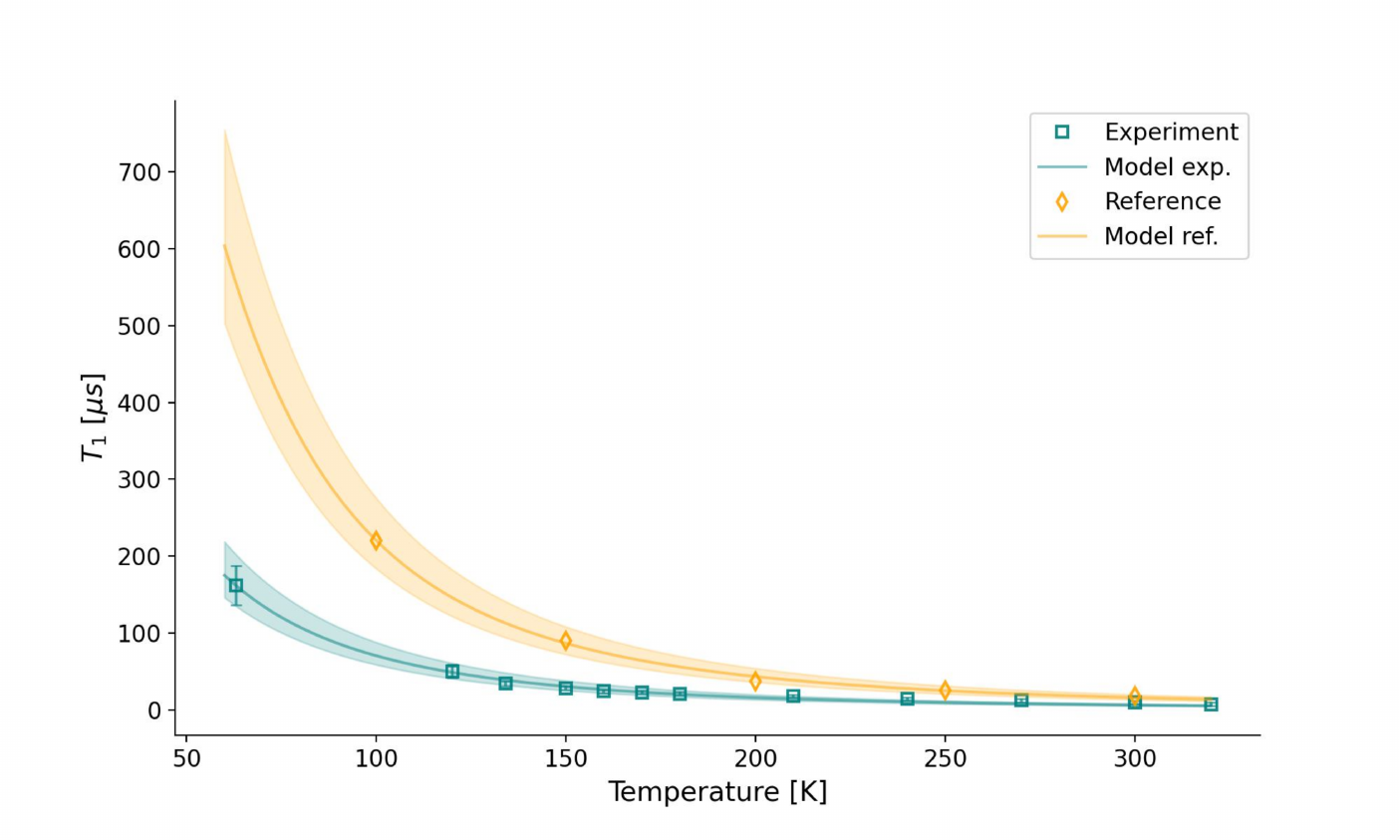}
\caption{Temperature dependence of $T_1$ of powder 2 (green squares) in comparison with the data from Ref.~\cite{GottschollSA2021} obtained with a neutron-irradiated bulk hBN crystal (orange diamonds). Solid lines are a power law fit, with the shading indicating one standard error.}
\label{Fig_TempDep}
\end{figure*}

As a part of our initial investigation regarding $T_1$ measurements in our nanopowder samples, we also measured the $T_1$ dependence on temperature. Measurements were performed using a custom-built wide-field fluorescence microscope which is constructed around a closed-cycle cryostat (Attocube attoDRY1000), described in detail in Ref. \cite{Lillie2020}. The base temperature of the cryostat is about $5$\,K, and the sample temperature can be increased to up to room temperature using a heater. 

The measurement conditions were similar to the room-temperature measurements as described in Sec.~\ref{sec:setup} and \ref{sec:acquisition} except for the laser power, 50 mW here, resulting in a smaller laser intensity. The powder was dropcast on to a PCB with a MW waveguide in similar fashion to our other samples before being loaded into the cryostat for measurement. 

The results of $T_1$ versus temperature are shown in Fig.~\ref{Fig_TempDep} over the range $T=70-325$\,K. In these measurements, $T_1$ was found to increase from $T_1=15 \pm 3$\,$\mu$s at room temperature to $T_1=160 \pm 30$\,$\mu$s at $T=70$\,K. We did not measure below 70 K because the required acquisition time became impractically long as $T_1$ got longer. This temperature dependence indicate that $T_1$ is governed by a thermally activated process, but further work will be required to elucidate the exact mechanism. For comparison, in Fig.~\ref{Fig_TempDep} we also plotted the data from Ref.~\cite{GottschollSA2021} which was obtained with a neutron-irradiated bulk hBN crystal. At low temperatures, the $T_1$ times we measure in our powders are in general shorter and do not exhibit as strong of an overall dependence on temperature below $250$\,K. However, at room temperature the $T_1$ of our powders is similar to the $T_1\approx 18$\,$\mu$s value reported in Ref.~\cite{GottschollSA2021}, and can even exceed this value as was the case for powder 1 as reported in the main text. 

We also note that the temperature dependence implies that any heating effect during the measurements (caused by laser or MW absorption by the sample) may affect the measured $T_1$ value. For instance, $T_1$ will be about 20\% shorter if the temperature is raised from 300\,K to 320 \,K, according to Fig.~\ref{Fig_TempDep}. Using pulsed-ODMR spectroscopy and monitoring the shift of the spin resonance frequency $\omega_0$, it is possible to estimate the temperature experienced by the V$_{\rm B}^-$ defects~\cite{Gottscholl2021NC,Liu2021}. In Fig. 4(a) of the main text, the temperature increase was estimated to be $\approx70$\,K. However, in conditions similar to those of our $T_1$ measurements, in particular with a minimum delay of $\tau\geq2\,\mu$s between the laser pulse and the MW probe pulse, we find the heating to be generally less than $\approx20$\,K above room temperature at $\tau=2\,\mu$s, with further cooling at longer $\tau$. Therefore, $T_1$ is expected to be shortened by 20\% at most as a result of heating, compared to the room-temperature value. This effect is one of the major sources of uncertainty in our experiments, and may explain the variability sometimes observed between $T_1$ measurements, since the local sample temperature will be sensitive to the specific experimental conditions.      

\section{Determining the expected $T_1$ quenching}

\begin{figure*}[b!]
\centering
\includegraphics[width=0.8\textwidth]{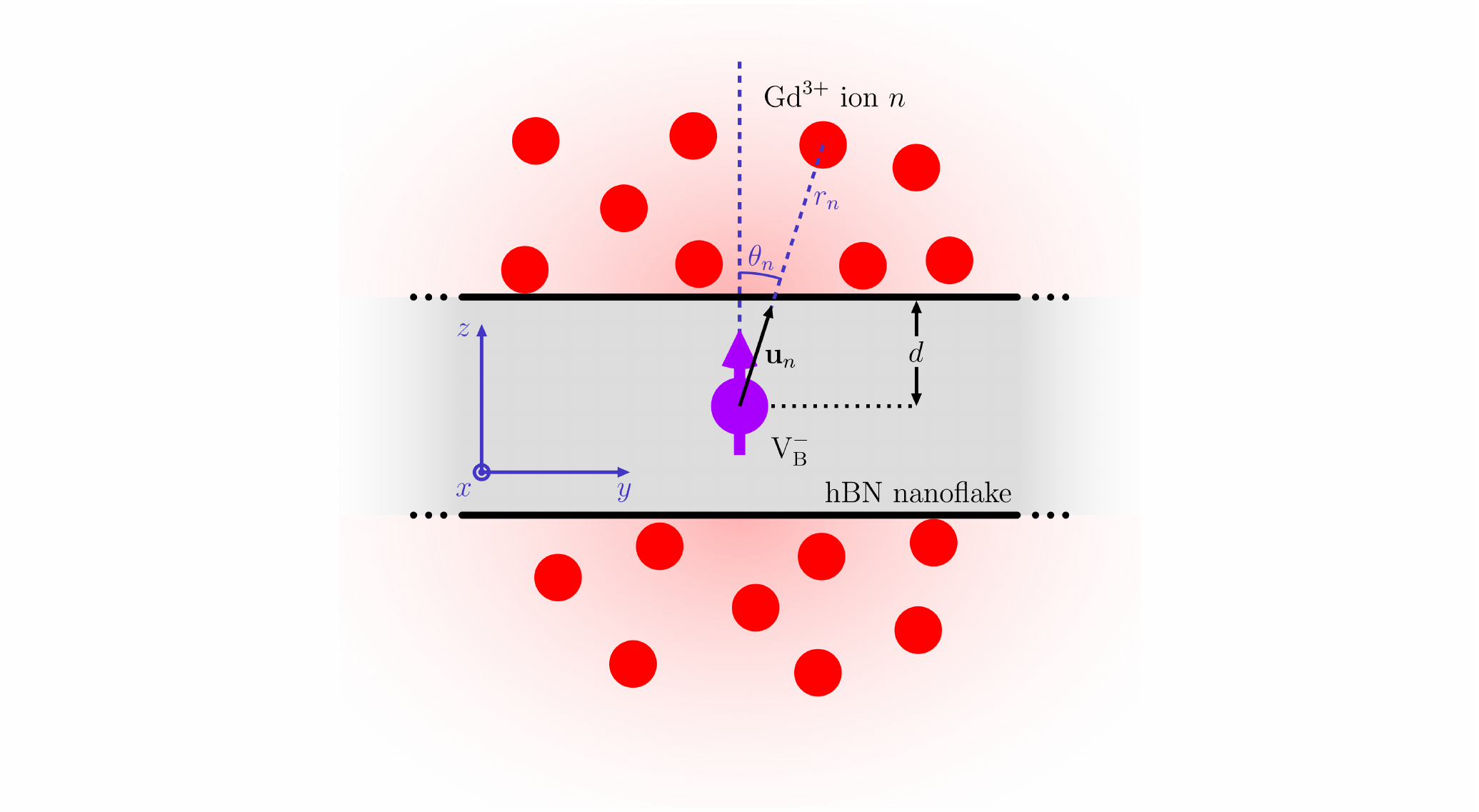}
\caption{Schematic representation of the geometry for which the variance $B_\perp^2$ of the transverse magnetic field experienced at the V$_{\rm B}^-$ from a bath of external paramagnetic spins (Gd$^{3+}$) is derived. The V$_{\rm B}^-$ is assumed to exist at the centre of a hBN nanoflake with thickness $2d$, which is much smaller than the lateral dimensions (i.e. an infinite slab of hBN with finite thickness). We define the principle axis of the V$_{\rm B}^-$ to be along the $z$-axis and also describe the locations of the Gd$^{3+}$ spins relative to the centre of the defect. The problem is assumed to be symmetric about the defect in the $x$ and $y$ directions.}
\label{Fig_QuenchMod}
\end{figure*}

Here we model the expected contribution to the total relaxation rate from the external paramagnetic ions. 
As stated in the main text, the total relaxation rate ($\Gamma_1$) is the sum of both the intrinsic ($\Gamma^{\rm int}_1$) and external magnetic noise ($\Gamma^{\rm ext}_1$) contributions. The latter is caused by randomly fluctuating magnetic fields and is thus described by~\cite{Steinert2013,Tetienne2013}
\begin{equation}
    \Gamma^{\rm ext}_1 = 3\gamma_e^2 B_{\perp}^2 \frac{\tau_c}{1 + \omega_0^2 \tau_c^2}
\end{equation}
where $\gamma_e$ is the electron gyromagnetic ratio, $B_{\perp}^2$ is the variance in the fluctuating transverse dipolar magnetic field  which constitutes the noise source,  $\tau_c$ is its correlation time, and $\omega_0$ is the resonant frequency of the V$_{\rm B}^-$ triplet ground state. We can use this equation to model the expected quenching effect by first finding an expression for $B_{\perp}^2$. The dipolar field radiated by a spin $\mathbf{S}_n$ at a point in space $\mathbf{r}_n$ is
\begin{equation}
    \mathbf{B}_n = \frac{\mu_0 \gamma_e \hbar}{4\pi r_n^3} \big[\mathbf{S}_n - 3(\mathbf{S}_n \cdot \mathbf{u}_n)\mathbf{u}_n \big],
\end{equation}
where $\mathbf{u}_n = \mathbf{r}_n/r_n$. For a fluctuating (paramagnetic) spin, we assume all projections of the spin are equally probable and so take the trace over a purely mixed state described by the density matrix $\rho = \frac{1}{2S+1}\mathbf{1}_{2S+1}$, 
\begin{equation}
    B_{\perp,n}^2 = \langle B_{x,n}^2 \rangle + \langle B_{y,n}^2 \rangle = \text{Tr} [\rho(B_{x,n}^2 + B_{y,n}^2)] = \bigg( \frac{\mu_0 \gamma_e \hbar}{4\pi} \bigg)^2 C_S \frac{2 + \sin^2{\theta_n}}{r_n^6}
\end{equation}
where $C_S={\rm Tr}[\rho S _{i,n}^2] = \frac{1}{2S+1} \sum_{m=-S}^S m^2 = \frac{S(S+1)}{3}$ with $i=x,y,z$, and $\theta_n$ is the angle between the $z$-axis (V$_{\rm B}^-$ quantisation axis) and $\mathbf{r}_n$. For a slab of hBN with thickness $2d$ and containing V$_{\rm B}^-$ at a depth $d$ [Fig.~\ref{Fig_QuenchMod}], assuming a density of external spins $\rho_{\text{ext}}$ occupying all space external to the slab of hBN, we can obtain the total transverse field by summing over all external paramagnetic ions, 
\begin{align}
    B_\perp^2 &= \sum_n B_{\perp,n}^2 = \bigg( \frac{\mu_0 \gamma_e \hbar}{4\pi} \bigg)^2 \rho_{\text{ext}} C_S \int_0^{2\pi} {\rm d}\phi \int_0^{\frac{\pi}{2}} {\rm d}\theta \sin{\theta} \int_{\frac{d}{\cos{\theta}}}^{\infty} {\rm d}r \frac{2 + \sin^2{\theta}}{r^4}\\
    &= \bigg( \frac{\mu_0 \gamma_e \hbar}{4\pi} \bigg)^2 \frac{\pi \rho_{\text{ext}} C_S}{3 d^3}.
\end{align}

In our initial quenching experiment, we assume the solution of GdCl$_3$ percolates through the hBN powder film and surrounds individual flakes of hBN before evaporation which leaves behind crystalline GdCl$_3$. The crystal has a density of $4520$\,kg\,m$^{-3}$ and molar mass $0.236$\,kg\,mol$^{-1}$ giving a Gd$^{3+}$ density of $\rho_{\text{ext}} = 1.03 \times 10^{28}$\,m$^{-3}$ with each Gd$^{3+}$ contributing $S = \frac{7}{2}$. In the optimal case where there is maximal coupling between the noise source and the V$_{\rm B}^-$ defect then $\tau_c = \frac{1}{\omega_0}$ with $\omega_0\approx 2\pi\times 3.45$~GHz. For a flake of thickness $2d = 10$\,nm, and given there is an effectively infinite volume of Gd$^{3+}$ ions on both sides of the flake (thus we double the magnetic noise) we calculate $\Gamma^{\rm ext}_1 = 27$\,kHz which approximately matches our experimentally measured value of $30$\,kHz. We note for this calculation we have chosen an idealised value for the correlation time $\tau_c$ of the noise source; however, we have compensated by using a conservative value for the thickness of the hBN flake which is over one standard deviation larger than what the AFM data in Sec.~\ref{sec:prep} indicates. If instead we assume the mean thickness of $2d=6$\,nm, we can relax the optimised choice on $\tau_c$ and reduce or increase its value by a factor of $\approx10$ to obtain the same $\Gamma^{\rm ext}_1$, i.e. $\tau_c\approx\frac{10}{\omega_0}$ or $\tau_c\approx\frac{1}{10\omega_0}$. Furthermore this model assumes a single V$_{\rm B}^-$ defect located at the middle of a flake which is the worst case scenario in terms of sensitivity. The electron irradiation process should populate a hBN flake with multiple defects distributed throughout and thus would allow us to further relax the conditions on $\tau_c$.

\end{document}